\documentclass[proof]{WileyASNA-v1}
\usepackage{amsmath}
\usepackage{hyperref}
\usepackage{xcolor}
\hypersetup{
    colorlinks=true,
    linkcolor=blue,
    citecolor=blue,
    filecolor=blue,      
    urlcolor=blue,
}

\title{Repeating Nuclear Transients from Repeating Partial Tidal Disruption Events}
\articletype{Proceedings}

\received{15 November 2025}
\revised{22 January 2026}
\accepted{26 January 2026}

\begin{document}

\author[1]{Ananya Bandopadhyay} 

\author[1]{Eric R.~Coughlin}

\author[1]{Julia Fancher}

\author[2]{C. J.~Nixon} 

\author[3]{Dheeraj R.~Pasham}

\authormark{Bandopadhyay \textsc{et al}}

\address[1]{\orgdiv{Department of Physics}, \orgname{Syracuse University}, \orgaddress{Syracuse, \state{NY 13210}, \country{USA}}}

\address[2]{\orgdiv{School of Physics and Astronomy}, \orgname{University of Leeds}, \orgaddress{ Sir William Henry Bragg Building, Woodhouse Ln, \state{Leeds LS2 9JT}, \country{UK}}}

\address[3]{\orgname{MIT Kavli Institute for Astrophysics and Space Research}, \orgaddress{Cambridge, \state{MA 02139}, \country{USA}}}

\corres{*Ananya Bandopadhyay \email{abandopa@syr.edu}}

\abstract{Extragalactic nuclear transients that exhibit repeating outbursts can be modeled as the repeated dynamical interaction between bound stars and supermassive black holes (SMBHs). A subset of these transients, with recurrence timescales of months-to-years, have been explained as accretion flares from the repeated tidal stripping of a star by an SMBH, in a repeating partial tidal disruption event (rpTDE). We outline the scope of the rpTDE model and discuss hydrodynamical simulations and analytical predictions for the stability of stars undergoing repeated mass loss, and the long-term evolution of these flares as a function of stellar type and orbital parameters. Our findings demonstrate that high-mass and centrally concentrated stars undergo negligible changes in structure in response to small amounts ($\sim 1-10\% M_\star$) of mass loss, and can survive many mass-stripping encounters with an SMBH. Contrarily, low-mass and less evolved stars are unstable to mass loss, and would be destroyed within a few orbits. We discuss the implications of these results for constraining the stellar type and orbital parameters of observed sources, such as ASASSN-14ko, for which $\gtrsim 20$ flares have been observed, and AT2020vdq, which exhibits a second flare that is brighter than its primary outburst.}


\maketitle 

\protect\jnlcitation{\cname{A. Bandopadhyay, E.~R.~Coughlin, J.~Fancher, C.~J.~Nixon, and D.~R.~Pasham} (\cyear{2026}), \ctitle{Repeating Nuclear Transients from Repeating Partial Tidal Disruption Events}, \cjournal{Astron. Nachr.}, \cvol{2026;00:1--6}.}

\section{Introduction}
\label{sec:intro}
The standard picture of a tidal disruption event (TDE) involves the disruption of a star that enters the tidal sphere of a supermassive black hole (SMBH), which has a radius $r_{\rm t} = R_{\star}\left(M_{\bullet}/M_{\star}\right)^{1/3}$ (where $R_{\star}$ is the stellar radius, and $M_{\star}$ and $M_{\bullet}$ are the stellar and SMBH masses respectively), from a distance comparable to the sphere of influence of the SMBH (e.g., \citealt{hills75, rees88}). The latter corresponds to $\sim$ pc-scale distances from the SMBH, while $r_{\rm t}$ is on the order of hundreds of stellar radii, and hence the stars involved in TDEs are on effectively parabolic orbits. It has been known that stars entering just outside (or within, depending on the stellar properties) the tidal sphere are not completely destroyed, but are instead partially disrupted with only their outer layers removed (e.g., \citealt{khokhlov93}), but the energy imparted to the star via tides -- at most its own binding energy -- is insufficient to tightly bind the star to the SMBH \citep{cufari23}. Therefore, TDEs are expected to generate one, and only one, bright accretion flare that rises and fades on timescales of $\sim$ months to years.

It is therefore surprising that time domain surveys have recently detected \emph{repeating} nuclear transients, which appear to consist of two or more TDE-like flares that are separated by months\footnote{There are other, recently detected repeating transients from galactic nuclei that repeat on $\sim$ single-day timescales and emit only in soft X-rays, known as quasi-periodic eruptions (QPEs; e.g., \citealp{miniutti19,giustini20,arcodia21,chakraborty21,miniutti23,miniutti23b,arcodia24}); we discuss the possible connection of these systems to rpTDEs in Section \ref{sec:qpes}.} to years. These are referred to as repeating partial tidal disruption events (rpTDEs), and have been speculated to arise from the repeated partial stripping of a star on a tightly bound orbit, with the star placed on that orbit by a dynamical exchange process (i.e., Hills capture; \citealt{hills88, cufari22}). This process can generate recurrent accretion flares for as long as the star remains relatively structurally unchanged after undergoing mass loss. The rapidly expanding class of rpTDE candidates includes ASASSN-14ko \citep{payne21}, AT2018fyk \citep{wevers23}, eRASSt-J045650 \citep{liu24}, AT2020vdq \citep{somalwar23}, Swift J0230 \citep{evans23} AT2022dbl \citep{lin24,hinkle24,makrygianni25} and AT2021aeuk \citep{sun25}.

The energetics of the ASASSN-14ko transient are consistent with a star losing $\sim {\rm few} \times 0.01 M_\odot$ per pericenter passage, assuming the luminosity is roughly $L=0.1 \dot{M} c^2$ \citep{cufari22}, which would enable a star with $M_\star \gtrsim 1M_\odot$ to survive $\sim 100$ encounters on its orbit. However, if the mass loss per orbit is $\sim 0.001 M_\odot$, and a slowly increasing function of time, then a sun-like star can survive many more encounters, and its orbital period can be explained through other mechanisms, such as gravitational-wave emission or hydrodynamical drag as the star interacts with an accretion disk \citep{linial24}. However, a different mechanism then has to be invoked in order to perturb the star to a pericenter distance of $\sim 2 r_{\rm t}$, as gravitational-wave emission and tides both effectively preserve the pericenter distance while shrinking the semimajor axis \citep{peters64,cufari22,bandopadhyay24}. \cite{yao25} studied the less extreme case of stars orbiting SMBHs in which the pericenter distance is initially not close enough to strip off any mass. However, they showed that in this regime, tidal heating can significantly structurally alter the star such that it eventually overflows its Roche lobe and undergoes mass transfer.  

While the various models for repeating nuclear transients successfully address certain properties of these sources and are able to make predictions regarding the evolution of flares (e.g., \citealp{wevers23} predicted the timescale on which AT2018fyk would exhibit a second rapid dimming, based on the inference of the orbital period of the star around the SMBH, which was later verified by \citealt{pasham24}), several open questions about the formation and evolution of these sources and the fate of stars orbiting SMBHs on tightly bound orbits remain unanswered. Several works have raised concerns about the survivability of stars undergoing drastic structural changes in response to the cumulative tidal heating over multiple encounters, which would inflate the star, leading to runaway mass transfer and possible destruction over only a few orbits~\citep{linial24,liu24}. While the energetics of sources such as ASASSN-14ko are broadly consistent with the luminosity of observed outbursts assuming an accretion luminosity of $L=0.1 \dot{M} c^2$, the observed period derivative $\dot{P}=-0.0026$ \citep{payne22} is not well understood in terms of a plausible energy sink for the dissipated orbital energy. Additionally, while observations suggest a possible connection between TDEs and QPEs, with the latter arising from the interaction of stars with a pre-existing TDE disk (e.g., \citealp{shu18,sheng21,miniutti23b,arcodia24,nicholl24,chakraborty25,hernandez25}) or the repeated stripping of a white dwarf (\citealt{king20}; see Section \ref{sec:qpes} for brief, additional discussion along these lines), a definitive link between these different classes of repeating nuclear transients is yet to be established.

To address some of these issues, specifically in the context of rpTDEs, we analyzed the stability of mass transfer, and the survivability of a star, using numerical hydrodynamical simulations of the repeated partial tidal disruption of a range of main-sequence star on an orbit around an SMBH in~\cite{bandopadhyay24} (see also \citealp{liu25} for a similar study based on hydrodynamical simulations of repeated tidal interactions). These works showed that in order for a star to survive many mass-stripping encounters with an SMBH, it must have a centrally concentrated core-like structure that remains relatively unaffected as a result of mass transfer. Consequently, high mass and evolved stars, which have a dense core and a diffuse outer envelope, are the most promising candidates for generating transients like ASASSN-14ko, which has flared $\gtrsim20$ times. On the other hand, low mass and less evolved stars undergo a more substantial change in their internal structure, making them increasingly susceptible to unstable mass transfer and can thus generate a few flares with increasing peak luminosity (consistent with the two observed outbursts of AT2020vdq) before being completely destroyed. 

Completely mapping out the parameter space of stellar, SMBH and orbital parameters that can explain the differences in the features of observed rpTDE lightcurves with hydrodyamical simulations is prohibitively computationally expensive, and subject to questions regarding the accuracy and self-consistency of tidal dissipation. To surmount these issues, we developed a hybrid approach, based on a spherically symmetric adiabatic mass loss model, to understand the response of stars to the loss of small amounts of mass from their outer layers~\citep{bandopadhyay25}. The predictions of this model for different stellar progenitors corroborate the results of numerical simulations of rpTDEs, and specifically, we find that (i) high mass ($M_\star\gtrsim1.5 M_\odot$) stars (low mass stars) contract (expand) in response to removal of small amounts of mass from their outer envelope, making them less susceptible (more susceptible) to mass transfer on subsequent encounters with an SMBH, (ii) the change in the energy of the system as a result of mass transfer is primarily determined by the difference in the binding energy of the core and that of the original star (core and envelope), and the kinetic energy contained in the oscillations, i.e. the tidal heating term is several orders of magnitude smaller than the total change in energy, as well as the individual changes in the gravitational and thermal energies of the core, thus suggesting that tidal heating likely does not excessively heat the star or pose a significant challenge to the rpTDE model, and (iii) the rotational energy of the core following tidal spin-up (which is included in the numerical simulations, but not in our spherically symmetric mass loss model) is second in importance (following the $p\,dV$ work done by the expansion of the core) in determining the energy balance, and is of the order of $10\%$ of the difference in binding energy of the core and that of the initial star. 

In this proceeding, we summarize the results presented in \cite{bandopadhyay24,bandopadhyay25} in the context of their application to rpTDE modeling, as follows. In Section~\ref{sec:hydro}, we discuss the Smoothed Particle hydrodynamics (SPH) simulations of the repeated tidal stripping of main-sequence stars by an SMBH presented in \cite{bandopadhyay24}, which show that, by varying the stellar type and the impact parameter characterizing the
distance of the closest approach between the star and the SMBH, the rpTDE model can be used to qualitatively reproduce the lightcurves of observed transients, including ASASSN-14ko, AT2018fyk and AT2020vdq. In Section~\ref{sec:mass-loss-model}, we outline the framework for the spherically symmetric Lagrangian model developed in \cite{bandopadhyay25} to study the response of stars to mass loss, and use it to make predictions for main-sequence stars undergoing mass loss, for the time dependent radial oscillations, the change in average density, and the energy imparted as a result of the mass transfer. We discuss the implications of these results in the context of survivability of stars in rpTDE systems, and summarize and conclude in Section~\ref{sec:conclusion}.

\section{hydrodynamical Simulations}
\label{sec:hydro}
\subsection{Simulation Setup}
\begin{figure*}
    \centering
    \includegraphics[height=10.5cm,width=0.995\textwidth]{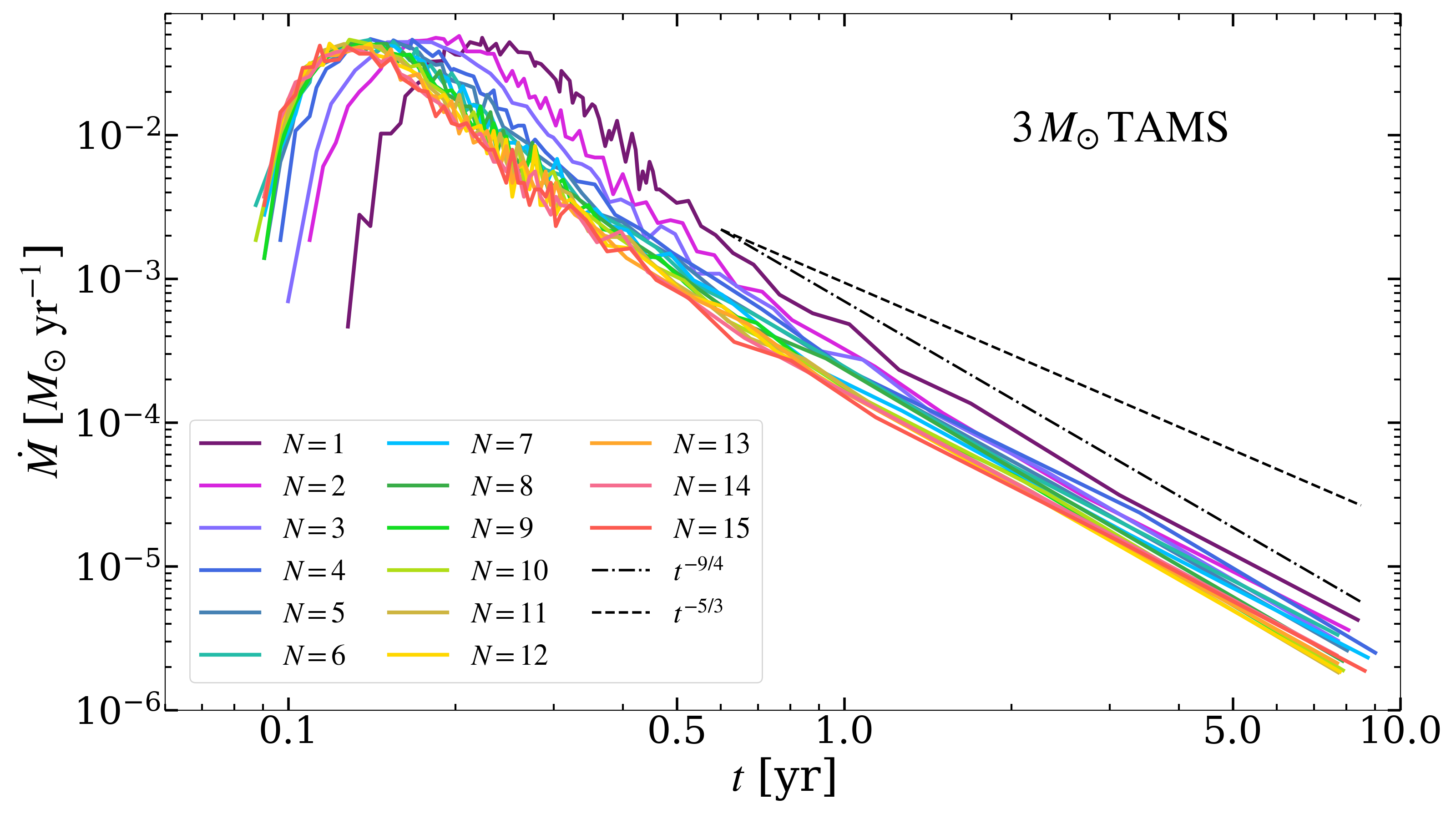}
    \caption{\hspace*{0.2cm}Fallback rates for the first fifteen pericenter passages of a $3M_\odot$ TAMS star on a $\beta=1$ orbit around a $10^6$ SMBH. The peak of the fallback rates remains relatively constant across all fifteen encounters, whereas the peak timescale $t_{\rm peak}$ progressively shifts towards earlier times.}
    \label{fig:3msun-fallbackrates}
\end{figure*}
To test the feasibility of the rpTDE model, and the survivability of stars being repeatedly stripped of mass, we performed hydrodynamical simulations of the repeated partial tidal disruption of a range of main-sequence stars by a $10^6 M_\odot$ SMBH~\citep{bandopadhyay24}, using the Smoothed-Particle hydrodynamics (SPH) code, {\sc phantom}~\citep{price18}. We used the stellar evolution code {\sc mesa}~\citep{paxton11,paxton13,paxton15,paxton18} to evolve stars from their zero-age main sequence (ZAMS; when the core hydrogen fraction is $\sim 0.7$) to the terminal-age main sequence (TAMS; when the core hydrogen fraction drops to $\lesssim 0.001$). The {\sc mesa} profiles were mapped to a three dimensional particle distribution in {\sc phantom} and relaxed using the routine implemented in~\cite{golightly19}. We used $10^6$ particles, and an adiabatic equation of state with an adiabatic index $\Gamma=5/3$ to model the star, with the details of numerical implementation being identical to that in~\cite{coughlin15}.

We initiated the disruption process by placing the center-of-mass (COM) of the star on a parabolic orbit around the SMBH, at an initial distance of $5 r_{\rm t}$ from the SMBH. The pericenter distance $r_{\rm p}$ (or equivalently, the impact parameter $\beta=r_{\rm t}/r_{\rm p}$) of the star was chosen such that the amount of mass stripped, $\Delta M$, is a small fraction of the stellar mass $M_\star$ (i.e., $\Delta M / M_\star <1$). Using the maximum gravity (MG) model developed in~\cite{coughlin22}, we can estimate the critical pericenter distance $r_{\rm t,c}$ at which a star is completely destroyed by the tides of an SMBH, by equating the maximum self-gravitational field that occurs in the interior of the star to the tidal field at that distance. The radius within the star at which its self-gravitational field is maximized is denoted as the ``core radius,'' $R_{\rm c}$. This gives as an estimate of the critical distance for complete disruption, $r_{\rm t,c}$, or equivalently, the critical impact parameter $\beta_{\rm c} \equiv r_{\rm t}/r_{\rm t,c}$ for any given stellar profile. Figure 4 in~\cite{bandopadhyay24b} shows the predicted $\beta_{\rm c}$ for a range of stellar structures. \cite{fancher25} tested the predictions of the MG model with an extensive set of hydrodynamical simulations, and found that the model predictions for $\beta_{\rm c}$ were adequate for the complete disruption of a variety of ZAMS stars, with masses ranging from $0.2-5.0 M_\odot$. For stars that are further along their main-sequence evolution, they found that the presence of a core modifies the debris dynamics and results in a partial disruption at $\beta = \beta_{\rm c}$ predicted by the model. Additionally, they argued that the failure to completely destroy a star at $\beta=\beta_{\rm c}$, as seen in their numerical simulations, indicates that the effective core radius for complete disruption is smaller than that obtained from the {\sc mesa} profile, as the tidal compression in the direction orthogonal to the orbital plane enhances the self-gravitational field of the star (cf. \citealp{nixon22}). This would imply that the actual value of $\beta_{\rm c}$ required to completely destroy such a star would be higher than the model prediction. \cite{fancher25} verified for a subset of their simulations (for middle-age main sequence (MAMS) stars) that this is indeed the case, and that all but one ($0.9 M_\odot$) of the MAMS stars considered in their work were completely destroyed at $\beta_{\rm c}+1$. 

Using the simulations presented in~\cite{fancher25} as a test of reliability of the MG model, we treat the $\beta_{\rm c}$ prediction from the model as an upper limit on the range of $\beta$'s considered in this work. However, we note that the actual upper limit on $\beta$ for the partial disruption of highly evolved stars is higher than that shown in Figure 4 of \cite{bandopadhyay24b}. To calculate the fallback rate of stellar debris onto the SMBH, we used the core-replacement procedure described in, e.g.,~\cite{miles20,nixon21} (since performing the hydrodynamical evolution of the core reduces the computational timestep but results in negligible differences in the fallback rate compared to that obtained by replacing the core with a sink particle), and tracked the rate at which material from the tidally disrupted debris stream returns to the accretion radius (set to $3 r_{\rm t}$ for any given star) of the SMBH. To simulate the subsequent pericenter passages, instead of evolving the star through the entirety of its highly eccentric orbit (which would be computationally intractible for realistic orbital eccentricities), we evolved the star on its orbit for $\sim 2$ days past pericenter passage, following which we translated the surviving core back to the initial position of the original star by subtracting the COM velocity and position from every SPH particle, and subsequently adding back (to each particle in the core) those of the original stellar orbit. This constitutes a faster and efficient way of computing the fallback rate of disrupted debris from multiple pericenter passages of a given star on its orbit around the SMBH, examples of which are presented in the following subsection.

\subsection{Results for a \texorpdfstring{$3M_{\odot}$}{Lg} TAMS star}
\label{subsec:3msun}
\begin{figure}
    \includegraphics[width=0.49\textwidth]{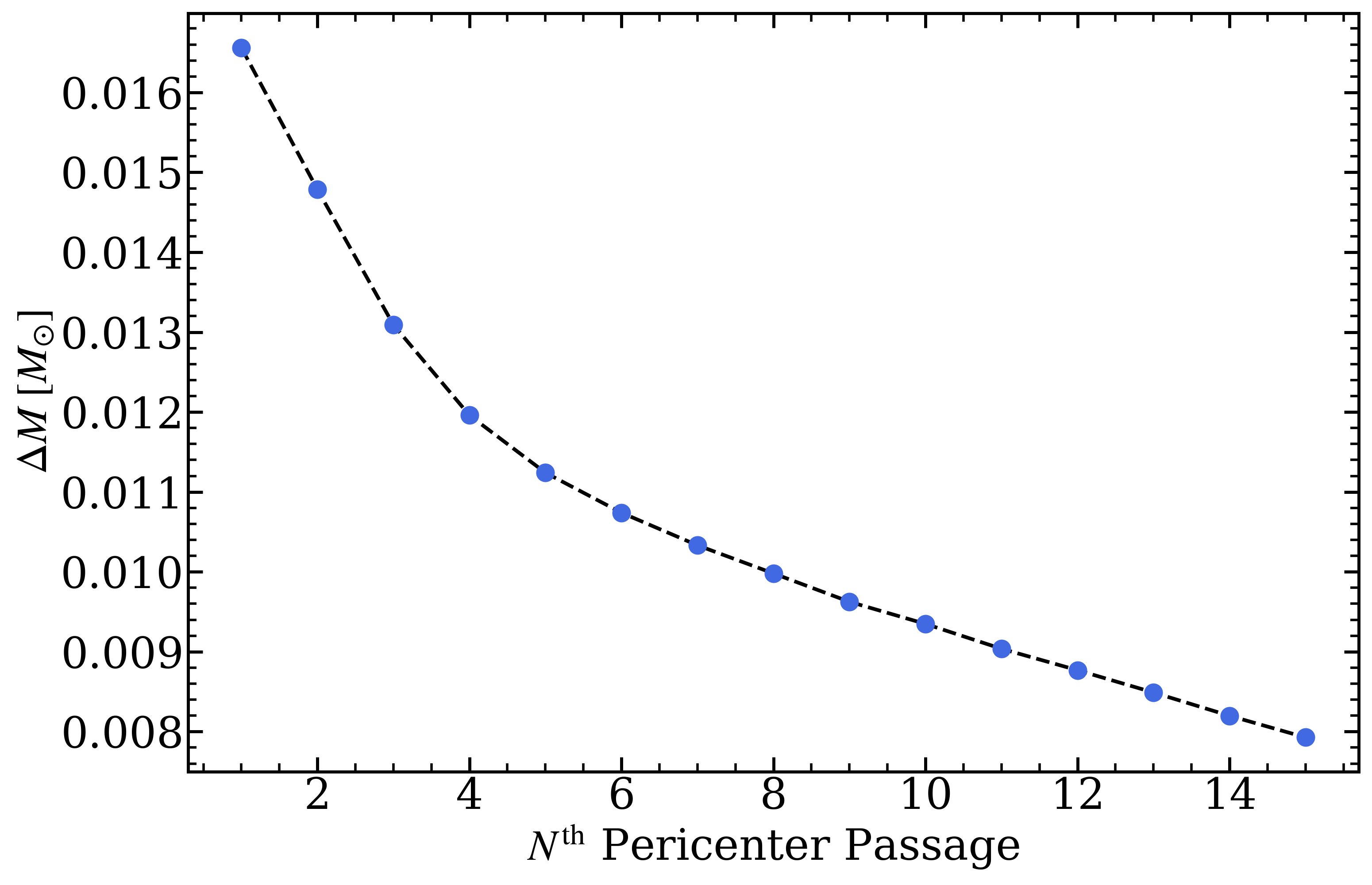}
     \caption{\hspace*{0.2cm}The amount of mass stripped $\Delta M$ for the first fifteen pericenter passages of the $3M_\odot$ TAMS star on a $\beta=1$ orbit.}
    \label{fig:mass_stripped_vsN}
\end{figure}

\begin{figure}
     \includegraphics[height=6cm,width=0.49\textwidth]{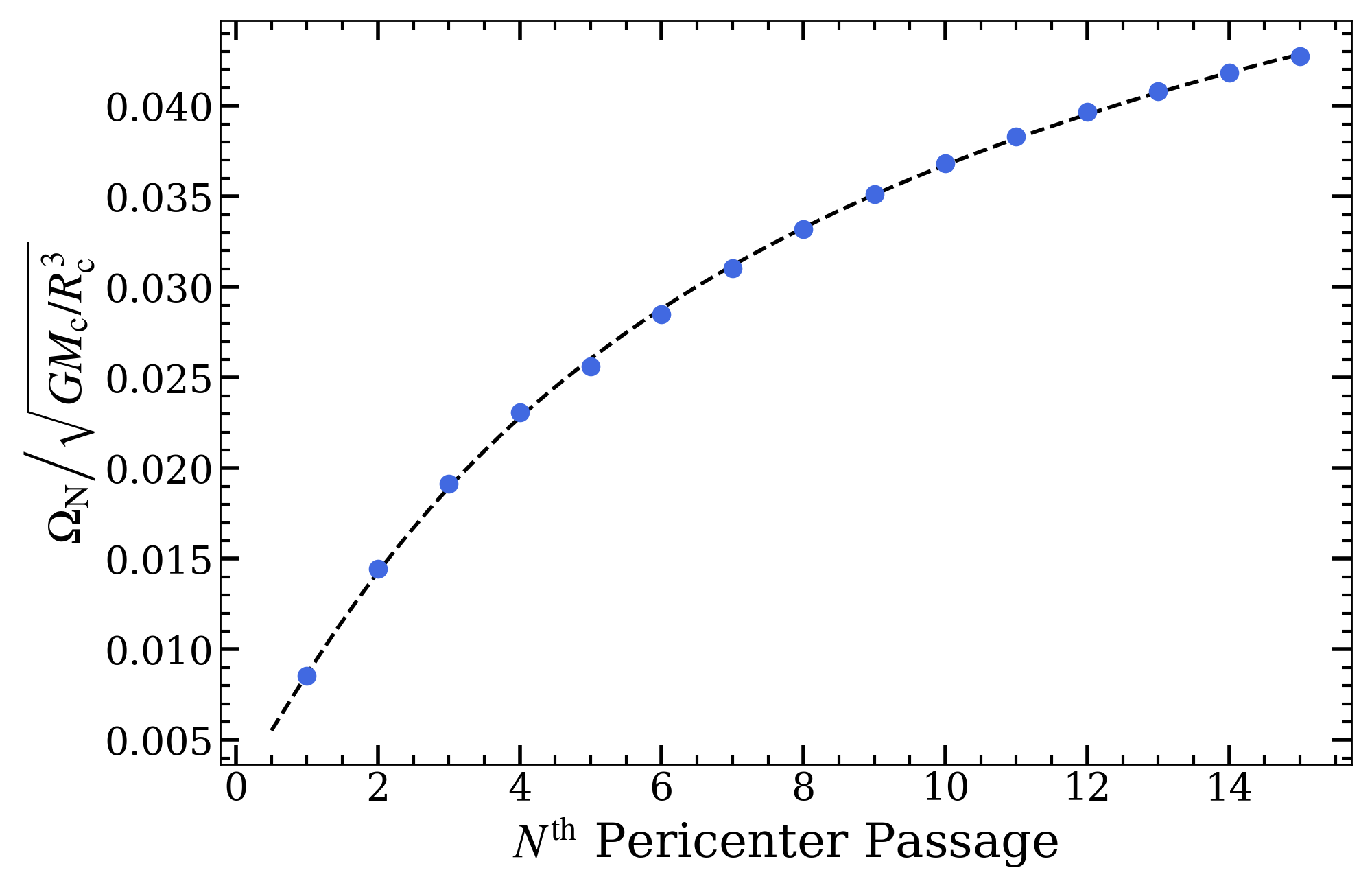}
    \caption{\hspace*{0.2cm}The angular velocity imparted to the core of the $3M_\odot$ TAMS star on a $\beta=1$ orbit, normalized by its break-up angular velocity $\sqrt{G M_{\rm c}/R_{\rm c}^3}$, for the first 15 pericenter passages.  }
    \label{fig:omegavsN}
\end{figure}
One of our main goals of this work was to analyze the feasibility of the rpTDE model in reproducing the observed characteristics of the lightcurve of ASASSN-14ko, which is one of the earliest discovered transients of this category. Specifically, to generate sustained flares with negligible evolution in their peak properties, over $\gtrsim20-30$ orbits of a star, the surviving core of the star would be required to remain relatively structurally unaffected over the entire duration over which the flares are observed. Thus, we simulated the disruption of a $3M_\odot$ TAMS star (which is one example of a high mass and highly evolved star, having a dense core and a diffuse outer envelope) on a $\beta=1$ orbit around a $10^6 M_\odot$ SMBH. The critical impact parameter required for the complete disruption of the star is $\beta_{\rm c} > 6$, and thus the per-orbit mass loss on a $\beta=1$ orbit is a negligible fraction of the stellar mass. Figure~\ref{fig:3msun-fallbackrates} shows the fallback rates from the first 15 pericenter passages of the star. The late time scaling of the fallback rates is proportional to $t^{-9/4}$, which is the expected trend for partial disruptions~\citep{coughlin19}. As seen in the figure, the peak magnitude of the fallback rates remains almost constant over all of the 15 encounters. However, the presence of a centrally concentrated core makes it increasingly difficult to strip off mass on subsequent pericenter passages, such that the amount of mass stripped per encounter is $\Delta M < 0.01 M_\star$, and is a declining function of the number of pericenter passages, $N$. The evolution of $\Delta M$ with $N$ is shown in Figure~\ref{fig:mass_stripped_vsN}. Despite the declining trend in the amount of mass stripped, the peak of the fallback rate remains roughly constant over the 15 encounters shown in Figure \ref{fig:3msun-fallbackrates}, owing to the fact that the surviving core of the star gets spun up (in a prograde sense with respect to the SMBH) through the tidal interaction. \cite{golightly19b} showed that, for a star spinning at a fraction of its breakup angular velocity, the rotation can be treated as a perturbation to the orbital energy of its COM. Using the Keplerian energy-period relation, they showed that this results in a change in the peak timescale for the fallback rate of stellar debris, which depends on the angular velocity as $t_{\rm peak} \propto (1+\sqrt{2}\Omega)^{-3/2}$ (where $\Omega$ is the angular velocity of the star, normalized by its breakup angular velocity). For a prograde spinning star, this results in a shorter $t_{\rm peak}$ value compared to a non-spinning star. Figure~\ref{fig:omegavsN} shows the average angular velocity of the stellar core (averaged over all the particles constituting the core), normalized by its breakup angular velocity, as a function of the number of pericenter passages, $N$, for the $3M_\odot$ TAMS star. As the core gets progressively spun up with an increase in $N$, this shifts the peak timescale of the fallback rate to shorter times, and maintains a roughly constant peak magnitude ($\dot{M}_{\rm peak} \sim \Delta M/t_{\rm peak}$) despite the declining amount of mass stripped.

\subsection{Results for a \texorpdfstring{$1M_{\odot}$}{Lg} ZAMS star}
\label{subsec:1msun}
\begin{figure}
    \includegraphics[width=0.49\textwidth]{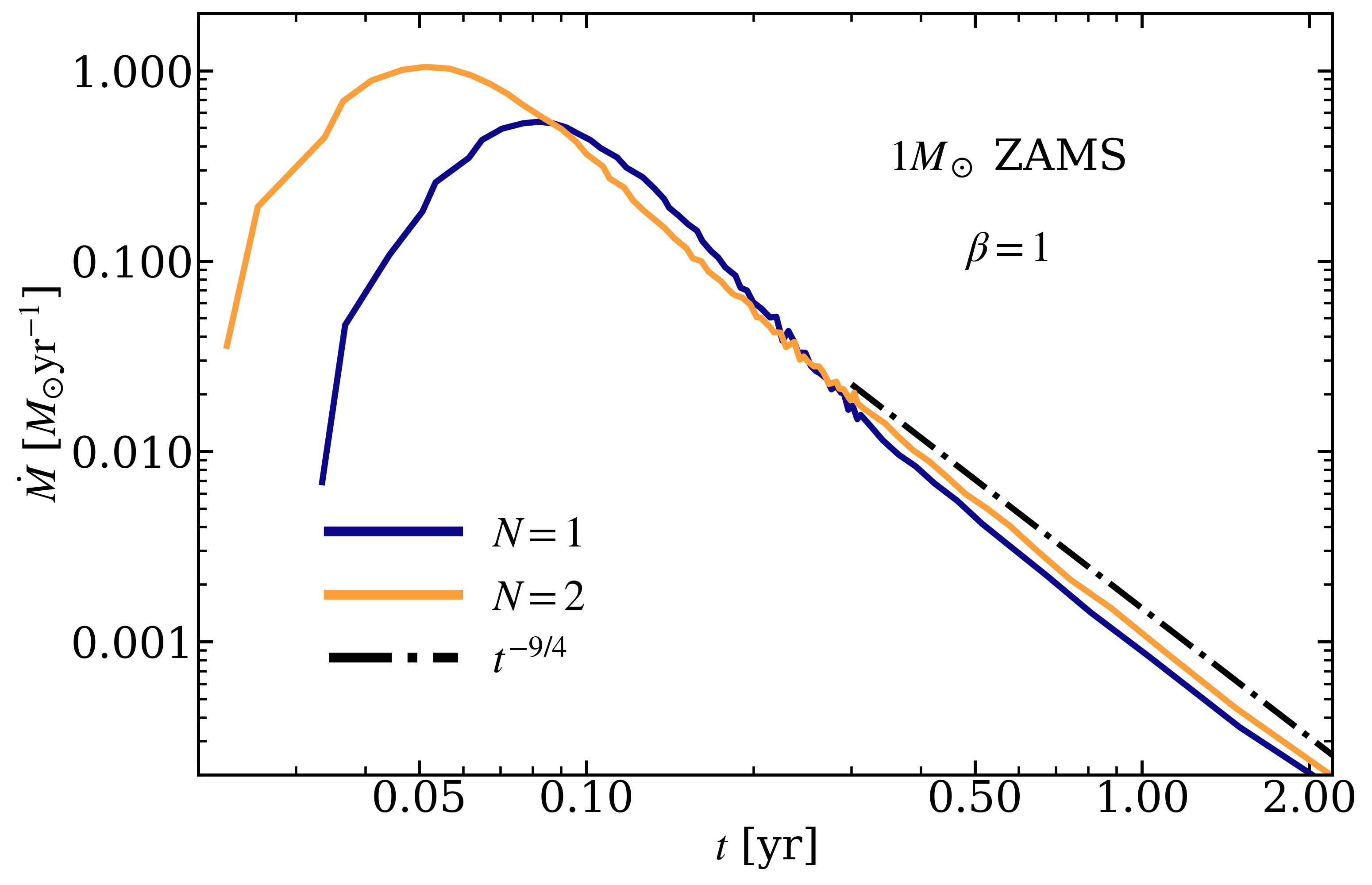}
     \caption{\hspace*{0.2cm}Fallback rates for the first two encounters of a $1 M_{\odot}$ ZAMS star on a $\beta=1$ orbit. The amount of mass stripped increases from $N=1$ to $N=2$, leading to a brighter second peak. }
    \label{fig:fbr-1msun-beta1}
\end{figure}

\begin{figure}
    \includegraphics[width=0.49\textwidth]{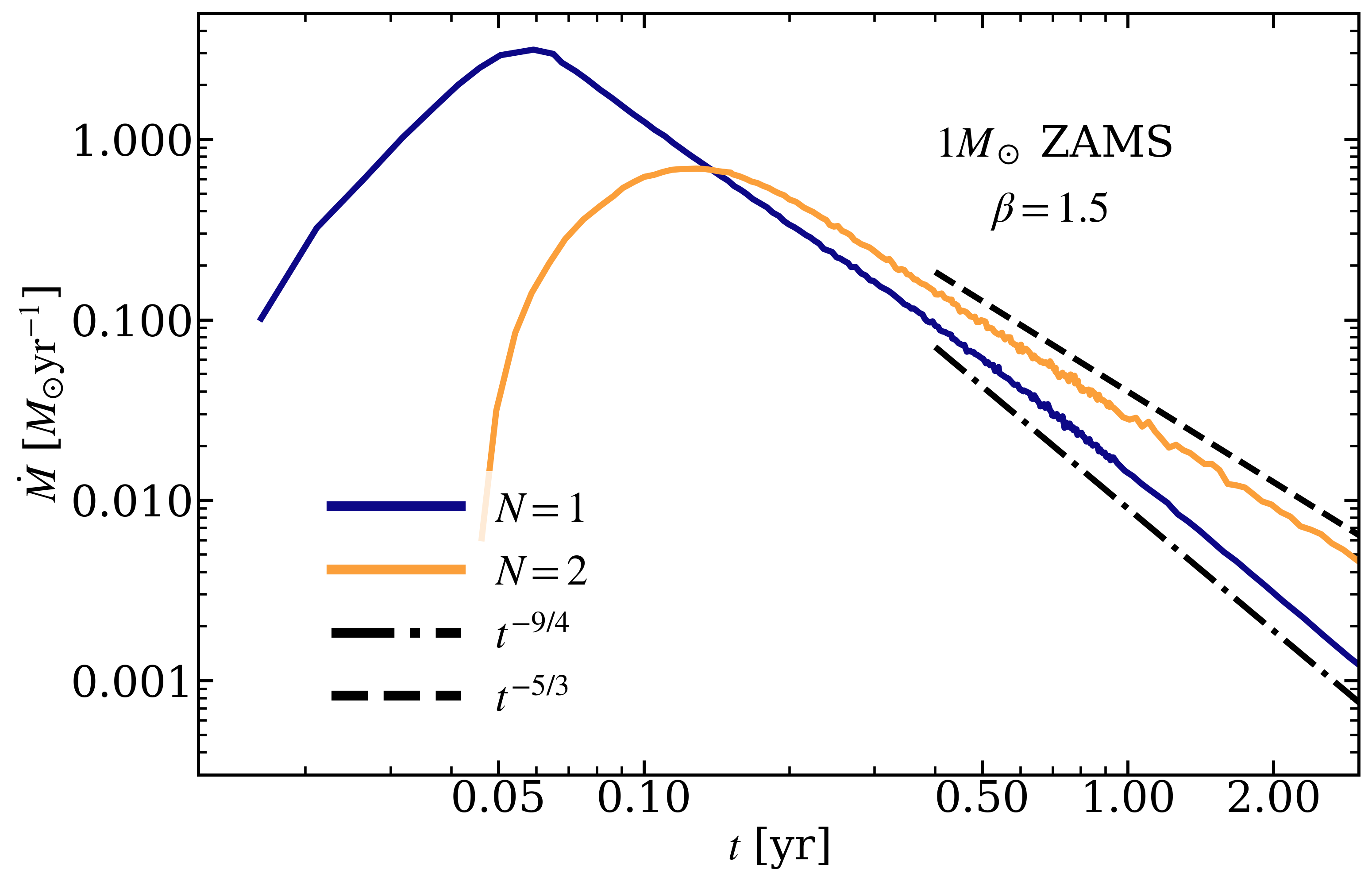}
     \caption{\hspace*{0.2cm}Fallback rates for the $1M_\odot$ ZAMS star on a $\beta=1.5$ orbit around a $10^6M_\odot$ SMBH. The star loses a significant fraction of its mass on its first pericenter passage ($\beta_{\rm c}$ for complete disruption being $\sim 1.8$), but yields a partial disruption, as indicated by the $\propto t^{-9/4}$ scaling of the fallback rate at late times. The star gets completely destroyed on its second encounter, (as indicated by the $\propto t^{-5/3}$ scaling) and yields a dimmer second peak.}
    \label{fig:fbr-1msun-beta1p5}
\end{figure}

The mass distribution for a sunlike star at its ZAMS stage is more uniform compared to the $3M_\odot$ TAMS star, described in the previous section, i.e., it does not have a distinct core-envelope separation. This makes it easier for the tidal interaction to strip off mass from the star, on an orbit having a comparable pericenter distance. Figure~\ref{fig:fbr-1msun-beta1} shows the fallback rates for the first 2 orbits of a $1M_\odot$ ZAMS star on a $\beta=1$ orbit around a $10^6 M_\odot$ SMBH. The tidal interaction strips off $\sim10\%$ of the star's mass on the first, pericenter passage, and an increasing amount of mass on subsequent pericenter passages, which is evident from the brighter second peak seen in the figure. Thus, the $1M_\odot$ star can only survive $\sim4$ encounters on a $\beta=1$ orbit, after which it is completely destroyed. The fallback rates shown in Figure~\ref{fig:fbr-1msun-beta1} are consistent with the observed lightcurve of AT2020vdq, for which the second peak was $\sim$half an order of magnitude brighter than the first~\citep{somalwar23}.

In contrast, Figure~\ref{fig:fbr-1msun-beta1p5} shows the fallback rates for the first two encounters of the $1M_\odot$ ZAMS star on a $\beta=1.5$ orbit around a $10^6 M_\odot$ SMBH. $\beta_{\rm c}$ for complete disruption of this star is $\sim1.80$ \citep{nixon21,coughlin22}, and thus a significant fraction of its mass is stripped during its first pericenter passage on a $\beta=1.5$ orbit. Consequently, the second pericenter passage yields a lower peak magnitude, and, as indicated by the late time scaling of $\dot{M} \propto t^{-5/3}$ for $N=2$, the star is completely destroyed in the second encounter. The dimmer second peak is consistent with the lightcurve of AT2018fyk \citep{wevers23}. However, the observation of a prompt shutoff in the second flare of AT2018fyk indicates that the star was likely not destroyed during the encounter, and suggests a possible rebrightening in 2027~\citep{pasham24}.

\subsection{Tidal Heating}
\begin{figure}
    \includegraphics[width=0.49\textwidth]{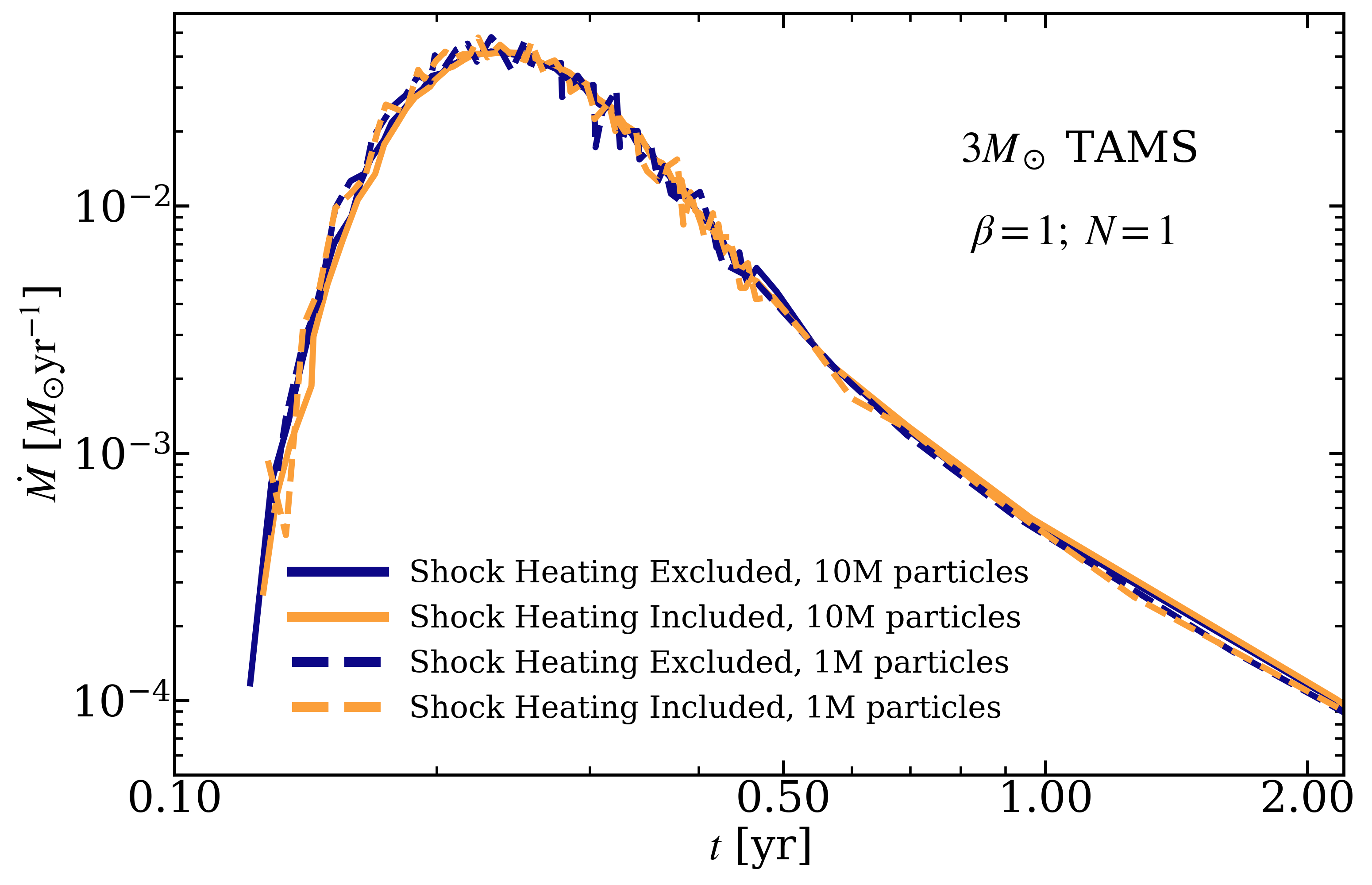} \\
    \includegraphics[width=0.49\textwidth]{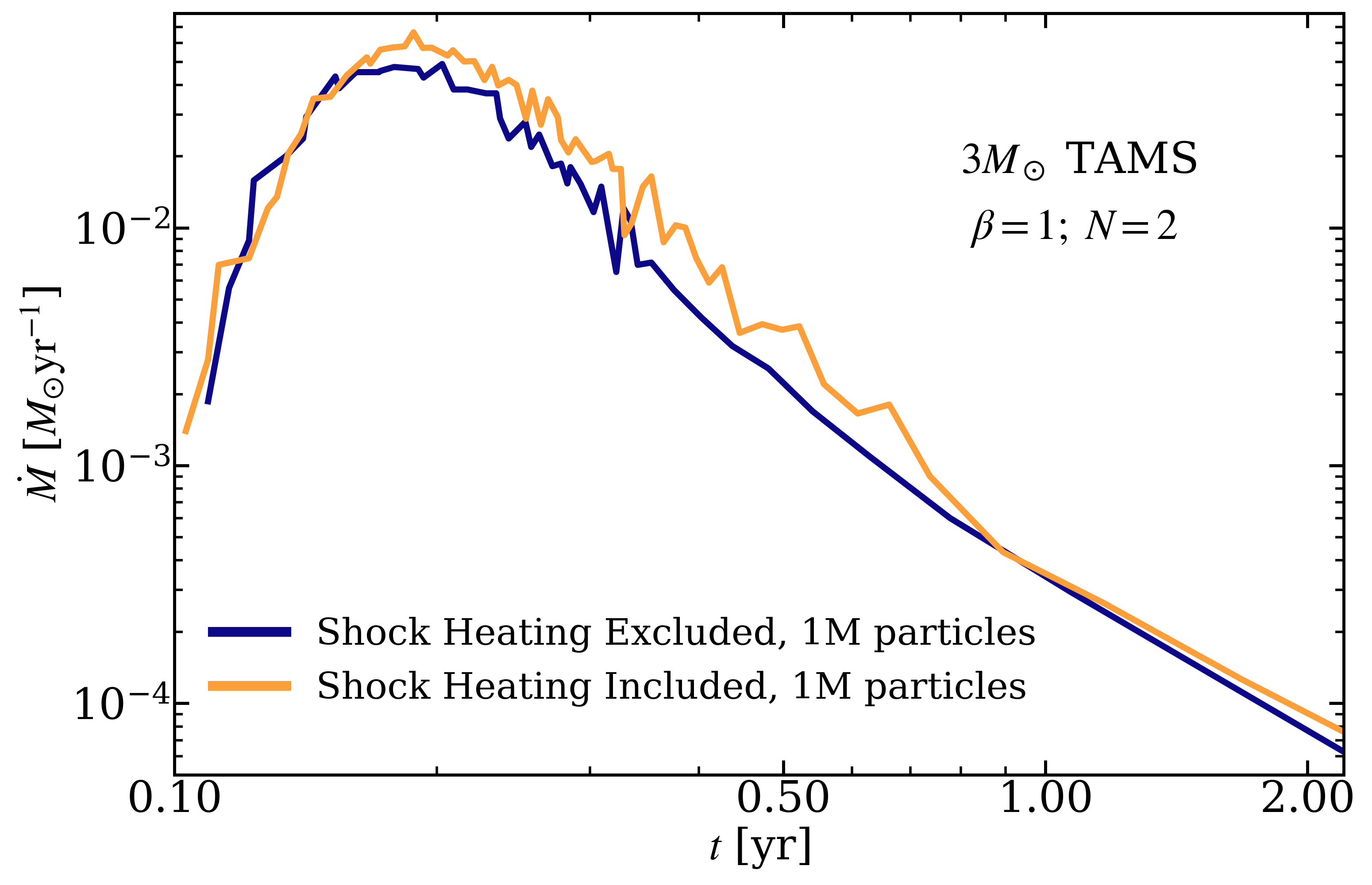} \\
    \includegraphics[width=0.49\textwidth]{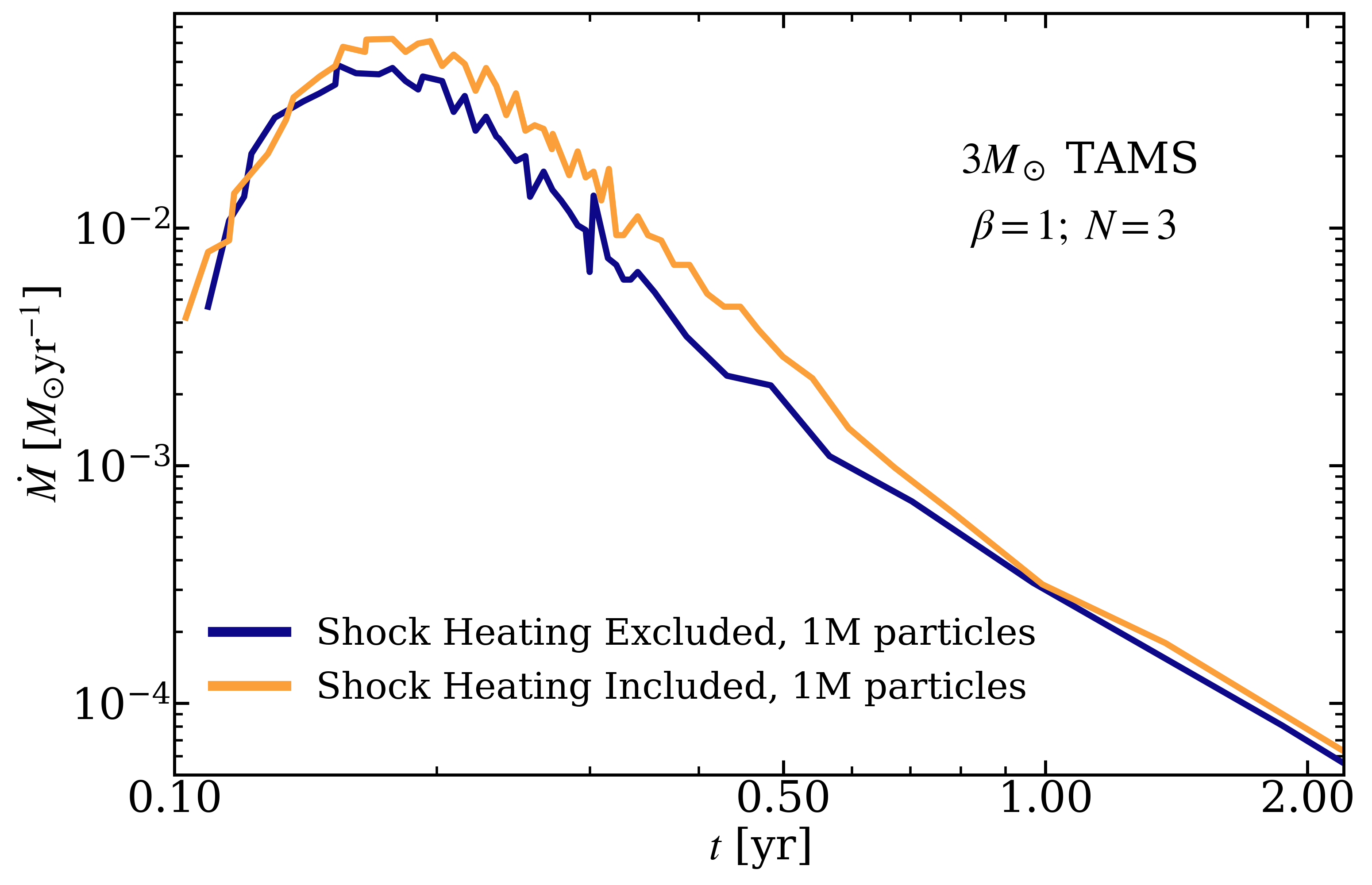}
     \caption{\hspace*{0.2cm} Top panel: Fallback rates with and without the inclusion of shock heating, for the first pericenter passage of the $3 M_{\odot}$ TAMS star at $\beta=1$, at a resolution of $10^6$ (dashed lines) and $10^7$ (solid lines) particles. Middle and bottom panels: Fallback rates for the second and third pericenter passages for the same star, at a resolution of $10^6$ particles.}
    \label{fig:fbr-3msun-shock-heating}
\end{figure}

To assess the impact of tidal heating on the structure and survivability of a star, and validate the choice of thermodynamic prescription adopted in our simulations, we performed a subset of the simulations for the $3M_\odot$ TAMS star with a different thermodynamic prescription, in which the thermal energy that is generated as a byproduct of viscous dissipation (here due to numerical viscosity; see \citealp{price18} for details in the context of PHANTOM specifically) is retained within the fluid, as opposed to the assumption that this additional heat is lost from the system which was employed in the simulations described in Sections \ref{subsec:3msun}-\ref{subsec:1msun}. Which of these prescriptions is a better representation of reality depends on a number of factors, including where in the star the energy is deposited, and the rate at which the oscillatory kinetic energy is dissipated, either radiatively, or through nonlinear damping, e.g., \citealt{mcmillan87, kochanek92, kumar96, weinberg12}), and the photon mean free path, but in general we expect the numerical heating to be artificially large in the outer layers of the star where the resolution is lowest (see \citealt{norman21, coughlin22b} specifically in the context of TDEs). 

\begin{figure*}
    \advance\leftskip-2cm
    \includegraphics[height=7.5cm,width=0.455\textwidth]{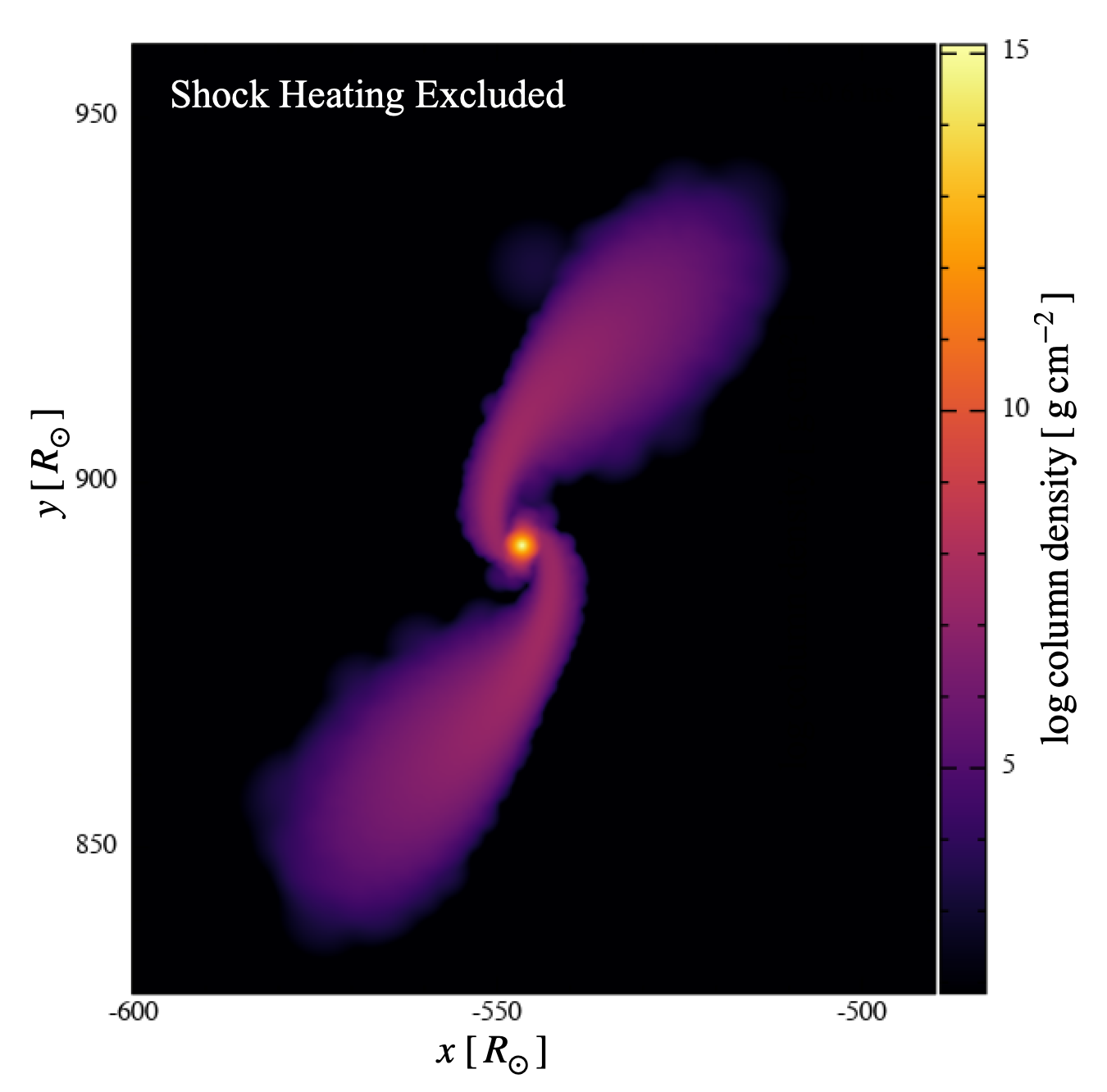} \hspace{0.1mm}
    \includegraphics[height=7.5cm,width=0.455\textwidth]{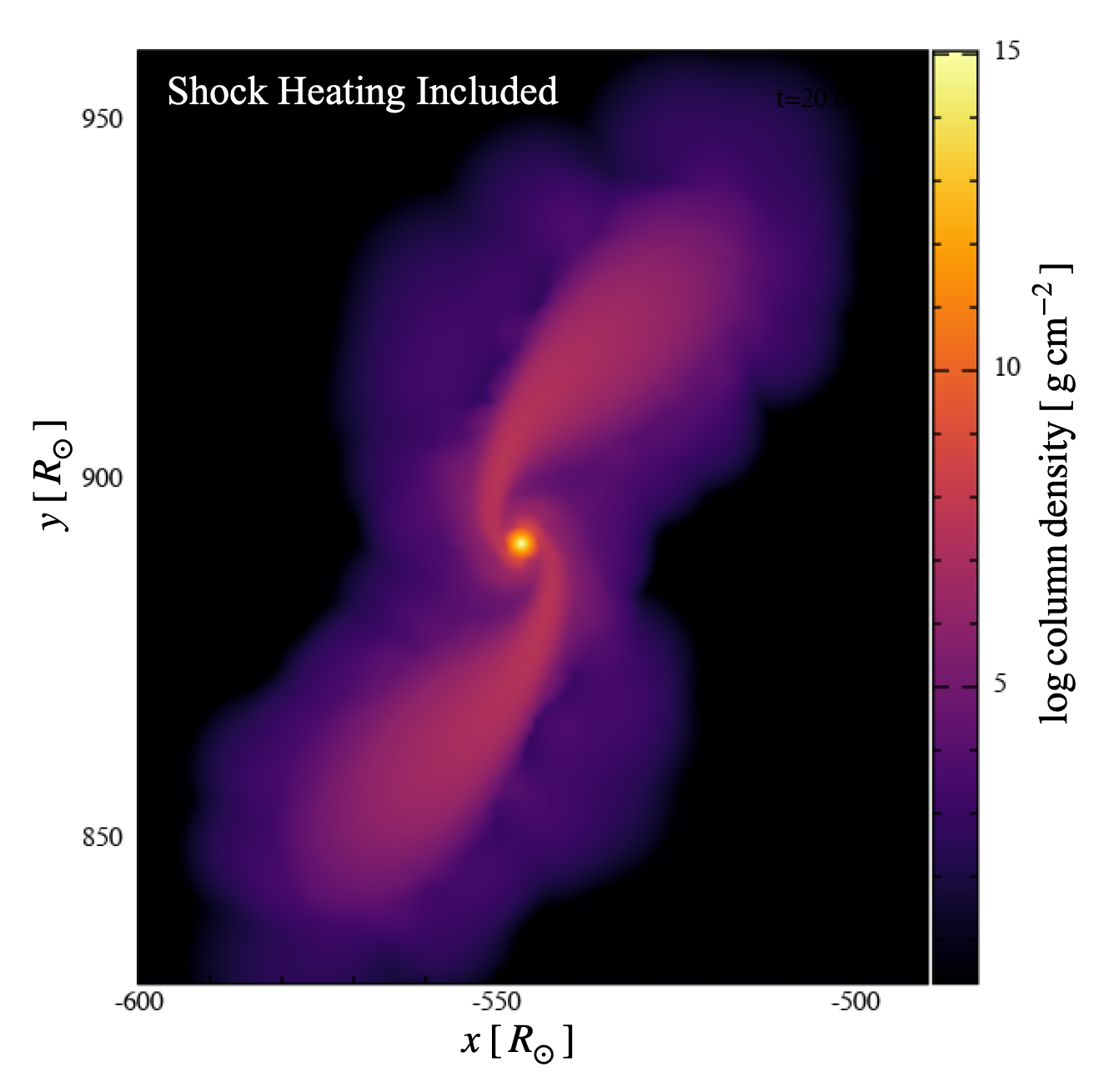} \\
    \centering
    \includegraphics[height=6.2cm,width=0.43\textwidth]{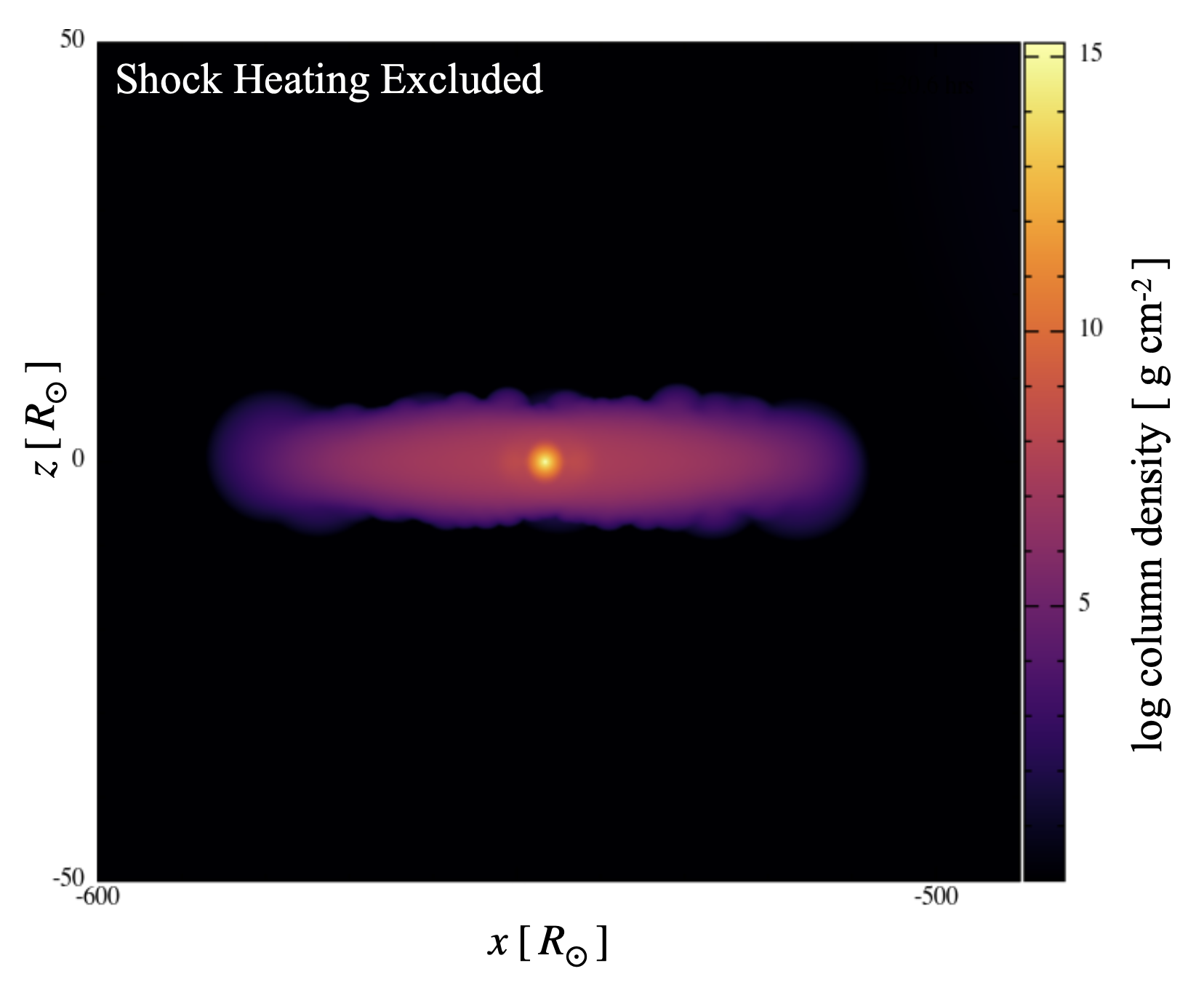} \hspace{5mm}
    \includegraphics[height=6.2cm,width=0.43\textwidth]{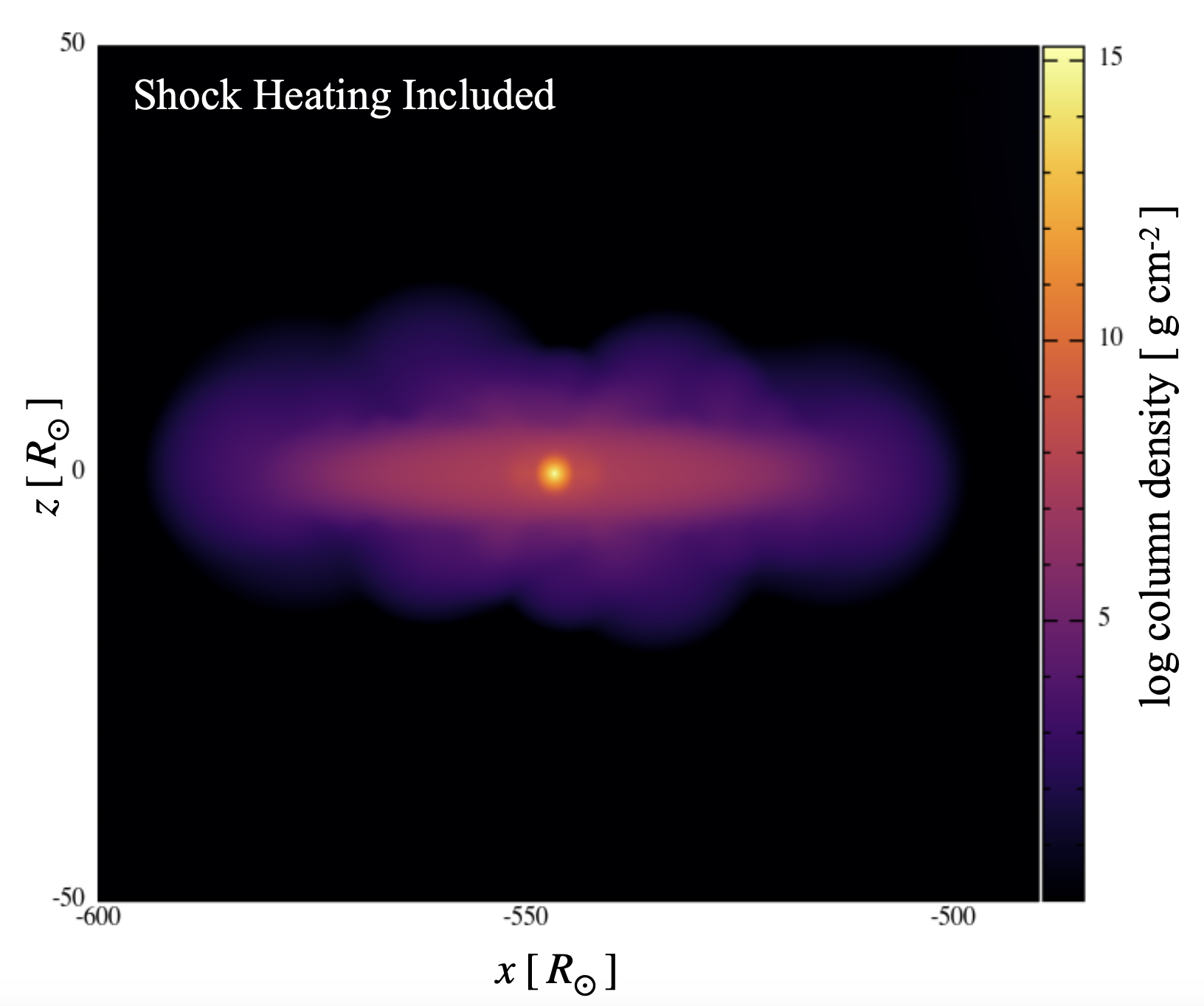}
    
     \caption{\hspace*{0.2cm} The column density profile of the $3M_\odot$ TAMS star at a resolution of $10^7$ particles, for its first pericenter passage on a $\beta=1$ orbit in the orbital plane (top panel) and perpendicular to the orbital plane (bottom panel). The left and right panel shows results with and without the inclusion of shock heating in the simulations. The outer layers of the star get inflated due to the inclusion of shock heating, but the core remains largely unmodified, thus allowing the star to survive many mass stripping encounters.}
     \label{fig:3msun-density}
\end{figure*}

The top panel of Figure~\ref{fig:fbr-3msun-shock-heating} shows the fallback rates at two different particle resolutions for the first encounter of the $3M_\odot$ TAMS star on a $\beta=1$ orbit, with and without the inclusion of ``shock heating,'' which is the term we adopt for numerical heating in the simulations. While the fallback rates are almost identical independent of the particle resolution and the thermodynamic prescription, the retention of heat in its outer layers causes the star to inflate, making it easier to remove these tidally heated layers on subsequent encounters. This causes an increased amount of mass loss in the second and third pericenter passages, and the fallback rates for these two encounters (shown in the middle and bottom panels of Figure~\ref{fig:fbr-3msun-shock-heating}) lie systematically above those from the corresponding simulations in which shock heating is ignored (see Figure 15 of \citealp{bandopadhyay24} for a comparison of the fallback rates for the first 6 pericenter passages). Once the tidally heated outermost layers of the star are stripped, shock heating is no longer significant, and the fallback rates are again indistinguishable. Figure \ref{fig:3msun-density} shows the fluid column density projected onto $(x-y)$ and perpendicular to $(x-z)$ the orbital plane. While the highest-density regions remain unaltered, the retention of heat causes the outer layers of the star to inflate, which are removed on subsequent encounters. These comparisons demonstrate the insensitivity of our results to the thermodynamic prescription adopted -- the retention of heat in the fluid inflates the outer layers of the star, which are mechanically removed on subsequent encounters. For systems such as ASASSN-14ko, which has a short orbital period of $\sim114$ days, with an observed period derivative $\dot{P}=-0.0026$, the rate at which energy has to be radiated away from the star in order to maintain a constant period derivative exceeds the Eddington limit of the star. The mechanical (as opposed to radiative) loss of heat from the tidally stripped outer layers can provide a possible outlet for energy that enables the system to maintain a roughly constant period derivative. 

\section{Adiabatic Mass Loss Model}
\label{sec:mass-loss-model}
\begin{figure*}
    \includegraphics[width=0.495\textwidth]{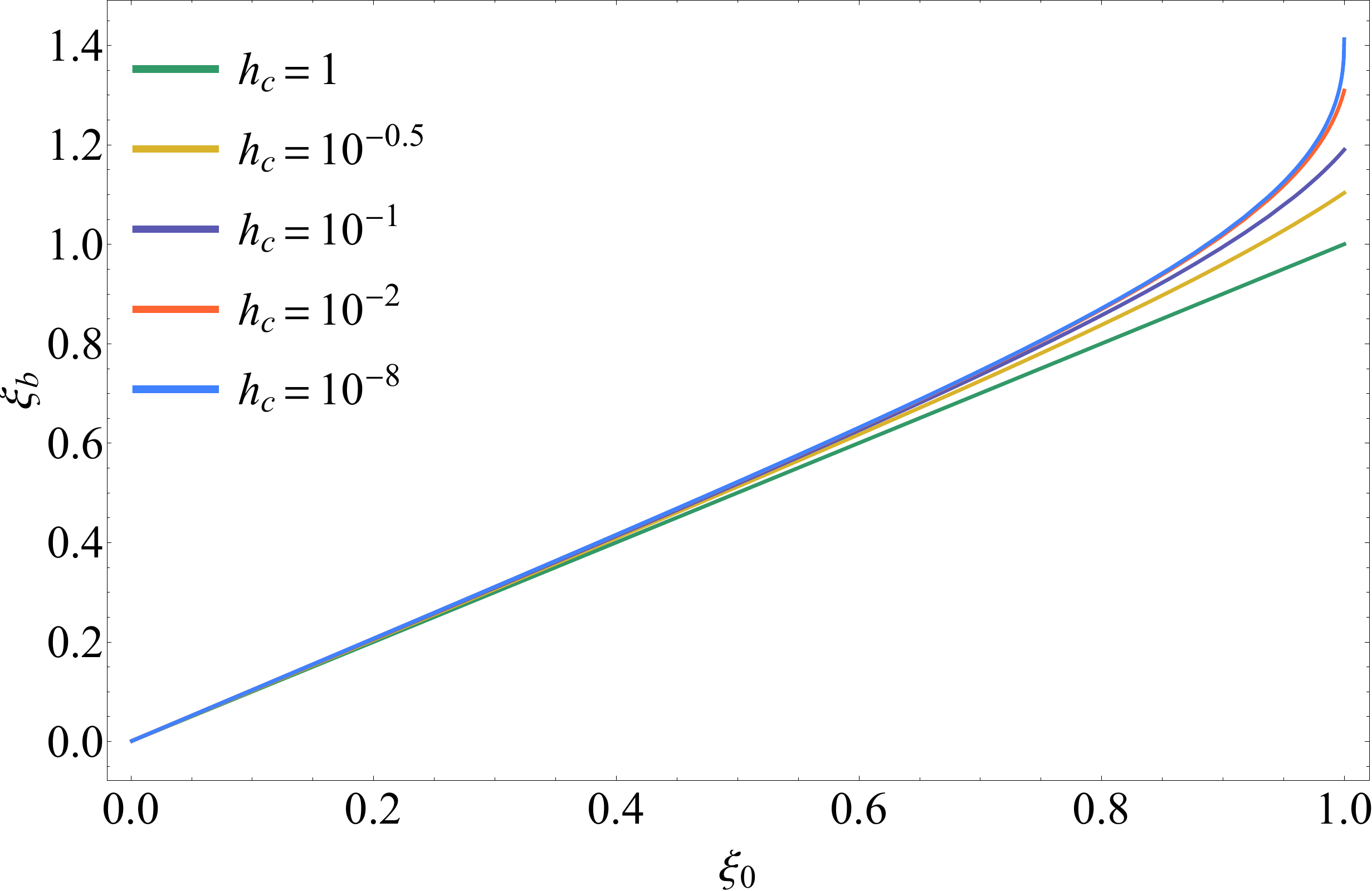}
    \includegraphics[width=0.495\textwidth]{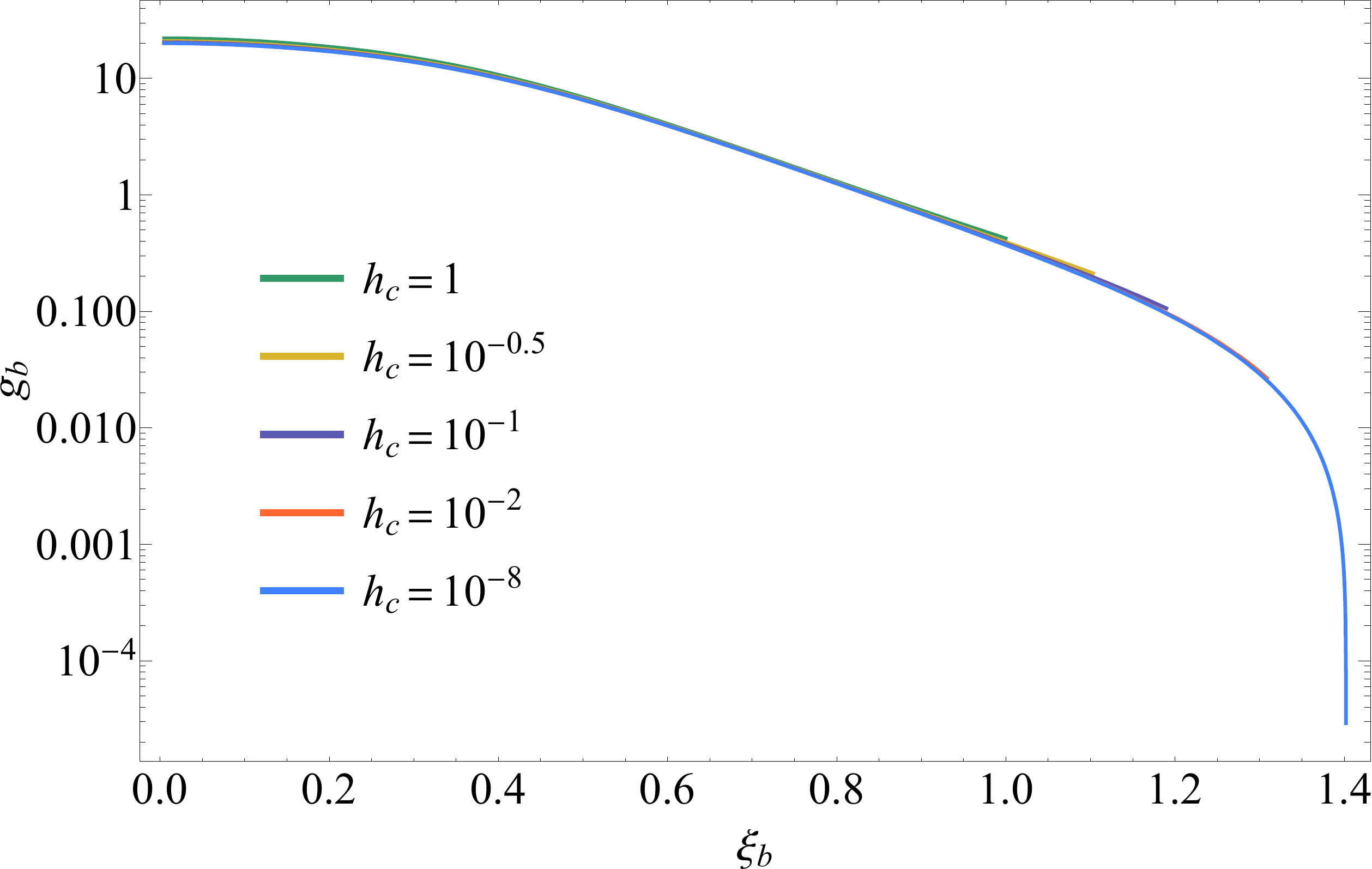}
    \caption{\hspace*{0.2cm}Left: The non-dimensionsionalized background state $\xi_{\rm b}$ as a function of the initial Lagrangian radial coordinate $\xi_0$ within the star. The legend indicates the surface pressure relative to its initial value. As the pressure at the surface declines $\xi_{\rm b}$ tends to an equilibrium configuration, with $\partial \xi_{\rm b}/\partial \xi_0 \rightarrow \infty$. Right: The dimensionless density $\xi_{\rm b}$ of the background state as a function of the current positions of the fluid elements $\xi_{\rm b}$, for relative surface pressures $h_{\rm c}$ as shown in the legend.}
    \label{fig:xib_of_xi0}
\end{figure*}

The hydrodynamical simulations described in the Section~\ref{sec:hydro} illustrate that stellar structure plays a vital role in determining the fate of a star being repeatedly stripped of mass in an rpTDE. While these simulations constitute an accurate model of rpTDEs, and allow us to constrain the effects of stellar and orbital properties on the lightcurves of these events, mapping out the entire parameter space of stellar and orbital properties is computationally intractible. Thus we developed an intermediate/hybrid approach to model the effects of stellar structure~\citep{bandopadhyay25}, wherein we treat the star as an initially spherically symmetric fluid configuration that undergoes pressure deconfinement as a result of mass loss. We divide the original star (of mass $M_\star$, radius $R_\star$) into two regions -- the core, which has a mass $M_{\rm c} = M_\star -\Delta M$, and initial radius $R_{0,\rm c}$, and an outer envelope of mass $\Delta M$, that extends from $R_{0,\rm c}$ to $R_\star$. The removal of the outer envelope of the star results in a loss of pressure support from the surrounding envelope, causing the core to expand. If the relative change in the pressure at the outer radius of the the star is small, the response of the star can me modeled self-consistently using linear perturbation theory. However, since we are interested in the limit where the star reaches an  equilibrium with the surrounding medium, the pressure at the outer radius of the star must eventually drop to zero. Linear perturbation theory can not be used to accurately model the response of the star in this regime.

To overcome the issue and obtain the response of the post-mass-loss core to the loss in pressure support, we use a sequence of quasi-steady states, each of which exactly matches the pressure of the surrounding medium at any instant in time. As the surface pressure gradually declines, the system approaches a final time-averaged state for which the pressure and density at the surface are zero. The final time-averaged radius which the star relaxes to is the radius at which the surface pressure of the time-averaged state goes to zero. In addition to the expansion, the core also exhibits time-dependent oscillatory behaviour, which we show can be related to the time-dependence of the background state. 
\begin{figure*}
    \includegraphics[width=0.49\textwidth]{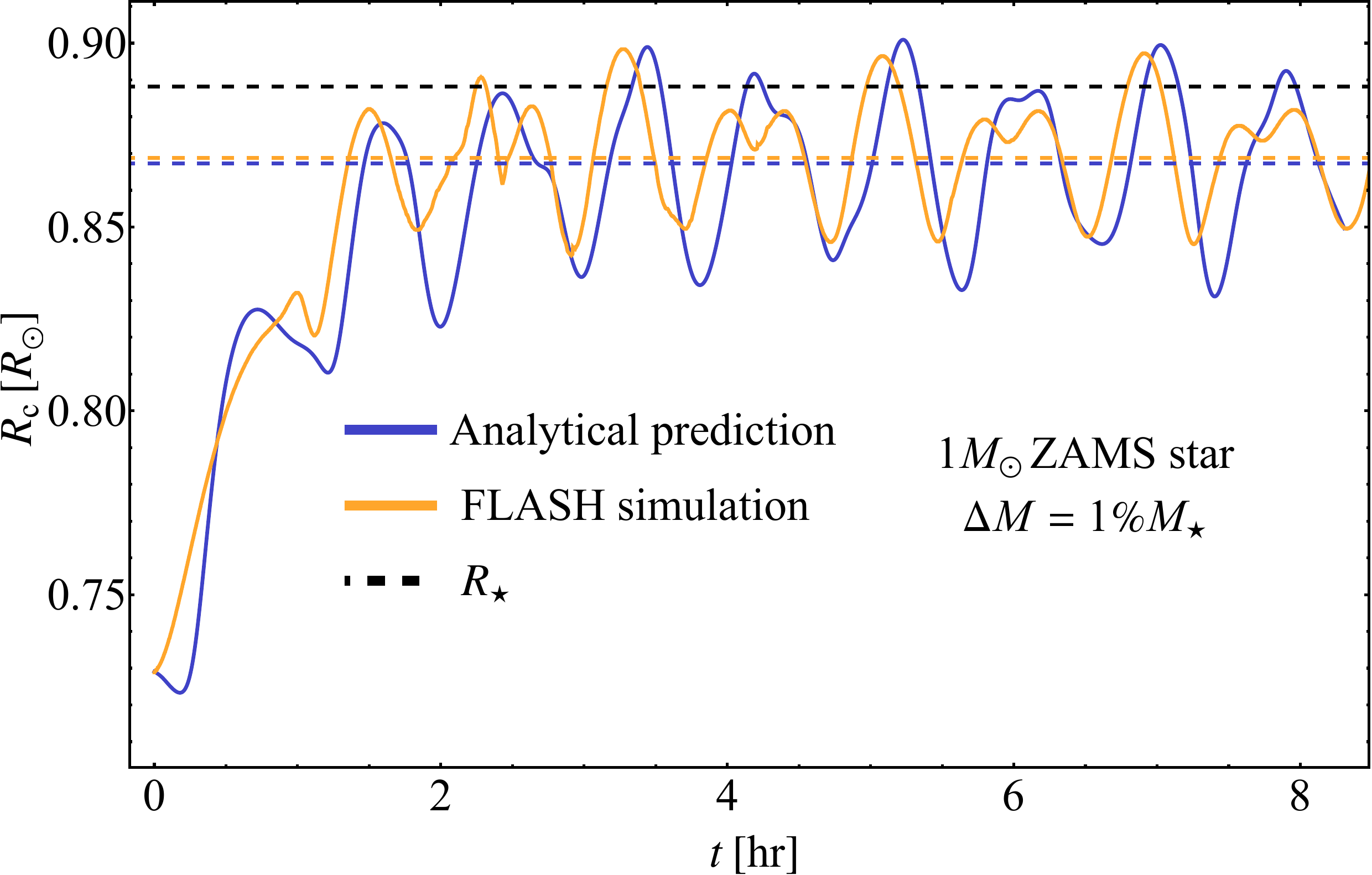}
    \includegraphics[width=0.48\textwidth]{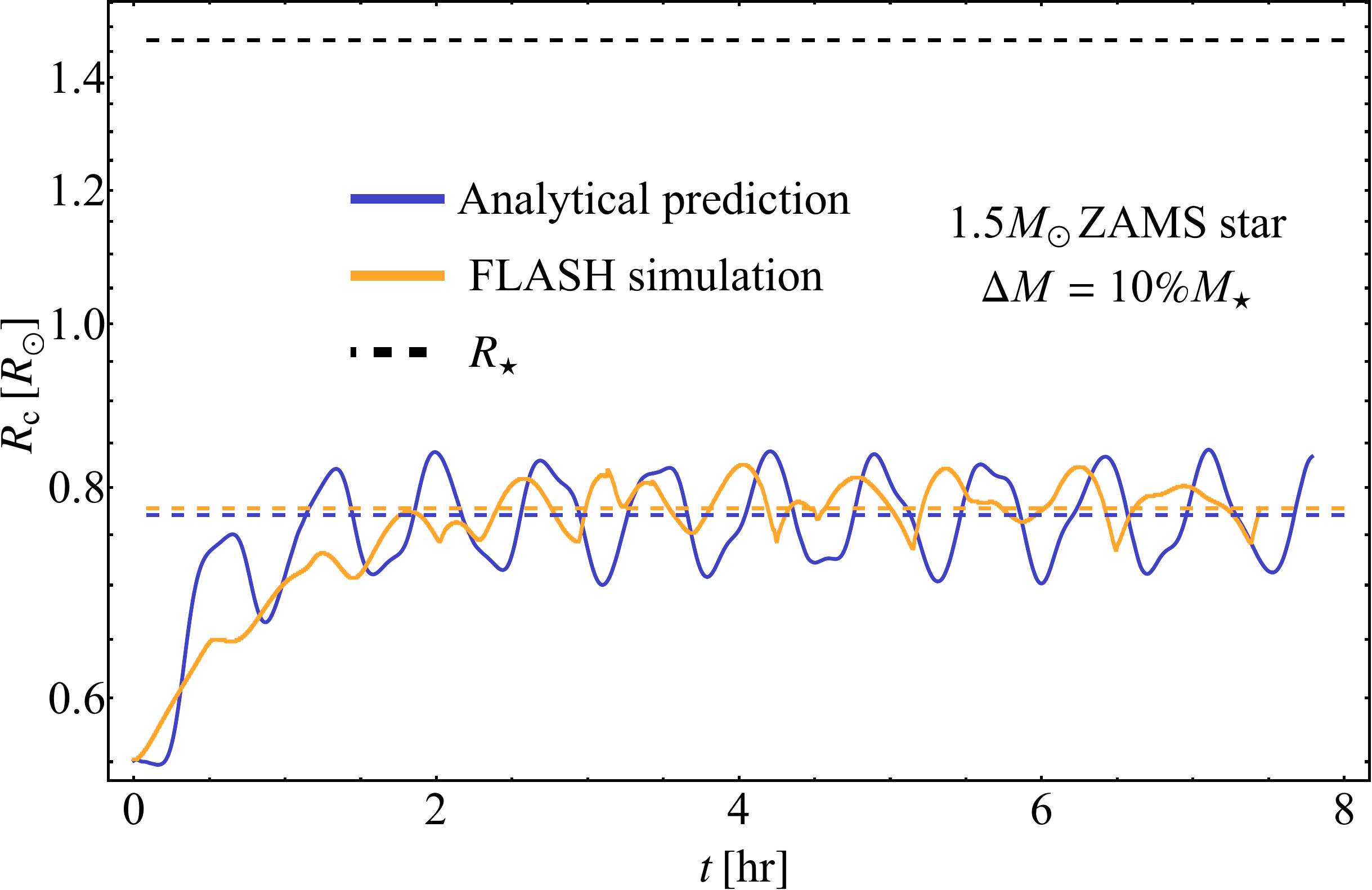}
    \caption{\hspace*{0.2cm} A comparison of the time-dependent outer radius of the post-mass-loss core for a $1M_\odot$ ZAMS star undergoing a $1\%$ mass loss (left) and a $1.5M_\odot$ ZAMS star undergoing a $10\%$ mass loss (right). The blue and orange curves show the analytical prediction and the result of a 1D hydrodynamical simulation performed using {\sc flash}. The post-mass-loss average radius (indicated by the blue and orange dashed lines for the model and simulation respectively) is smaller than the original radius of the star (black dashed line), suggesting that these stars undergo an increase in their average density as a result of mass loss.} \label{fig:ZAMS1_radius}
\end{figure*}
\subsection{Background State}
\label{sec:background-state}
To model the temporal evolution of the core, we first solved the spherically symmetric fluid equations in Lagrangian form (continuity, momentum and entropy conservation) for the core, treating it as a fluid having an adiabatic equation of state with an adiabatic index $\Gamma=5/3$. We do not model the fluid dynamics of the envelope, but allow the pressure in this region to decline over time. The time-dependent fluid variables such as pressure and density can be denoted as $p(r_0,t)$, $\rho(r_0,t)$ where $r_0$ is the initial Lagrangian radius of a fluid shell within the core. Treating the declining pressure at the surface of the core as a time-dependent boundary condition imposed at the surface, we model the evolution of the core as a sequence of background states $r_{\rm b}(r_0,t)$, which can be obtained by solving the radial momentum equation. We assert that the evolution of the background-state is ``quasi-steady,'' i.e., the background state is always in equilibrium with the surrounding medium. Thus, we ignore the acceleration of the background state $\partial^2 r_{\rm b} /\partial t^2$ in solving the radial momentum equation. Once we solve for the background state, $r_{\rm b}(r_0,t)$, we can use the continuity equation and the entropy conservation equation to obtain the pressure and density profiles as a function of the background state. The left panel of Figure~\ref{fig:xib_of_xi0} shows an example of the solution for the normalized (dimensionless; $\xi \equiv r/R_{0,\rm c} $) background state, $\xi_{\rm b}(\xi_0,h_{\rm c})$ for a $1M_\odot$ ZAMS star undergoing a $10\%$ mass loss, which implicitly depends on time through the declining surface pressure, where $h_{\rm c} \equiv p_{\rm c}(t)/p_0(R_{0,\rm c})$ is the pressure at the surface of the core, normalized by its initial value $p_0(R_{0,\rm c})$. The original radius of the star, prior to mass-loss, is $R_\star = 0.888 R_\odot$, while the radius within which $90\%$ of its mass is enclosed is $R_{0,\rm c} = 0.494 R_\odot$. For values of the dimensionless surface pressure $h_{\rm c} \lesssim 10^{-3},$ the solution $\xi_b(\xi_0,h_{\rm c})$ approaches the asymptotic time-averaged state. The final dimensionless radius of the star is $\xi_{\rm b} \simeq 1.41$, such that its time averaged radius is $R_{\rm c} \simeq 1.41 R_{0,\rm c} = 0.70 R_\odot$, which is smaller than the original radius of the star. Thus, the final configuration of the star post-mass-loss has a higher average density compared to the original star. The right panel of the star shows the dimensionless density profile of the background state $g_{\rm b}$ ($g = (M_{\rm c}/4\pi R_{0,\rm c}^3)^{-1} \rho $ is the non-dimensionalized density), for different values of the surface pressure $h_{\rm c}$. Here $g_{\rm b}$ is proportional to the initial density but scaled by $\left(\xi_{\rm b}/\xi_0\right)^{-2}\left(\partial\xi_{\rm b}/\partial \xi_0\right)^{-1}$, i.e., it is the density that is derived from the initial mass profile of the star with the radii of fluid elements stretched according to $\xi_{\rm b}(\xi_0)$, and its value at the surface approaches zero at the pressure drops.

\subsection{Total Solution}
The background state described in Section~\ref{sec:background-state} depends on the surface pressure, which introduces an implicit time dependence (that is externally imposed through $h_{\rm c}(\tau)$ in our model). The fluid shells thus have a non-zero velocity as the pressure drops over a finite timescale, that is ignored in the definition of the background state. We can solve for the explicit time dependence of the background state by re-introducing the acceleration term ($\partial^2 r_{\rm b}/\partial t^2$) in the radial momentum equation. Expressing the Lagrangian position of a fluid element as
\begin{equation}
    \xi = \xi_{\rm b}(\xi_0,p_{\rm c}(\tau) ) + \xi_1(\xi_0,\tau), \label{eq:total-sol}
\end{equation}
we can treat $\xi_1$ as a small correction to the background state arising due to the neglect of the core velocity, and linearize the radial momentum equation, i.e. retain up to first-order terms in $\xi_1$, to obtain a second-order partial differential equation for $\xi_1(\xi_0,\tau)$ (see Section 2.3 of~\citealp{bandopadhyay25} for details of the equations). Taking a Laplace transform of the radial momentum equation yields a Strum-Liouville form for the spatial operator, which can be integrated with the appropriate boundary conditions to obtain the eigenmodes of any given star. Using the orthogonality of the eigenmodes, we can then obtain the time-dependent perturbation $\xi_1(\xi_0,\tau)$ and use Equation~\ref{eq:total-sol} to construct the time-dependent response of a star for any given mass-loss. The time-dependent radius of the core $R_{\rm c}(\tau)$ can be obtained by evaluating the expression in Equation~\ref{eq:total-sol} at $\xi_0=1$, i.e.,
\begin{equation}
    R_{\rm c}(\tau) = R_{0,\rm c} (\xi_{\mathrm{b}}(\xi_0=1,\tau) + \xi_1(1,\tau)), \label{eq:core-radius}
\end{equation}
Figure~\ref{fig:ZAMS1_radius} shows the core-radius for two {\sc mesa}-generated stellar models, a $1M_\odot$ ZAMS star and a $1.5 M_\odot$ ZAMS star, undergoing $1\%$ and $10\%$ mass losses respectively. To vet the accuracy of our model, we also simulated the mass loss process using the finite-volume hydrodynamics code {\sc flash}~( V4.7; \citealp{fryxell00}). The core of the star was mapped onto a uniform spherical grid comprised of $2^{14}=16384$ cells. We performed additional simulations with 8192 cells, to verify that the results are independent of the numerical resolution. We used an adiabatic equation of state with $\Gamma=5/3$, and the self-gravity of the core was included. We imposed a reflecting boundary condition for all the fluid variables at the inner edge of the domain, as well as for the velocity and density at the outer boundary. To simulate the mass-loss and subsequent pressure de-confinement of the core, the pressure at the outer edge of the domain was reduced from its initial value at $R_{\rm 0,c}$ to $\sim$ zero as $e^{-\tau}$. For each panel of the Figure, the orange curve is the result obtained from {\sc flash}, and is in very good agreement with the analytical solution (shown by the blue curve). For the $1M_\odot$ ZAMS star undergoing $1\%$ mass loss, the average radius of the core (depicted by the dashed blue/orange line for the analytical/numerical model) is $\sim 0.867 R_{\odot}$, which is slightly smaller than the original stellar radius $R_{\star} = 0.888 R_{\odot}$, indicating that the core has a higher average density compared to the original star (despite the reduction in its mass). For the $1.5 M_\odot$ ZAMS star, which has a radius $R_\star \approx 1.5 R_\odot$, a removal of $10\%$ of its outer envelope results in $R_{0,\rm c} \approx 0.55 R_\odot$. The mean radius about which the star oscillates is  $\langle R_{\rm c} \rangle_{\rm t} \approx 0.77 R_\odot$ from the analytical model, and  $\langle R_{\rm c} \rangle_{\rm t} \approx 0.78 R_\odot$ from the FLASH simulation, demonstrating excellent agreement between the model and the simulation. Since the time averaged radius of the core following mass loss is significantly smaller than its original radius, the average density of the star increases by a substantial amount, being $\sim 6$ times the average density of the original star.

\subsection{Average Densities}
\begin{figure}
    \includegraphics[width=0.49\textwidth]{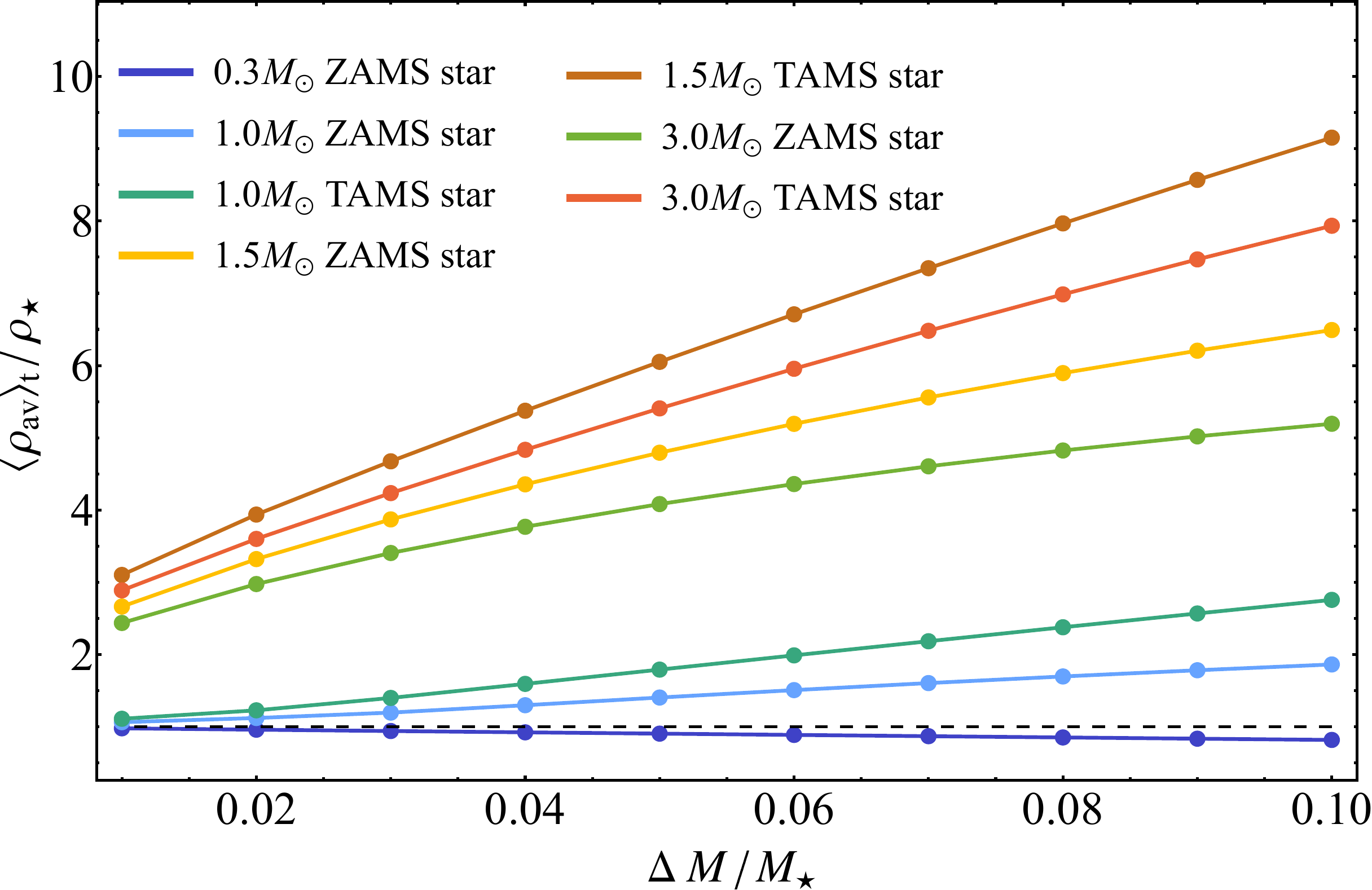}
    \caption{\hspace*{0.2cm}The spatially and temporally averaged density as a function of fractional mass lost $\Delta M / M_\star$ for seven different stellar structures, as shown in the legend. The horizontal black dashed line depicts $\langle \rho_{\rm av} \rangle_{\rm t}/\rho_\star=1$, where the post-mass-loss average density equals the original density of the star.} \label{fig:mesa_avg_densities}
\end{figure}

Figure \ref{fig:mesa_avg_densities} shows the average density of the post-mass-loss core, $\langle \rho_{\rm av} \rangle_{\rm t}$, normalized by that of the original star, $\rho_\star=3M_\star/4\pi R_\star^3$, as a function of the fractional mass loss, for a range of different stars shown in the legend. Low mass stars, such as the $0.3 M_\odot$ ZAMS star shown in the Figure, undergo a monotonic increase in their radius, and thus a monotonic decrease in their average density as a function of the fractional mass loss, making them more susceptible to tidal stripping. Contrarily, higher mass stars $\gtrsim1M_\odot$ exhibit a monotonic increase in their average density as a function of fractional mass loss. The average density of the $1M_\odot$ ZAMS star increases by a factor of $\sim 1-1.8$ times its original value for a $1-10\%$ fractional mass loss. The average density of higher mass and evolved stars increases by a significant factor of its original value, with the post-mass-loss average density ranging between $\sim3-9$ ($\sim3-8$) times that of the original star for the $1.5M_\odot$ ($3.0M_\odot$) TAMS star, making such stars highly stable against further mass loss.

\begin{figure*}
    \includegraphics[width=0.49\textwidth]{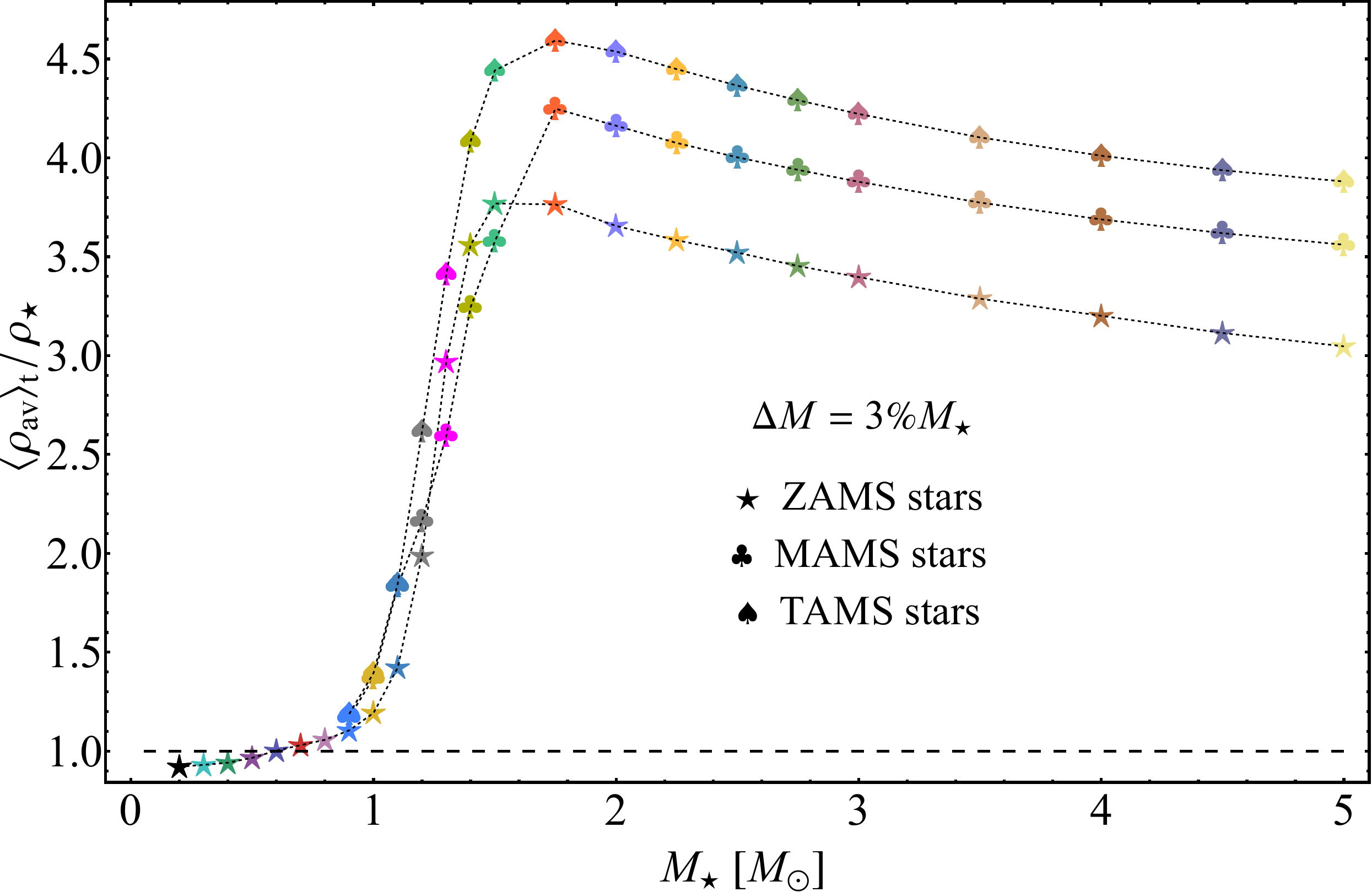}
    \includegraphics[width=0.48\textwidth]{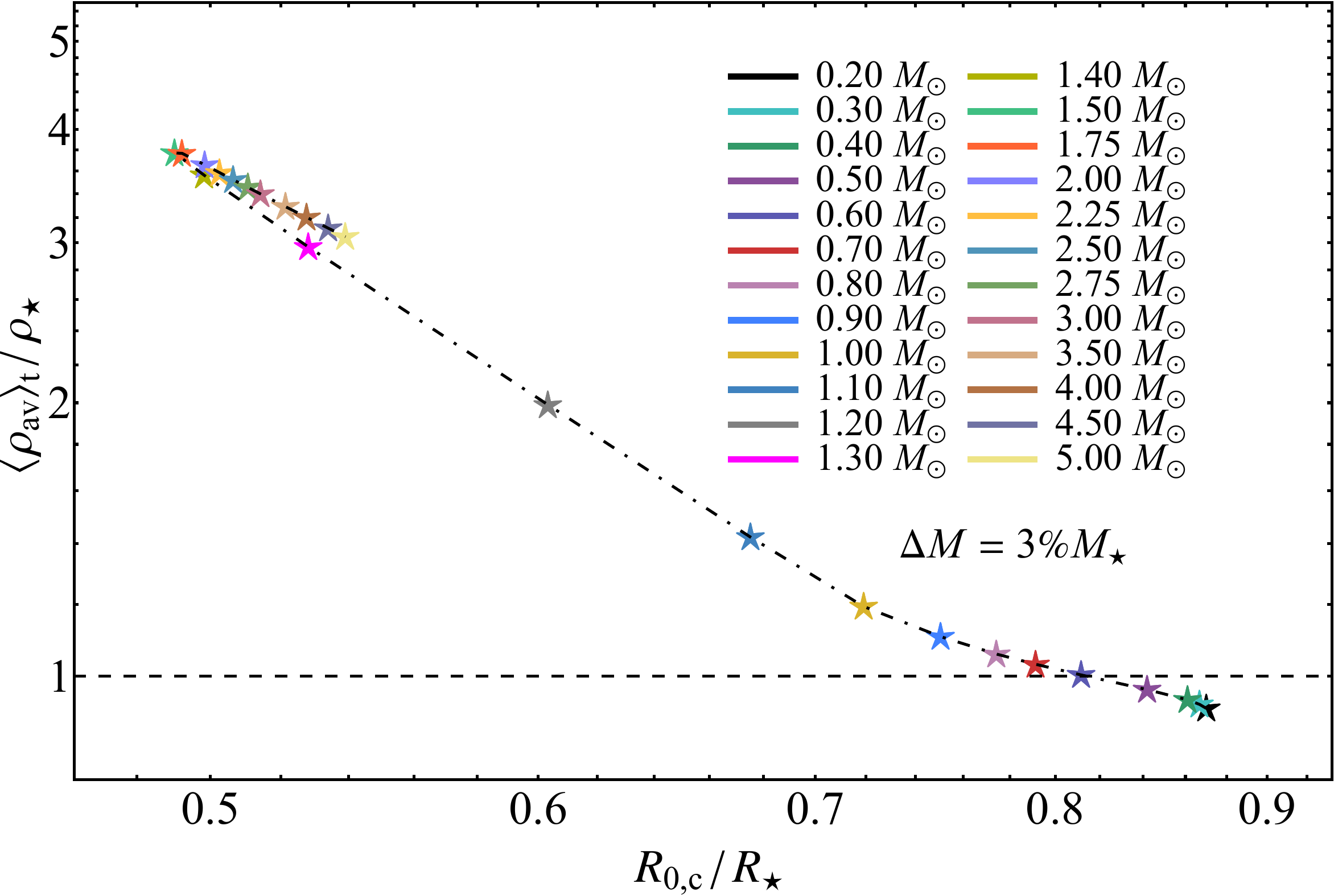}
    \caption{\hspace*{0.2cm}Left: The post-mass-loss average density, normalized by the original average density of a star for $\Delta M = 3\%M_\star$, as a function of stellar mass. The different symbols represent different stellar ages, as shown in the legend. Right:  The post-mass-loss average density, normalized by the original average density for a $3\%$ mass loss, as a function of $R_{\rm 0,c}/R_\star$, i.e., the initial radius within the star (relative to its outer radius) within which $97\%$ of its mass is contained. } \label{fig:mesa_stars_density_trend}
\end{figure*}

For a given fractional mass loss $\Delta M/M_\star$, the increase in the average density of a star depends on the location of $R_{0,\rm c}$ relative to the original stellar radius $R_\star$. The left panel of Figure \ref{fig:mesa_stars_density_trend} shows $\langle \rho_{\rm av} \rangle_{\rm t}/\rho_\star$ as a function of stellar mass, for three different stellar ages, and for a fixed mass loss of $\Delta M = 3\% M_\star$. We use different symbols to denote the different stellar ages, namely ZAMS (stars having a core hydrogen fraction of $\sim0.7$, depicted by $5-$pointed stars in the figure), MAMS  (determined by a core hydrogen fraction of $\sim0.2$ and depicted with clovers), and TAMS (when the core hydrogen fraction drops below $\sim 0.001$, depicted with spades). As seen in the figure, a relative maximum in $\langle \rho_{\rm av} \rangle_{\rm t}/\rho_\star$ is attained at a stellar mass of $M_\star=1.5 M_\odot$ ($M_\star=1.75 M_\odot$) for the ZAMS (MAMS and TAMS) stars. This is because for a given mass loss ($3\%$ for the purpose of the figure) the relative location of $R_{0,\rm c}$ with respect to the stellar radius $R_\star$ reaches a minimum at $M_\star=1.5 M_\odot$ ($M_\star=1.75 M_\odot$) for the ZAMS (MAMS and TAMS) stars, i.e., stars with $M_\star \sim 1.5-1.75 M_\odot$ have the most centrally concentrated cores, and are least susceptible to mass loss. To further illustrate the correlation between relative increase in the average density $\langle \rho_{\rm av} \rangle_{\rm t}/\rho_\star$ and the relative location of $R_{0,\rm c}$, the right panel of Figure~\ref{fig:mesa_stars_density_trend} shows $\langle \rho_{\rm av} \rangle_{\rm t}/\rho_\star$ as a function of $R_{0,\rm c}/R_\star$ for all of the ZAMS stars shown in the left panel, for a $3\%$ mass loss. As seen in the figure, $\langle \rho_{\rm av} \rangle_{\rm t}/\rho_\star$ increases monotonically with a decrease in $R_{0,\rm c}$, and is maximized for $M_\star=1.5M_\odot$ (for which the value of $R_{0,\rm c}/R_\star$ reaches a minimum), beyond which the relative change in the average density is less pronounced. 

\begin{figure}
    \includegraphics[width=0.48\textwidth]{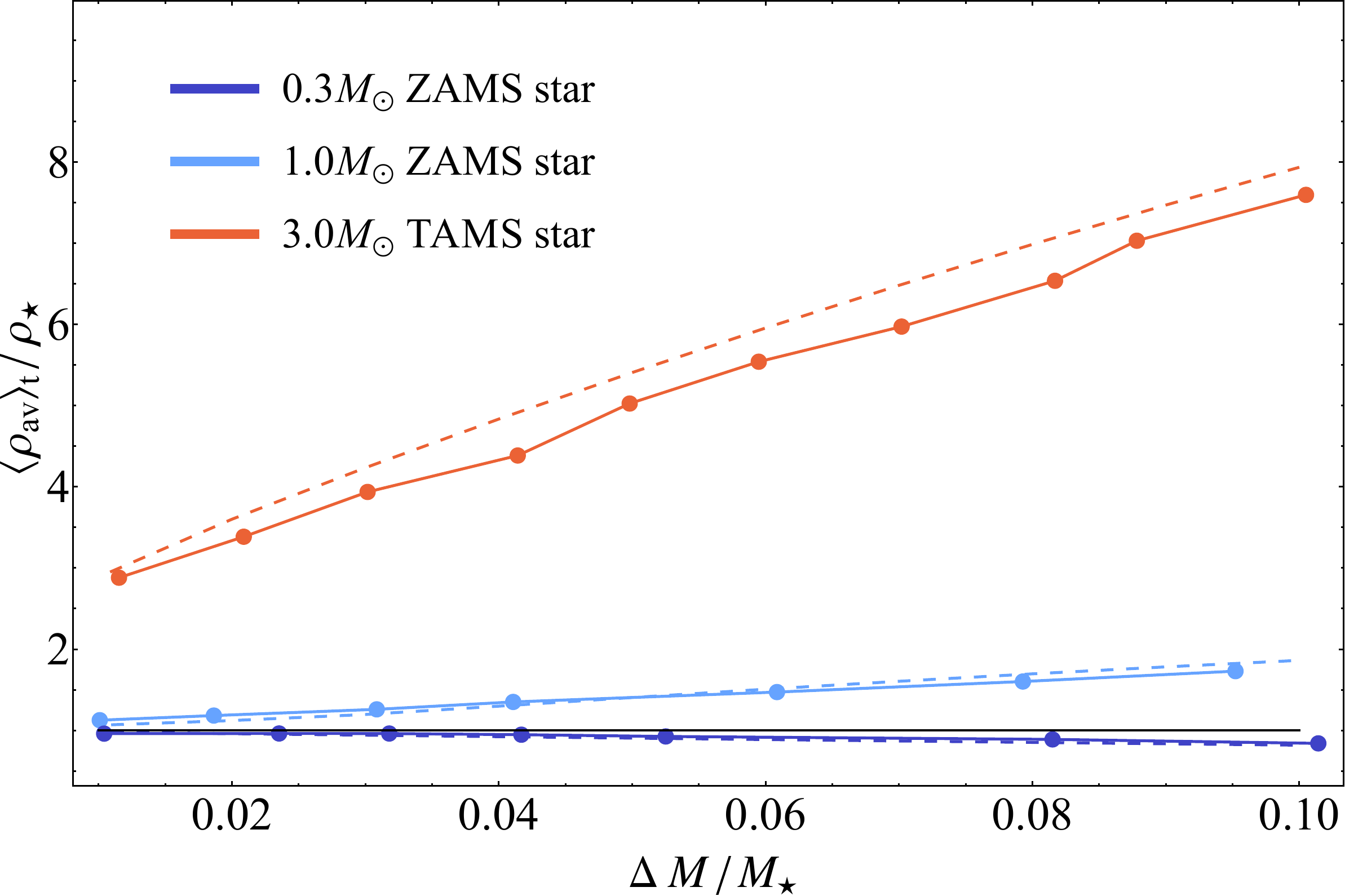}
    \caption{\hspace*{0.2cm}A comparison of $\langle \rho_{\rm av} \rangle_{\rm t}/\rho_\star$ as predicted by the model (dashed lines) against results of $3$D hydrodynamical simulations of TDEs (solid lines), for the three stellar types shown in the legend. The black solid line demarcates the line where $\langle \rho_{\rm av} \rangle_{\rm t} = \rho_\star$, i.e., where the star undergoes no change in its average density. The average density of low mass stars, like the $0.3M_\odot$ ZAMS star, progressively declines in response to increasing amounts of mass lost, whereas that of the $1M_\odot$ ZAMS and the $3M_\odot$ TAMS star increases, with a more pronounced change seen for the higher mass and evolved star.} \label{fig:phantom_densities}
\end{figure}

As a test of the efficacy of the preceding formalism in describing the response of stars being tidally stripped by SMBHs in repeating partial TDEs, we compared the results of our model to the results of hydrodynamical simulations of the partial disruption of 3 stars -- the $0.3M_\odot$ ZAMS star, the $1M_\odot$ ZAMS star and the $3M_\odot$ TAMS star, performed using {\sc phantom}. The numerical setup is identical to that described in Section~\ref{sec:hydro}. The amount of mass stripped in the tidal interaction is determined by the impact parameter $\beta \equiv r_{\rm t}/r_{\rm p}$. Figure \ref{fig:phantom_densities} shows the average density of the post-mass-loss stars for a fractional mass loss of $\sim1-10\%M_\star$ (for details of the computation of the average density, see Section 3 of \citealp{bandopadhyay25}). We show the analytical predictions alongside the numerical results, using dot-dashed
lines to depict the former. As seen in the Figure, the results of the hydrodynamical simulations are in agreement with our analytical
model, and the removal of small amounts of mass from the outer envelope of low mass star makes them more susceptible to mass loss, whereas for massive stars it stabilizes them against further mass loss, making such stars good candidates for surviving repeated tidal encounters with an SMBH.

\subsection{Energy Imparted to the Star}
Using this formalism, we can calculate the energy imparted to a post-mass-loss stellar core as a result of the removal of mass.  The core energy $E_{\rm c} = \mathcal{T}+\mathcal{V}+\mathcal{U}$ is the sum of the kinetic, $\mathcal{T}$, gravitational, $\mathcal{V}$, and thermal energies, $\mathcal{U}$, which are individually
\begin{equation}
    \mathcal{T} = 4\pi\int_0^{R_{\rm 0, c}}\frac{1}{2}v^2\rho_0(r_0)r_0^2dr_0, 
\end{equation}
\begin{equation}
    \mathcal{V} = -4\pi\int_0^{R_{\rm 0, c}}\frac{GM_0(r_0)}{r}\rho_0 r_0^2 dr_0,
\end{equation}
\begin{equation}
    \mathcal{U} = 4\pi \int_0^{R_{\rm 0, c}}\frac{1}{\Gamma-1}\frac{p}{\rho}\rho_0(r_0) r_0^2 dr_0,
\end{equation}
where we have used the continuity equation in it Lagrangian form to simplify the energy integrals in terms of the initial (i.e., pre-mass-loss mass) mass ($M_0$) and density profiles ($\rho_0$) profiles. Alternatively, we can evaluate the time rate of change of the core energy $E_{\rm c}$ as the $p\,dV$ work done by the expansion of the core as the pressure at the surface drops:
\begin{equation}
    \frac{\partial E_{\rm c}}{\partial t} = -4\pi p(R_{\rm c})R_{\rm c}^2\frac{\partial R_{\rm c}}{\partial t}. \label{dEtotdt}
\end{equation}
\begin{figure}
    \includegraphics[width=0.495\textwidth]{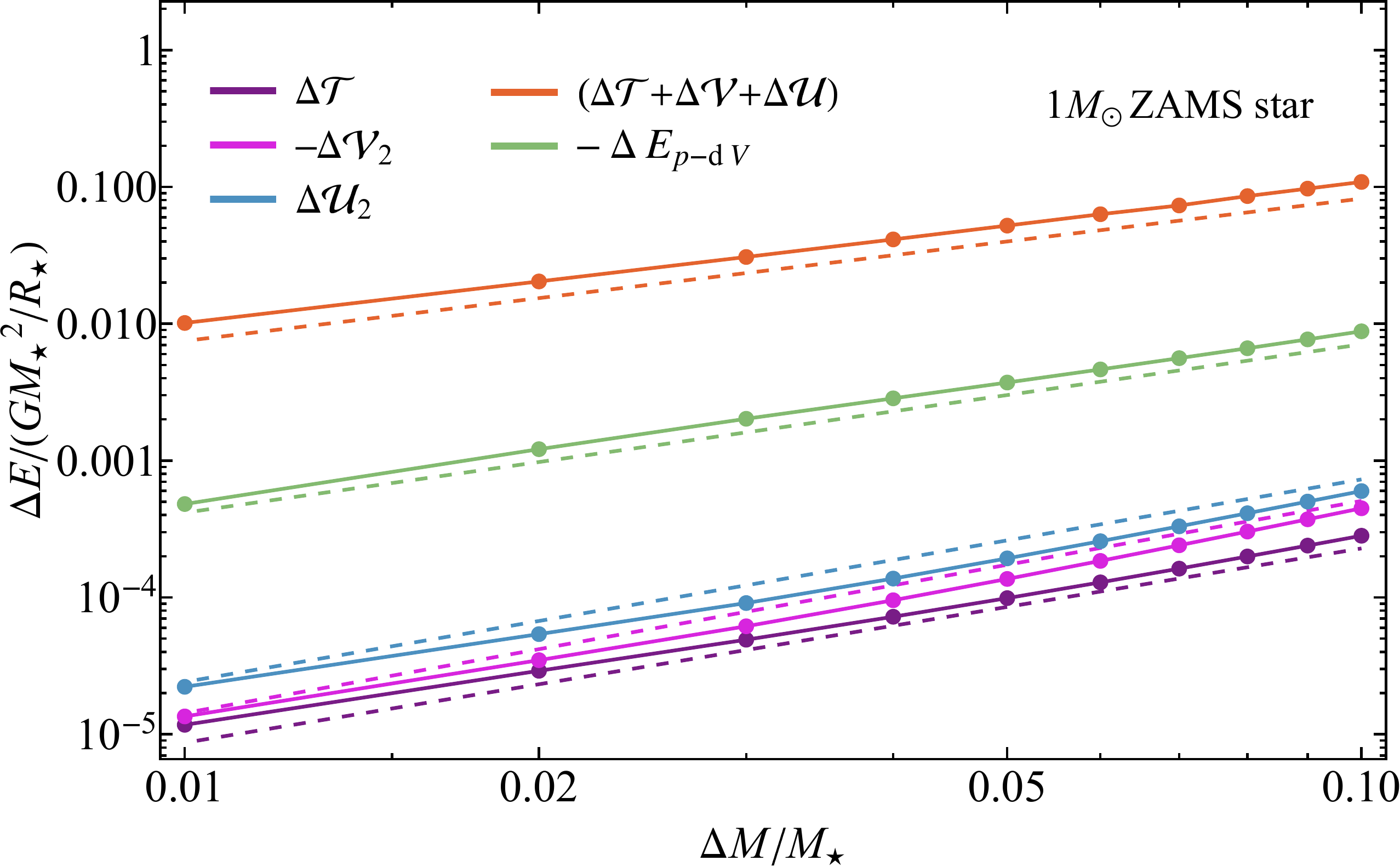}
     \caption{\hspace*{0.2cm} The time-averaged change in the kinetic energy (violet), total energy (orange), and p\,dV work done by the expansion of the core (green) as a function of the fractional mass lost. The second-order changes to the internal energy (blue) and potential energy (magenta) are also shown. The different energy contributions scale as approximate power laws in $\Delta M/M_\star$. The difference between the binding energy of the core prior to mass loss and that of the entire star largely determines the overall change in the star's energy.} 
     \label{fig:average-energies}
\end{figure}

The time-dependent energy of the core is then obtained by integrating Equation \eqref{dEtotdt} with respect to time, such that the final energy of the core (i.e., after the surface pressure drops to zero at asymptotically late times) is
\begin{equation}
    E_{\rm c, final} = E_{\rm c, 0}-4\pi\int_0^{\infty}p(R_{\rm c})R_{\rm c}^2\frac{\partial R_{\rm c}}{\partial \tau}\,d\tau, \label{DeltaE}
\end{equation}
where $E_{\rm c, 0}$ is the energy of the core prior to the mass loss. We define the second term in the right hand side of Equation \ref{DeltaE} as $\Delta E_{\rm p-dV}$, which is the $p-\mathrm{d}V$ work done by the surface. As the core radius expands in response to the drop in surface pressure, $\Delta E_{\rm p-dV}$ is a negative quantity. Thus, it follows that $E_{\rm c, final} < E_{\rm c, 0}$, and hence the expansion of the surface  results in the core becoming {more bound} to itself. 

However, this does not necessarily imply that the final core is more bound than the initial (core+envelope) star. In particular, the specific energy of the gas is generally negative near the stellar surface, such that by removing the outer stellar layers the core is less gravitationally bound than the original star. The total change in the energy relative to the initial energy of the star is then
\begin{equation}
    \Delta E = E_{\rm c, final}-E_{\star},
\end{equation}
such that if $\Delta E > 0$, the core is less gravitationally bound compared to the original star. We emphasize, however, that this change is due entirely to the removal of mass, and is independent of any additional energy imparted to the system via oscillatory modes. 

Figure \ref{fig:average-energies} shows the contributions to the temporally averaged change in the energy of the $1M_\odot$ ZAMS star as a function of the fractional mass lost, from the kinetic term (which is manifestly second order in the perturbation $\xi_1$), the gravitational and potential energy terms, as well as the $p-\mathrm{d}V$ work done by the expansion of the core and the total change in the energy ($\Delta \mathcal{T}+\Delta \mathcal{U}+\Delta \mathcal{V}$). The change in the energies are plotted relative to the energy of the original star, and normalized by $G M_\star^2/R_\star$. They scale approximately as power-laws in the fractional mass loss $\Delta M/M_{\star}$. As seen in the figure, the kinetic energy, which is the conventional ``tidal heating'' term is subdominant to the change in the gravitational and thermal energies, and has a negligible role in determining the total energy balance of the surviving stellar core. The net change in energy is positive, which, as noted above, indicates that the final state of the star is slightly less gravitationally bound than the original star, despite the core being more bound to itself as a result of the mass loss. 

\begin{figure*}
    \includegraphics[width=0.495\textwidth]{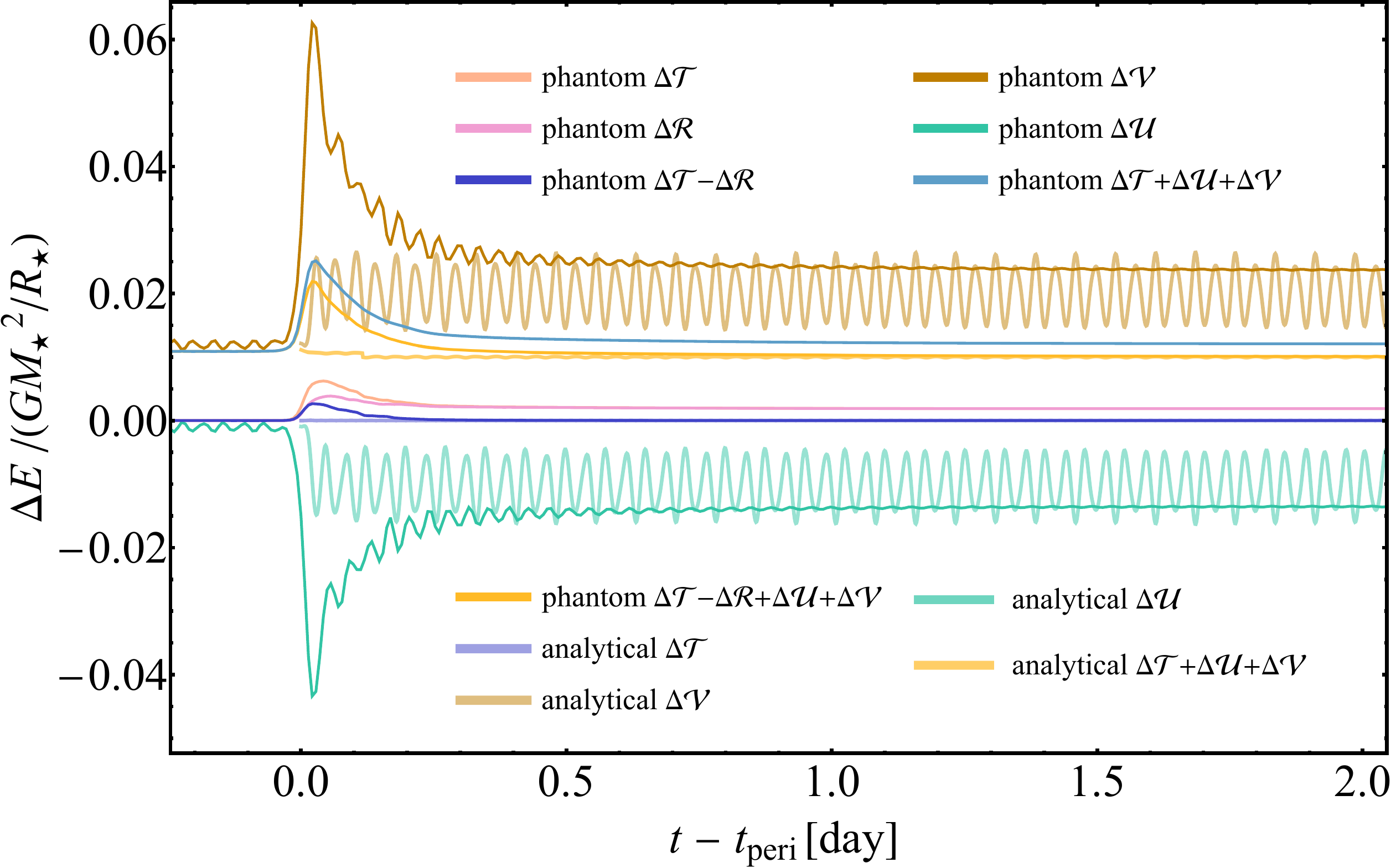}
     \includegraphics[width=0.495\textwidth]{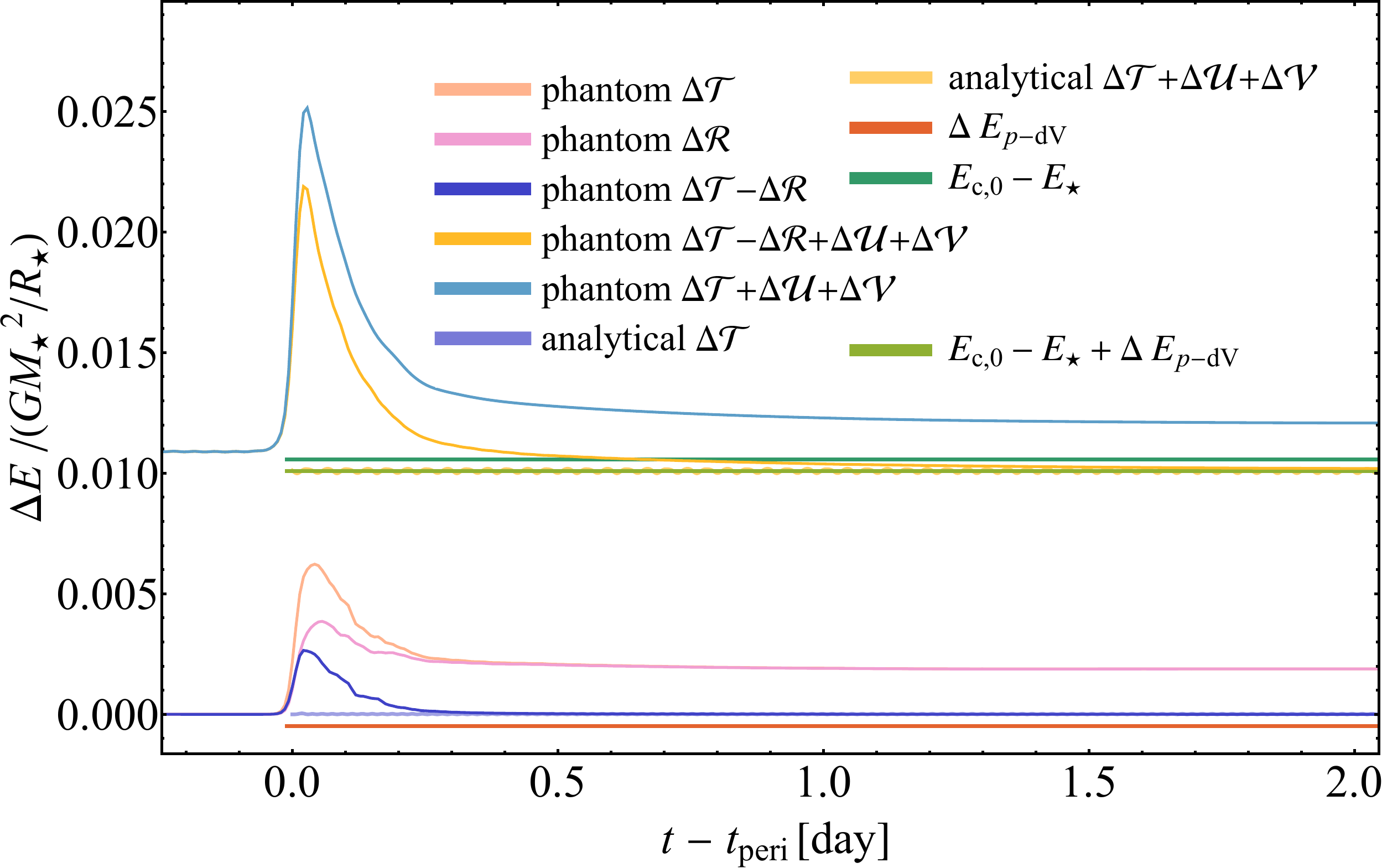} 
    \caption{\hspace*{0.2cm}Left: A comparison of the change in the different energy contributions, as measured from a hydrodynamical simulation of a $1M_\odot$ ZAMS star on a $\beta=0.7$ orbit around a $10^6 M_\odot$ SMBH (for which it loses $\sim 2\%$ of its mass), and the model predictions of the different energy contributions for a similar mass loss. The kinetic energy and the rotational energy contributions, as well as the total change in energy, are $\sim 1-2$ orders of magnitude smaller than the individual changes in the internal and potential energies.  Right: a zoomed-in version of the left panel. We also show the change in the binding energy resulting from mass loss, $E_{\rm c,0}-E_\star$, the work done by the expansion of the core, $\Delta E_{p-\mathrm{d}V}$, and their sum, which is comparable to the total change in the core energy.}
    \label{fig:ZAMS1-sph-energies}
\end{figure*}

Figure \ref{fig:ZAMS1-sph-energies} shows a comparison of the time-dependent changes in the various contributions to the energy as predicted by the model for the $1M_\odot$ ZAMS star undergoing a mass loss of $2\%M_\star$, against the results of the {\sc phantom} simulation of the star on a $\beta=0.7$ orbit (for which it loses comparable amount of mass). The contributions to the energy for the {\sc phantom} simulations are calculated as sums over all the particles comprising the core. The total kinetic energy $\Delta \mathcal{T}$ is a sum over the oscillatory and rotational energies (the rotational energy is imparted as a result of the tidal torque, and is not applicable for the spherically symmetric model of the star undergoing mass loss). The oscillatory component of the kinetic energy is given by the difference $\Delta \mathcal{T}-\Delta \mathcal{R}$, and, as seen in the Figure, is comparable to the kinetic energy of the modes calculated using the spherically symmetric mass loss model. The gravitational potential and internal energies are $\gtrsim2$ orders of magnitude larger than the kinetic energy of the modes, and the change in the potential energy outweighs that in the internal energy, thus $\Delta \mathcal{U}+\Delta \mathcal{V}>0$ at all times. The different contributions to the energy as calculated from the analytical model and from the numerical simulations agree to within $a \, few \% $. As discussed above, the total change in the energy being positive indicates that the post-mass-loss core is less self-bound than the original star (core+envelope). 

\section{Discussion and Conclusions}
\label{sec:conclusion}
Over the last $\sim10$ years, a number of TDE flares have been observed to rebrighten on timescales of months-to-years following the initial peak, leading to various differences in the observed lightcurves relative to those of standard (parabolic) TDEs. In the sequence of works summarized here, we tested the viability of the rpTDE model through hydrodynamical simulations of main-sequence stars being tidally stripped by an SMBH \citep{bandopadhyay24}, and using a spherically symmetric adiabatic mass loss model \citep{bandopadhyay25} to analyze the role of stellar structure in determining the stability of a star to mass loss. The main implications of these works are summarized in the following subsections.

\subsection{Effect of Stellar Structure on Mass Transfer Stability and Survivability}

The hydrodynamical simulations of the repeated partial disruption of stars described in Section~\ref{sec:hydro} demonstrate that the survivability of a star that is being repeatedly tidally stripped of mass is determined by its ability to remain structurally relatively unaffected. The specific case of a $3M_\odot$ TAMS star on a $\beta=1$ orbit illustrates that high mass ($M_\star\gtrsim1.5M_\odot$) and evolved stars having centrally concentrated cores can survive many encounters on a grazing orbit around an SMBH, losing $\lesssim1\%$ of its mass per encounter. The presence of a centrally concentrated core makes it increasingly difficult to strip off mass, and, as shown in Figure~\ref{fig:mass_stripped_vsN}, the amount of mass stripped for the $3M_\odot$ TAMS star on a $\beta=1$ orbit is a monotonically decreasing function of the number of pericenter passages, suggesting that the system cound give rise to $\gtrsim100$ flares. Contrarily, low mass and less evolved stars, such as the $1M_\odot$ ZAMS star on a $\beta=1$ orbit around an SMBH, loses an increasing amount of mass on each subsequent pericenter passages, and gets completely destroyed in $\sim4$ pericenter passages. Tidal heating does not play an important role in determining the survivability of a star in rpTDEs, and, as we directly demonstrated for a subset of our simulations  with the ``shock heating'' prescription, the tidally heated outer layers of the star are mechanically removed on subsequent encounters, thus resulting in negligible differences between the fallback rates obtained using the two different thermodynamic prescriptions, i.e., with and without the retention of heat in the stellar fluid. These results are in agreement with other works that studied the stability of stars in rpTDEs, e.g., \cite{liu23} concluded from the results of their adiabatic mass loss model that a centrally concentrated high mass star having a diffuse outer envelope is necessary to generate ASASSN-14ko-like flares, in which the system survives for $\gtrsim20$ outbursts. \cite{liu25} also studied the hydrodynamical evolution of a sunlike star undergoing repeated tidal disruption on $\beta=0.5,0.6,1.0$ orbits, and found that the star loses increasing amounts of mass and is destroyed within a few encounters.

The results of our spherically symmetric mass-loss model corroborate those of the numerical simulations of TDEs. As shown in Figures \ref{fig:mesa_avg_densities}-\ref{fig:phantom_densities}, for small amounts of mass lost, high mass and evolved stars undergo a monotonic increase in their average density. The average density of the $1.5M_\odot$ TAMS ($3.0M_\odot$ TAMS) star increases by a factor of $\sim4-8$ ($\sim4-10$) times the original average density of the star prior to mass loss, making such stars excellent candidates for surviving many encounters and giving rise to ASASSN-14ko-like flares. On the other hand, low mass stars ($M_\star\lesssim0.7M_\odot$) expand in response to mass loss, and the average denisty of such stars is a monotonically decreasing function of the amount of mass loss. Thus, such stars are likely to lose increasing amounts of mass on each subsequent pericenter passage, making them good candidates to explain transients like AT2020vdq (for which the second observed flare was brighter than the first).

\subsection{Spin-up of the stellar core}
Our discussion of the $3M_\odot$ TAMS star on a $\beta=1$ orbit around a $10^6M_\odot$ SMBH, and specifically Figure~\ref{fig:omegavsN} shows that the surviving stellar core gets spun up by a small fraction of its breakup angular velocity as a result of the tidal interaction, and the average angular velocity of the core is a monotonically increasing function of the number of pericenter passages. In the limit of a large number of pericenter passages, the core gets spun up to $0.07 \Omega_{\rm c}$ (where $\Omega_{\rm c}$ is its breakup angular velocity). In terms of the specific angular velocity of the COM orbit, $\Omega_{\rm p} = \sqrt{GM_\bullet(1+e)/r_{\rm p}^3}$, this corresponds to an angular valocity of $1.4\Omega_{\rm p}$ in the asymptotic limit. However, we note that once the star is spun up to $\sim \Omega_{\rm p}$, the black hole appears stationary in its co-rotating frame, implying that the star will not be torqued beyond this value, and we expect the angular velocity of the star to asymptote to a value of $\Omega_{\rm p}$. The tidal spin-up of the star has important implications for the survivability of the star on subsequent orbits. Since the star is spun up in a prograde sense through the tidal interaction, its effective tidal radius increases, (see Equation 6 of \citealp{golightly19b}) making it more susceptible to mass loss on subsequent encounters. This counter-balances the enhanced stability that results from mass loss to some degree, and explains, for e.g., the increased susceptibility of the $1M_\odot$ ZAMS star to mass loss on the $\beta=1$ orbit (in which it loses $\sim10\%$ of its mass during the first encounter) despite the model prediction being an increase in the average density of the star by a factor of $\sim2$ relative to its original average density. The prograde spin up of the stellar core also shifts the peak timescale $t_{\rm peak}$ of the mass fallback rates to shorter times ($t_{\rm peak} \propto(1+\Omega/\Omega_{\rm p})^{-3/2}$; \citealp{golightly19b}). Thus, if an rpTDE is observed during its first few orbits, it can exhibit a significant variation in its peak timescale, as illustrated in Figure \ref{fig:3msun-fallbackrates} for the $3M_\odot$ TAMS star.

To test the resolution independence of the tidal spin-up of the stellar core, we performed additional simulations for the $3M_\odot$ TAMS star and the $1M_\odot$ ZAMS star at lower and higher resolutions (see Appendix B of \citealp{bandopadhyay24} and Figure 9 of \citealp{bandopadhyay25} for resolution tests), which corroborate the origin of the spin up proposed in \cite{kochanek92}, namely, that the bulk rotation of the core arises as a result of the non-linear coupling between the $\ell=2,m=\pm2$ modes. The decay in the oscillatory component of the kinetic energy and the increase in the rotational kinetic energy of the star, which is independent of particle resolution (see Figure 10 of \citealp{bandopadhyay25}), is consistent with the interpretation of the dissipation of the kinetic energy of the tidally excited modes giving rise to bulk rotation of the stellar core.

\subsection{Energetics of the Surviving Core and Connection to ASASSN-14ko's flares}
\begin{figure}
    \includegraphics[width=0.48\textwidth]{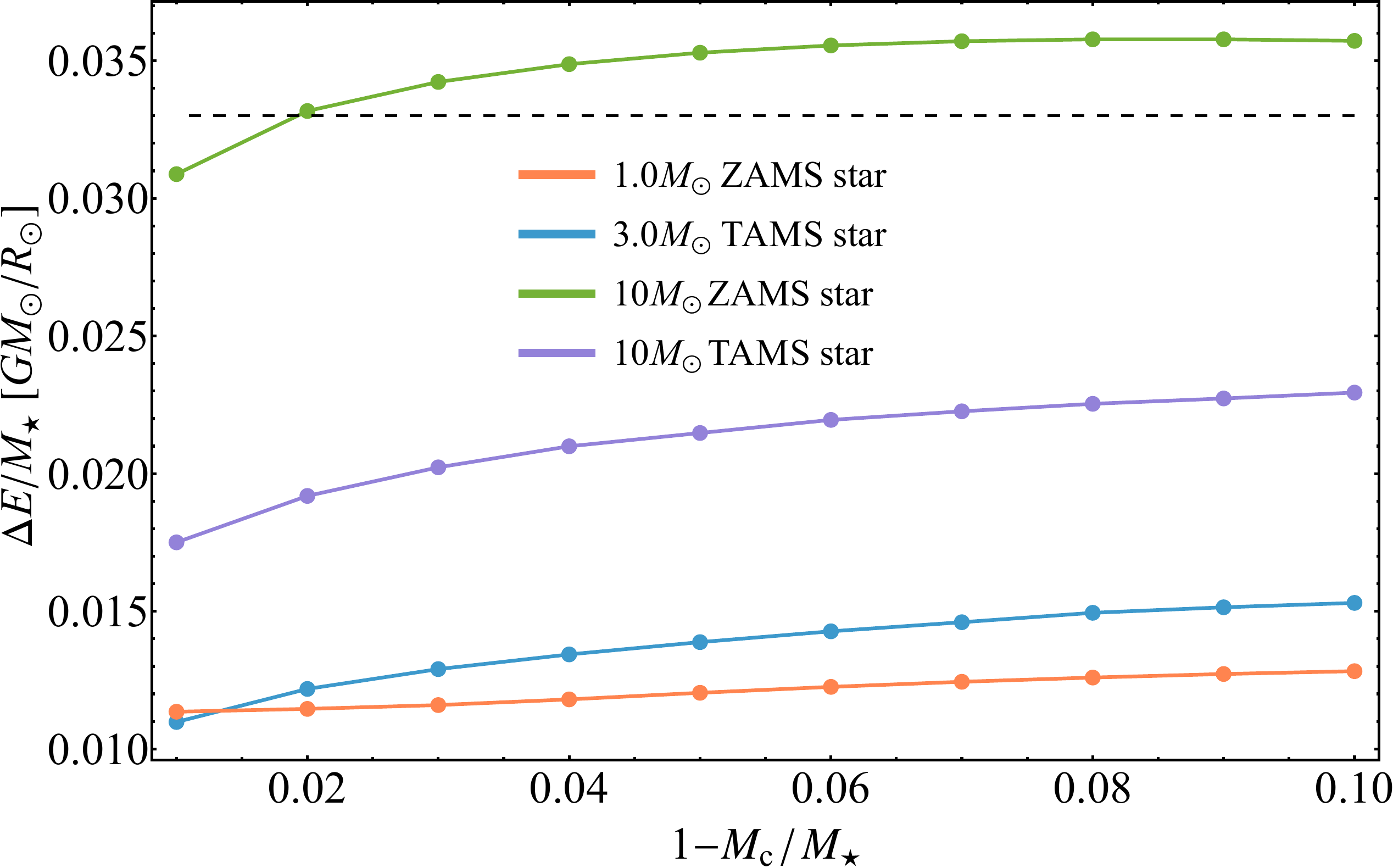}
    \caption{\hspace*{0.2cm} The per-orbit change in the specific energy of the core (in solar units, $GM_\odot/R_\odot)$ as a function of the cumulative amount of mass lost, for stars losing $1\%$ of their original mass on each encounter. The black dashed line shows the per orbit change in energy required to yield a period derivative $\dot{P} = -0.001$, on a $114$ day orbit around a $10^6M_\odot$ SMBH.} \label{fig:deltae-per-orbit}
\end{figure}

The average energy calculated using the eigenmodes of the asymptotic background state in Section~\ref{sec:mass-loss-model} (which are consistent with those imparted to the hydrodynamical simulations of the tidal disruption of a star) demonstrate that the individual contributions from the change in the gravitational and thermal energies of the core exceed the kinetic energy of oscillations by $\gtrsim2-3$ orders of magnitude. The total change in the energy of a star undergoing small fractional mass loss in an rpTDE is a small positive fraction ($\lesssim10\%$) of its binding energy, the exact value being determined by the difference between the binding energy of the original star and the sum of the binding energy of the core and the $p\,dV$ work done by the expansion of the core (in a rpTDE, the rotational kinetic energy of the tidally spun up core also contributes to the total energy change, which our model neglects).

In the standard tidal dissipation picture \citep{fabian75,press77}, the excitation and subsequent dissipation of non-radial oscillatory modes leads to the tidal heating of a star, causing it to become inflated, and thus more susceptible to unstable mass transfer in subsequent encounters \citep{ray87,mcmillan87,gu04,linial24,koenigsberger24}. The agreement between the different energy contributions calculated using the model, and those obtained from the {\sc phantom} simulations (as shown in Figure \ref{fig:ZAMS1-sph-energies}) demonstrates that this picture does not extend to the limit in which a star loses some fraction of its mass, and the kinetic energy of the oscillatory modes is orders of magnitude smaller than the total change in the energy (and the individual contributions from the change in the gravitational and thermal energies), which is primarily determined by the difference between the binding energies of the post-mass-loss core and the original star, the $p \,dV$ work done by the expansion of the core, and its rotational kinetic energy. 

Our results suggest that the binding energy of the envelope can add as an additional sink for the orbital energy, that in the rpTDE case corresponds to the work that needs to be done by the tidal field on the star to remove the envelope. Since this energy comes at the expense of the orbital energy, we can use the Keplerian energy-period relation to estimate the ensuing change in the orbital period. Figure \ref{fig:deltae-per-orbit} shows the change in the specific energy $\Delta \epsilon = \Delta E / M_\star$ of a star losing $1\%$ of its original mass per-orbit, as a function of the total amount of mass lost. For a $10^6M_\odot$ SMBH and a $114$ day orbital period (akin to that of ASASSN-14ko), the change in specific energy per-orbit required to yield a period derivative $\dot{P}=-0.001$ is $\Delta \epsilon = 0.03 GM_\odot/R_\odot$. The figure shows that for a $10^6M_\odot$ SMBH, the required change in specific energy can be achieved by a high mass star losing $\gtrsim1\%$ of its mass per orbit. For a higher mass SMBH, the required change in specific energy increases ($\Delta \epsilon \propto M_\bullet$), requiring a higher fraction of mass to be lost per orbit to maintain a constant period derivative $\dot{P}\approx-0.001$, thus posing a challenge to using this energy sink as an explanation for the observed period derivative of ASASSN-14ko ($\dot{P}=-0.0026$), and requiring suggesting the need to consider additional sources of dissipation, such as hydrodynamical drag as a star interacts with an accretion disk \citep{linial24}.

\section{Possible relation to quasi-periodic eruptions}
\label{sec:qpes}
As noted in the Introduction \ref{sec:intro}, a second type of repeating nuclear transient was discovered roughly contemporaneously with rpTDEs, the class known as quasi-periodic eruptions, or QPEs \citep{miniutti19,giustini20,arcodia21}. Unlike rpTDEs, which have recurrence times on the order of months to individual years and exhibit outbursts in the optical and/or X-ray (both soft-thermal and hard-nonthermal), QPEs recur over $\sim 1$-day periods (with durations $\sim$ single-hours, or duty cycles $\sim$ tens of percent) and emit exclusively in the soft X-ray, with temperatures between $0.1-1$ keV. QPEs have now been definitely linked to TDEs, the first of which being AT2019qiz, which was first discussed by \citet{nicholl20} as a TDE in a nearby galaxy, and later was shown to exhibit QPEs \citep{nicholl24}. Other TDEs to show subsequent QPEs include AT2019vcb \citep{quintin23,bykov25}, AT2022upj \citep{chakraborty25}, and ZTF19acnskyy (also known as ``Ansky''; \citealp{hernandez25}).

That TDEs and QPEs are causally related is thus indubitable, but whether they stem from the same -- or at least similar -- underlying phenomenon is less clear. The original QPE sources were detected without associated optical/UV TDEs, and models invoked a star interacting with a pre-existing accretion flow \citep{linial23, franchini23}, an orbiter transferring mass through Roche lobe overflow \citep{krolik24, lu23}, gravitational lensing by a binary \citep{ingram21}, the late-time return of gravitationally condensed ``knots'' from a past TDE \citep{coughlin20}, and accretion disc instabilities \citep{sniegowska20,raj21,pan25}. The model proposed by\footnote{Note that \citet{zalamea10} studied this scenario for white dwarfs on circular orbits near the tidal disruption radius, having been brought in to such close distances through gravitational-wave emission.} \citet{king20} was perhaps most closely related to the rpTDE paradigm, and suggested that a white dwarf orbiting a massive black hole being repeatedly stripped of mass (and yielding associated accretion flares) powered the outburst. An aspect of this model that lends to its plausibility is that the return time of the debris in such a system is $\sim$ hours (being $\sim$ the square root of the mass ratio times the dynamical time of the star; see, e.g., the simulations in \citealt{garain24, oconnor25}), which is comparable to the flare duration from QPEs.

However, there are some issues with this interpretation, the most relevant of which (given the content contained in this proceedings) is that white dwarfs are effectively $\gamma=5/3$ polytropes, expand in response to mass loss, and are thus unstable to mass transfer \citep{hjellming87,zalamea10,dai13,ge15,bandopadhyay25,yao25}. Such systems should therefore rapidly undergo tidal disruption, and flares should become brighter with time, which is not observed (see \citet{lu23, pasham24b, pasham24c} for additional discussion of problems with the white dwarf disruption picture). Other models have also come under some level of scrutiny and have difficulty explaining one or more features of newly detected systems, e.g., \citet{guo25} pointed out that the energetics involved in star-disc collisions is orders of magnitude too small to explain AT2022upj and Ansky. As noted in \citet{guolo24}, it is also not clear why the knots formed from the gravitational instability of the debris stream from a past TDE would produce regularly spaced outbursts, given the stochastic nature of the instability and the prediction that their spacing should lengthen with time \citep{coughlin-metzger-barnes20b}. Contrarily, it seems difficult to produce highly \emph{irregular} outbursts, such as those seen in eRO-QPE2 \citep{arcodia21}, RX J1301.9+2747 \citep{giustini24} and Ansky, with a model as periodic as an object orbiting a black hole (even with the various degrees of precession that can be induced by relativistic gravity). Models based on gravitational lensing by a binary companion are disfavored due to their inability to reproduce the asymmetry (fast rise, slow decay) and flux magnification in observed QPE lightcurves, and the strong energy dependence (lensing would predict achromatic; \citealp{arcodia22,miniutti23b}). Finally, disc-instability models are not yet sufficiently developed to enable direct comparison with observations. It is therefore, at present, difficult to assess whether they represent a compelling scenario to explain the observations \citep{pasham24b}. 

Thus, while the relation between TDEs and QPEs is now firmly and observationally established, the underlying mechanism powering the latter remains unclear.

\section{Summary and Future Prospects}
\label{sec:summary}
In this work, we reviewed the current observational status and theoretical understanding of the relatively nascent field of repeating nuclear transients arising from the continued interactions between bound stars and SMBHs in galactic centers, focusing primarily on rpTDEs, which are produced by the repeated tidal stripping of bound stars by SMBHs. Through a combinantion of hydrodynamical simulations \citep{bandopadhyay24} and analytical estimates for the stability of a star to mass loss \citep{bandopadhyay25}, we showed that events that exhibit tens to hundreds of outbursts (e.g., ASASSN-14ko; \citealp{payne21}) are most likely produced by high mass and evolved stars, which have a centrally concentrated core and contract in response to mass loss, making them promising candidates for surviving many mass stripping encounters. Contrarily, low mass and less evolved stars expand in response to mass loss, and can thus power a few flares of progressively growing magnitudes (as seen for the case of AT2020vdq; \citealt{somalwar23}) or (in cases where the impact parameter $\beta$ is close to its critical value for complete disruption) $\sim 2$ flares, with the second being dimmer than the first.

While these simulations of the hyrdodynamical interactions between a single (initially non-rotating) star and an SMBH can explain the nature of the observed lightcurves for some rpTDEs, they do not reproduce all possible outcomes, in particular, that of generating multiple progressively dimming flares, as seen in, e.g., eRASSt-J045650 \citep{liu24}. Other aspects of rpTDE systems, such as a possible sink of orbital energy that could yield the observed period derivative of ASASSN-14ko, also remain elusive. Since rpTDE systems have been shown to form through the Hills capture of a star and the simultaneous ejection of its binary companion \citep{cufari22}, studying the detailed hydrodynamical evolution of the Hills breakup of binary systems and the possible initial conditions it can lead to for the captured star can lend possible explanations for some of these issues, and improve our theoretical understanding of how rpTDE systems form and evolve.

Several methods have been proposed to increase our discovery potential for these repeating sources, and further explore the connection between TDEs and QPEs. Following up observed TDEs to track possible QPE connections would serve as a challenging but promising way to detect more QPE sources. \cite{pasham25} demonstrated an alternative and novel approach of studying the long term infrared lightcurve of a TDE to observe the reprocessed dust emission that reveals the presence of a QPE in AT2019qiz, suggesting that following up optical TDEs with wide-field infrared surveys could reveal more such sources. As the number of observed TDEs is set to undergo a significant increase, in the age of the Vera Rubin observatory \citep{bricman20}, discovering more of these sources would pave the pathway for an improved understanding of the SMBH population in our universe, as well as the population of stars giving rise to repeating nuclear transients.

\section*{Acknowledgments}
A.B.~acknowledges support from \fundingAgency{NASA} through the \fundingAgency{FINESST} program, grant \fundingNumber{80NSSC24K1548}. E.R.C.~acknowledges support from \fundingAgency{NASA} through the \fundingAgency{Astrophysics Theory Program}, grant \fundingNumber{80NSSC24K0897}. 
J.F.~acknowledges support from \fundingAgency{Syracuse University SOURCE program} and an \fundingNumber{Emerging Research Fellows award}. C.J.N.~acknowledges support from \fundingAgency{The Leverhulme Trust} (\fundingNumber{grant No. RPG-2021-380 }). This research was supported in part by grant \fundingAgency{NSF} \fundingNumber{PHY-2309135} to the Kavli Institute for Theoretical Physics (KITP).


\begin{thebibliography}{}

\bibitem [\protect \citeauthoryear {%
{Arcodia}%
\ \protect \BOthers {.}}{%
{Arcodia}%
\ \protect \BOthers {.}}{%
{\protect \APACyear {2024}}%
}]{%
arcodia24}
\APACinsertmetastar {%
arcodia24}%
\begin{APACrefauthors}%
{Arcodia}, R.%
, {Liu}, Z.%
, {Merloni}, A.%
\ et al.\end{APACrefauthors}%
\unskip\
\newblock
\APACrefYearMonthDay{2024}{{\APACmonth{04}}}{},
\newblock
\unskip
\newblock
\APACjournalVolNumPages{\aap}{684}{}{A64}.
\newblock
\begin{APACrefDOI} 10.1051/0004-6361/202348881 \end{APACrefDOI}
\PrintBackRefs{\CurrentBib}

\bibitem [\protect \citeauthoryear {%
{Arcodia}%
\ \protect \BOthers {.}}{%
{Arcodia}%
\ \protect \BOthers {.}}{%
{\protect \APACyear {2021}}%
}]{%
arcodia21}
\APACinsertmetastar {%
arcodia21}%
\begin{APACrefauthors}%
{Arcodia}, R.%
, {Merloni}, A.%
, {Nandra}, K.%
\ et al.\end{APACrefauthors}%
\unskip\
\newblock
\APACrefYearMonthDay{2021}{{\APACmonth{04}}}{},
\newblock
\unskip
\newblock
\APACjournalVolNumPages{\nat}{592}{7856}{704-707}.
\newblock
\begin{APACrefDOI} 10.1038/s41586-021-03394-6 \end{APACrefDOI}
\PrintBackRefs{\CurrentBib}

\bibitem [\protect \citeauthoryear {%
{Arcodia}%
\ \protect \BOthers {.}}{%
{Arcodia}%
\ \protect \BOthers {.}}{%
{\protect \APACyear {2022}}%
}]{%
arcodia22}
\APACinsertmetastar {%
arcodia22}%
\begin{APACrefauthors}%
{Arcodia}, R.%
, {Miniutti}, G.%
, {Ponti}, G.%
\ et al.\end{APACrefauthors}%
\unskip\
\newblock
\APACrefYearMonthDay{2022}{{\APACmonth{06}}}{},
\newblock
\unskip
\newblock
\APACjournalVolNumPages{\aap}{662}{}{A49}.
\newblock
\begin{APACrefDOI} 10.1051/0004-6361/202243259 \end{APACrefDOI}
\PrintBackRefs{\CurrentBib}

\bibitem [\protect \citeauthoryear {%
{Bandopadhyay}%
, {Coughlin}%
\BCBL {}\ \BBA {} {Nixon}%
}{%
{Bandopadhyay}%
\ \protect \BOthers {.}}{%
{\protect \APACyear {2025}}%
}]{%
bandopadhyay25}
\APACinsertmetastar {%
bandopadhyay25}%
\begin{APACrefauthors}%
{Bandopadhyay}, A.%
, {Coughlin}, E\BPBI R.%
\BCBL {}\ \BBA {} {Nixon}, C\BPBI J.%
\end{APACrefauthors}%
\unskip\
\newblock
\APACrefYearMonthDay{2025}{{\APACmonth{07}}}{},
\newblock
\unskip
\newblock
\APACjournalVolNumPages{\apj}{987}{1}{16}.
\newblock
\begin{APACrefDOI} 10.3847/1538-4357/add9a5 \end{APACrefDOI}
\PrintBackRefs{\CurrentBib}

\bibitem [\protect \citeauthoryear {%
{Bandopadhyay}%
, {Coughlin}%
, {Nixon}%
\BCBL {}\ \BBA {} {Pasham}%
}{%
{Bandopadhyay}%
, {Coughlin}%
\BCBL {}\ \protect \BOthers {.}}{%
{\protect \APACyear {2024}}%
}]{%
bandopadhyay24}
\APACinsertmetastar {%
bandopadhyay24}%
\begin{APACrefauthors}%
{Bandopadhyay}, A.%
, {Coughlin}, E\BPBI R.%
, {Nixon}, C\BPBI J.%
\BCBL {}\ \BBA {} {Pasham}, D\BPBI R.%
\end{APACrefauthors}%
\unskip\
\newblock
\APACrefYearMonthDay{2024}{{\APACmonth{10}}}{},
\newblock
\unskip
\newblock
\APACjournalVolNumPages{\apj}{974}{1}{80}.
\newblock
\begin{APACrefDOI} 10.3847/1538-4357/ad6a5a \end{APACrefDOI}
\PrintBackRefs{\CurrentBib}

\bibitem [\protect \citeauthoryear {%
{Bandopadhyay}%
, {Fancher}%
\BCBL {}\ \protect \BOthers {.}}{%
{Bandopadhyay}%
, {Fancher}%
\BCBL {}\ \protect \BOthers {.}}{%
{\protect \APACyear {2024}}%
}]{%
bandopadhyay24b}
\APACinsertmetastar {%
bandopadhyay24b}%
\begin{APACrefauthors}%
{Bandopadhyay}, A.%
, {Fancher}, J.%
, {Athian}, A.%
\ et al.\end{APACrefauthors}%
\unskip\
\newblock
\APACrefYearMonthDay{2024}{{\APACmonth{01}}}{},
\newblock
\unskip
\newblock
\APACjournalVolNumPages{\apjl}{961}{1}{L2}.
\newblock
\begin{APACrefDOI} 10.3847/2041-8213/ad0388 \end{APACrefDOI}
\PrintBackRefs{\CurrentBib}

\bibitem [\protect \citeauthoryear {%
{Bricman}%
\ \BBA {} {Gomboc}%
}{%
{Bricman}%
\ \BBA {} {Gomboc}%
}{%
{\protect \APACyear {2020}}%
}]{%
bricman20}
\APACinsertmetastar {%
bricman20}%
\begin{APACrefauthors}%
{Bricman}, K.%
\BCBT {}\ \BBA {} {Gomboc}, A.%
\end{APACrefauthors}%
\unskip\
\newblock
\APACrefYearMonthDay{2020}{{\APACmonth{02}}}{},
\newblock
\unskip
\newblock
\APACjournalVolNumPages{\apj}{890}{1}{73}.
\newblock
\begin{APACrefDOI} 10.3847/1538-4357/ab6989 \end{APACrefDOI}
\PrintBackRefs{\CurrentBib}

\bibitem [\protect \citeauthoryear {%
{Bykov}%
, {Gilfanov}%
, {Sunyaev}%
\BCBL {}\ \BBA {} {Medvedev}%
}{%
{Bykov}%
\ \protect \BOthers {.}}{%
{\protect \APACyear {2025}}%
}]{%
bykov25}
\APACinsertmetastar {%
bykov25}%
\begin{APACrefauthors}%
{Bykov}, S\BPBI D.%
, {Gilfanov}, M\BPBI R.%
, {Sunyaev}, R\BPBI A.%
\BCBL {}\ \BBA {} {Medvedev}, P\BPBI S.%
\end{APACrefauthors}%
\unskip\
\newblock
\APACrefYearMonthDay{2025}{{\APACmonth{06}}}{},
\newblock
\unskip
\newblock
\APACjournalVolNumPages{\mnras}{540}{1}{30-36}.
\newblock
\begin{APACrefDOI} 10.1093/mnras/staf686 \end{APACrefDOI}
\PrintBackRefs{\CurrentBib}

\bibitem [\protect \citeauthoryear {%
{Chakraborty}%
\ \protect \BOthers {.}}{%
{Chakraborty}%
\ \protect \BOthers {.}}{%
{\protect \APACyear {2025}}%
}]{%
chakraborty25}
\APACinsertmetastar {%
chakraborty25}%
\begin{APACrefauthors}%
{Chakraborty}, J.%
, {Kara}, E.%
, {Arcodia}, R.%
\ et al.\end{APACrefauthors}%
\unskip\
\newblock
\APACrefYearMonthDay{2025}{{\APACmonth{04}}}{},
\newblock
\unskip
\newblock
\APACjournalVolNumPages{\apjl}{983}{2}{L39}.
\newblock
\begin{APACrefDOI} 10.3847/2041-8213/adc2f8 \end{APACrefDOI}
\PrintBackRefs{\CurrentBib}

\bibitem [\protect \citeauthoryear {%
{Chakraborty}%
\ \protect \BOthers {.}}{%
{Chakraborty}%
\ \protect \BOthers {.}}{%
{\protect \APACyear {2021}}%
}]{%
chakraborty21}
\APACinsertmetastar {%
chakraborty21}%
\begin{APACrefauthors}%
{Chakraborty}, J.%
, {Kara}, E.%
, {Masterson}, M.%
, {Giustini}, M.%
, {Miniutti}, G.%
\BCBL {}\ \BBA {} {Saxton}, R.%
\end{APACrefauthors}%
\unskip\
\newblock
\APACrefYearMonthDay{2021}{{\APACmonth{11}}}{},
\newblock
\unskip
\newblock
\APACjournalVolNumPages{\apjl}{921}{2}{L40}.
\newblock
\begin{APACrefDOI} 10.3847/2041-8213/ac313b \end{APACrefDOI}
\PrintBackRefs{\CurrentBib}

\bibitem [\protect \citeauthoryear {%
{Coughlin}%
\ \BBA {} {Nixon}%
}{%
{Coughlin}%
\ \BBA {} {Nixon}%
}{%
{\protect \APACyear {2015}}%
}]{%
coughlin15}
\APACinsertmetastar {%
coughlin15}%
\begin{APACrefauthors}%
{Coughlin}, E\BPBI R.%
\BCBT {}\ \BBA {} {Nixon}, C.%
\end{APACrefauthors}%
\unskip\
\newblock
\APACrefYearMonthDay{2015}{{\APACmonth{07}}}{},
\newblock
\unskip
\newblock
\APACjournalVolNumPages{\apjl}{808}{1}{L11}.
\newblock
\begin{APACrefDOI} 10.1088/2041-8205/808/1/L11 \end{APACrefDOI}
\PrintBackRefs{\CurrentBib}

\bibitem [\protect \citeauthoryear {%
{Coughlin}%
\ \BBA {} {Nixon}%
}{%
{Coughlin}%
\ \BBA {} {Nixon}%
}{%
{\protect \APACyear {2019}}%
}]{%
coughlin19}
\APACinsertmetastar {%
coughlin19}%
\begin{APACrefauthors}%
{Coughlin}, E\BPBI R.%
\BCBT {}\ \BBA {} {Nixon}, C\BPBI J.%
\end{APACrefauthors}%
\unskip\
\newblock
\APACrefYearMonthDay{2019}{{\APACmonth{09}}}{},
\newblock
\unskip
\newblock
\APACjournalVolNumPages{\apjl}{883}{1}{L17}.
\newblock
\begin{APACrefDOI} 10.3847/2041-8213/ab412d \end{APACrefDOI}
\PrintBackRefs{\CurrentBib}

\bibitem [\protect \citeauthoryear {%
{Coughlin}%
\ \BBA {} {Nixon}%
}{%
{Coughlin}%
\ \BBA {} {Nixon}%
}{%
{\protect \APACyear {2020}}%
}]{%
coughlin20}
\APACinsertmetastar {%
coughlin20}%
\begin{APACrefauthors}%
{Coughlin}, E\BPBI R.%
\BCBT {}\ \BBA {} {Nixon}, C\BPBI J.%
\end{APACrefauthors}%
\unskip\
\newblock
\APACrefYearMonthDay{2020}{{\APACmonth{04}}}{},
\newblock
\unskip
\newblock
\APACjournalVolNumPages{\apjs}{247}{2}{51}.
\newblock
\begin{APACrefDOI} 10.3847/1538-4365/ab77c2 \end{APACrefDOI}
\PrintBackRefs{\CurrentBib}

\bibitem [\protect \citeauthoryear {%
{Coughlin}%
\ \BBA {} {Nixon}%
}{%
{Coughlin}%
\ \BBA {} {Nixon}%
}{%
{\protect \APACyear {2022}}%
{\protect \APACexlab {{\protect \BCnt {1}}}}}]{%
coughlin22}
\APACinsertmetastar {%
coughlin22}%
\begin{APACrefauthors}%
{Coughlin}, E\BPBI R.%
\BCBT {}\ \BBA {} {Nixon}, C\BPBI J.%
\end{APACrefauthors}%
\unskip\
\newblock
\APACrefYearMonthDay{2022{\protect \BCnt {1}}}{{\APACmonth{11}}}{},
\newblock
\unskip
\newblock
\APACjournalVolNumPages{\mnras}{517}{1}{L26-L30}.
\newblock
\begin{APACrefDOI} 10.1093/mnrasl/slac106 \end{APACrefDOI}
\PrintBackRefs{\CurrentBib}

\bibitem [\protect \citeauthoryear {%
{Coughlin}%
\ \BBA {} {Nixon}%
}{%
{Coughlin}%
\ \BBA {} {Nixon}%
}{%
{\protect \APACyear {2022}}%
{\protect \APACexlab {{\protect \BCnt {2}}}}}]{%
coughlin22b}
\APACinsertmetastar {%
coughlin22b}%
\begin{APACrefauthors}%
{Coughlin}, E\BPBI R.%
\BCBT {}\ \BBA {} {Nixon}, C\BPBI J.%
\end{APACrefauthors}%
\unskip\
\newblock
\APACrefYearMonthDay{2022{\protect \BCnt {2}}}{{\APACmonth{02}}}{},
\newblock
\unskip
\newblock
\APACjournalVolNumPages{\apj}{926}{1}{47}.
\newblock
\begin{APACrefDOI} 10.3847/1538-4357/ac3fb9 \end{APACrefDOI}
\PrintBackRefs{\CurrentBib}

\bibitem [\protect \citeauthoryear {%
{Coughlin}%
, {Nixon}%
, {Barnes}%
, {Metzger}%
\BCBL {}\ \BBA {} {Margutti}%
}{%
{Coughlin}%
\ \protect \BOthers {.}}{%
{\protect \APACyear {2020}}%
}]{%
coughlin-metzger-barnes20b}
\APACinsertmetastar {%
coughlin-metzger-barnes20b}%
\begin{APACrefauthors}%
{Coughlin}, E\BPBI R.%
, {Nixon}, C\BPBI J.%
, {Barnes}, J.%
, {Metzger}, B\BPBI D.%
\BCBL {}\ \BBA {} {Margutti}, R.%
\end{APACrefauthors}%
\unskip\
\newblock
\APACrefYearMonthDay{2020}{{\APACmonth{06}}}{},
\newblock
\unskip
\newblock
\APACjournalVolNumPages{\apjl}{896}{2}{L38}.
\newblock
\begin{APACrefDOI} 10.3847/2041-8213/ab9a4e \end{APACrefDOI}
\PrintBackRefs{\CurrentBib}

\bibitem [\protect \citeauthoryear {%
{Cufari}%
, {Coughlin}%
\BCBL {}\ \BBA {} {Nixon}%
}{%
{Cufari}%
\ \protect \BOthers {.}}{%
{\protect \APACyear {2022}}%
}]{%
cufari22}
\APACinsertmetastar {%
cufari22}%
\begin{APACrefauthors}%
{Cufari}, M.%
, {Coughlin}, E\BPBI R.%
\BCBL {}\ \BBA {} {Nixon}, C\BPBI J.%
\end{APACrefauthors}%
\unskip\
\newblock
\APACrefYearMonthDay{2022}{{\APACmonth{04}}}{},
\newblock
\unskip
\newblock
\APACjournalVolNumPages{\apjl}{929}{2}{L20}.
\newblock
\begin{APACrefDOI} 10.3847/2041-8213/ac6021 \end{APACrefDOI}
\PrintBackRefs{\CurrentBib}

\bibitem [\protect \citeauthoryear {%
{Cufari}%
, {Nixon}%
\BCBL {}\ \BBA {} {Coughlin}%
}{%
{Cufari}%
\ \protect \BOthers {.}}{%
{\protect \APACyear {2023}}%
}]{%
cufari23}
\APACinsertmetastar {%
cufari23}%
\begin{APACrefauthors}%
{Cufari}, M.%
, {Nixon}, C\BPBI J.%
\BCBL {}\ \BBA {} {Coughlin}, E\BPBI R.%
\end{APACrefauthors}%
\unskip\
\newblock
\APACrefYearMonthDay{2023}{{\APACmonth{03}}}{},
\newblock
\unskip
\newblock
\APACjournalVolNumPages{\mnras}{520}{1}{L38-L41}.
\newblock
\begin{APACrefDOI} 10.1093/mnrasl/slad001 \end{APACrefDOI}
\PrintBackRefs{\CurrentBib}

\bibitem [\protect \citeauthoryear {%
{Dai}%
, {Blandford}%
\BCBL {}\ \BBA {} {Eggleton}%
}{%
{Dai}%
\ \protect \BOthers {.}}{%
{\protect \APACyear {2013}}%
}]{%
dai13}
\APACinsertmetastar {%
dai13}%
\begin{APACrefauthors}%
{Dai}, L.%
, {Blandford}, R\BPBI D.%
\BCBL {}\ \BBA {} {Eggleton}, P\BPBI P.%
\end{APACrefauthors}%
\unskip\
\newblock
\APACrefYearMonthDay{2013}{{\APACmonth{10}}}{},
\newblock
\unskip
\newblock
\APACjournalVolNumPages{\mnras}{434}{4}{2940-2947}.
\newblock
\begin{APACrefDOI} 10.1093/mnras/stt1208 \end{APACrefDOI}
\PrintBackRefs{\CurrentBib}

\bibitem [\protect \citeauthoryear {%
{Evans}%
\ \protect \BOthers {.}}{%
{Evans}%
\ \protect \BOthers {.}}{%
{\protect \APACyear {2023}}%
}]{%
evans23}
\APACinsertmetastar {%
evans23}%
\begin{APACrefauthors}%
{Evans}, P\BPBI A.%
, {Nixon}, C\BPBI J.%
, {Campana}, S.%
\ et al.\end{APACrefauthors}%
\unskip\
\newblock
\APACrefYearMonthDay{2023}{{\APACmonth{11}}}{},
\newblock
\unskip
\newblock
\APACjournalVolNumPages{Nature Astronomy}{7}{}{1368-1375}.
\newblock
\begin{APACrefDOI} 10.1038/s41550-023-02073-y \end{APACrefDOI}
\PrintBackRefs{\CurrentBib}

\bibitem [\protect \citeauthoryear {%
{Fabian}%
, {Pringle}%
\BCBL {}\ \BBA {} {Rees}%
}{%
{Fabian}%
\ \protect \BOthers {.}}{%
{\protect \APACyear {1975}}%
}]{%
fabian75}
\APACinsertmetastar {%
fabian75}%
\begin{APACrefauthors}%
{Fabian}, A\BPBI C.%
, {Pringle}, J\BPBI E.%
\BCBL {}\ \BBA {} {Rees}, M\BPBI J.%
\end{APACrefauthors}%
\unskip\
\newblock
\APACrefYearMonthDay{1975}{{\APACmonth{08}}}{},
\newblock
\unskip
\newblock
\APACjournalVolNumPages{\mnras}{172}{}{15}.
\newblock
\begin{APACrefDOI} 10.1093/mnras/172.1.15P \end{APACrefDOI}
\PrintBackRefs{\CurrentBib}

\bibitem [\protect \citeauthoryear {%
{Fancher}%
, {Bandopadhyay}%
, {Coughlin}%
\BCBL {}\ \BBA {} {Nixon}%
}{%
{Fancher}%
\ \protect \BOthers {.}}{%
{\protect \APACyear {2025}}%
}]{%
fancher25}
\APACinsertmetastar {%
fancher25}%
\begin{APACrefauthors}%
{Fancher}, J.%
, {Bandopadhyay}, A.%
, {Coughlin}, E\BPBI R.%
\BCBL {}\ \BBA {} {Nixon}, C\BPBI J.%
\end{APACrefauthors}%
\unskip\
\newblock
\APACrefYearMonthDay{2025}{{\APACmonth{09}}}{},
\newblock
\unskip
\newblock
\APACjournalVolNumPages{\apj}{990}{2}{104}.
\newblock
\begin{APACrefDOI} 10.3847/1538-4357/adf33c \end{APACrefDOI}
\PrintBackRefs{\CurrentBib}

\bibitem [\protect \citeauthoryear {%
{Franchini}%
\ \protect \BOthers {.}}{%
{Franchini}%
\ \protect \BOthers {.}}{%
{\protect \APACyear {2023}}%
}]{%
franchini23}
\APACinsertmetastar {%
franchini23}%
\begin{APACrefauthors}%
{Franchini}, A.%
, {Bonetti}, M.%
, {Lupi}, A.%
\ et al.\end{APACrefauthors}%
\unskip\
\newblock
\APACrefYearMonthDay{2023}{{\APACmonth{07}}}{},
\newblock
\unskip
\newblock
\APACjournalVolNumPages{\aap}{675}{}{A100}.
\newblock
\begin{APACrefDOI} 10.1051/0004-6361/202346565 \end{APACrefDOI}
\PrintBackRefs{\CurrentBib}

\bibitem [\protect \citeauthoryear {%
{Fryxell}%
\ \protect \BOthers {.}}{%
{Fryxell}%
\ \protect \BOthers {.}}{%
{\protect \APACyear {2000}}%
}]{%
fryxell00}
\APACinsertmetastar {%
fryxell00}%
\begin{APACrefauthors}%
{Fryxell}, B.%
, {Olson}, K.%
, {Ricker}, P.%
\ et al.\end{APACrefauthors}%
\unskip\
\newblock
\APACrefYearMonthDay{2000}{{\APACmonth{11}}}{},
\newblock
\unskip
\newblock
\APACjournalVolNumPages{\apjs}{131}{1}{273-334}.
\newblock
\begin{APACrefDOI} 10.1086/317361 \end{APACrefDOI}
\PrintBackRefs{\CurrentBib}

\bibitem [\protect \citeauthoryear {%
{Garain}%
\ \BBA {} {Sarkar}%
}{%
{Garain}%
\ \BBA {} {Sarkar}%
}{%
{\protect \APACyear {2024}}%
}]{%
garain24}
\APACinsertmetastar {%
garain24}%
\begin{APACrefauthors}%
{Garain}, D.%
\BCBT {}\ \BBA {} {Sarkar}, T.%
\end{APACrefauthors}%
\unskip\
\newblock
\APACrefYearMonthDay{2024}{{\APACmonth{06}}}{},
\newblock
\unskip
\newblock
\APACjournalVolNumPages{\apj}{967}{2}{167}.
\newblock
\begin{APACrefDOI} 10.3847/1538-4357/ad3dfa \end{APACrefDOI}
\PrintBackRefs{\CurrentBib}

\bibitem [\protect \citeauthoryear {%
{Ge}%
, {Webbink}%
, {Chen}%
\BCBL {}\ \BBA {} {Han}%
}{%
{Ge}%
\ \protect \BOthers {.}}{%
{\protect \APACyear {2015}}%
}]{%
ge15}
\APACinsertmetastar {%
ge15}%
\begin{APACrefauthors}%
{Ge}, H.%
, {Webbink}, R\BPBI F.%
, {Chen}, X.%
\BCBL {}\ \BBA {} {Han}, Z.%
\end{APACrefauthors}%
\unskip\
\newblock
\APACrefYearMonthDay{2015}{{\APACmonth{10}}}{},
\newblock
\unskip
\newblock
\APACjournalVolNumPages{\apj}{812}{1}{40}.
\newblock
\begin{APACrefDOI} 10.1088/0004-637X/812/1/40 \end{APACrefDOI}
\PrintBackRefs{\CurrentBib}

\bibitem [\protect \citeauthoryear {%
{Giustini}%
\ \protect \BOthers {.}}{%
{Giustini}%
\ \protect \BOthers {.}}{%
{\protect \APACyear {2024}}%
}]{%
giustini24}
\APACinsertmetastar {%
giustini24}%
\begin{APACrefauthors}%
{Giustini}, M.%
, {Miniutti}, G.%
, {Arcodia}, R.%
\ et al.\end{APACrefauthors}%
\unskip\
\newblock
\APACrefYearMonthDay{2024}{{\APACmonth{12}}}{},
\newblock
\unskip
\newblock
\APACjournalVolNumPages{\aap}{692}{}{A15}.
\newblock
\begin{APACrefDOI} 10.1051/0004-6361/202450861 \end{APACrefDOI}
\PrintBackRefs{\CurrentBib}

\bibitem [\protect \citeauthoryear {%
{Giustini}%
, {Miniutti}%
\BCBL {}\ \BBA {} {Saxton}%
}{%
{Giustini}%
\ \protect \BOthers {.}}{%
{\protect \APACyear {2020}}%
}]{%
giustini20}
\APACinsertmetastar {%
giustini20}%
\begin{APACrefauthors}%
{Giustini}, M.%
, {Miniutti}, G.%
\BCBL {}\ \BBA {} {Saxton}, R\BPBI D.%
\end{APACrefauthors}%
\unskip\
\newblock
\APACrefYearMonthDay{2020}{{\APACmonth{04}}}{},
\newblock
\unskip
\newblock
\APACjournalVolNumPages{\aap}{636}{}{L2}.
\newblock
\begin{APACrefDOI} 10.1051/0004-6361/202037610 \end{APACrefDOI}
\PrintBackRefs{\CurrentBib}

\bibitem [\protect \citeauthoryear {%
{Golightly}%
, {Coughlin}%
\BCBL {}\ \BBA {} {Nixon}%
}{%
{Golightly}%
, {Coughlin}%
\BCBL {}\ \BBA {} {Nixon}%
}{%
{\protect \APACyear {2019}}%
}]{%
golightly19b}
\APACinsertmetastar {%
golightly19b}%
\begin{APACrefauthors}%
{Golightly}, E\BPBI C\BPBI A.%
, {Coughlin}, E\BPBI R.%
\BCBL {}\ \BBA {} {Nixon}, C\BPBI J.%
\end{APACrefauthors}%
\unskip\
\newblock
\APACrefYearMonthDay{2019}{{\APACmonth{02}}}{},
\newblock
\unskip
\newblock
\APACjournalVolNumPages{\apj}{872}{2}{163}.
\newblock
\begin{APACrefDOI} 10.3847/1538-4357/aafd2f \end{APACrefDOI}
\PrintBackRefs{\CurrentBib}

\bibitem [\protect \citeauthoryear {%
{Golightly}%
, {Nixon}%
\BCBL {}\ \BBA {} {Coughlin}%
}{%
{Golightly}%
, {Nixon}%
\BCBL {}\ \BBA {} {Coughlin}%
}{%
{\protect \APACyear {2019}}%
}]{%
golightly19}
\APACinsertmetastar {%
golightly19}%
\begin{APACrefauthors}%
{Golightly}, E\BPBI C\BPBI A.%
, {Nixon}, C\BPBI J.%
\BCBL {}\ \BBA {} {Coughlin}, E\BPBI R.%
\end{APACrefauthors}%
\unskip\
\newblock
\APACrefYearMonthDay{2019}{{\APACmonth{09}}}{},
\newblock
\unskip
\newblock
\APACjournalVolNumPages{\apjl}{882}{2}{L26}.
\newblock
\begin{APACrefDOI} 10.3847/2041-8213/ab380d \end{APACrefDOI}
\PrintBackRefs{\CurrentBib}

\bibitem [\protect \citeauthoryear {%
{Gu}%
, {Bodenheimer}%
\BCBL {}\ \BBA {} {Lin}%
}{%
{Gu}%
\ \protect \BOthers {.}}{%
{\protect \APACyear {2004}}%
}]{%
gu04}
\APACinsertmetastar {%
gu04}%
\begin{APACrefauthors}%
{Gu}, P\BHBI G.%
, {Bodenheimer}, P\BPBI H.%
\BCBL {}\ \BBA {} {Lin}, D\BPBI N\BPBI C.%
\end{APACrefauthors}%
\unskip\
\newblock
\APACrefYearMonthDay{2004}{{\APACmonth{06}}}{},
\newblock
\unskip
\newblock
\APACjournalVolNumPages{\apj}{608}{2}{1076-1094}.
\newblock
\begin{APACrefDOI} 10.1086/420867 \end{APACrefDOI}
\PrintBackRefs{\CurrentBib}

\bibitem [\protect \citeauthoryear {%
{Guo}%
\ \BBA {} {Shen}%
}{%
{Guo}%
\ \BBA {} {Shen}%
}{%
{\protect \APACyear {2025}}%
}]{%
guo25}
\APACinsertmetastar {%
guo25}%
\begin{APACrefauthors}%
{Guo}, W.%
\BCBT {}\ \BBA {} {Shen}, R\BHBI F.%
\end{APACrefauthors}%
\unskip\
\newblock
\APACrefYearMonthDay{2025}{{\APACmonth{04}}}{},
\newblock
\unskip
\newblock
\APACjournalVolNumPages{arXiv e-prints}{}{}{arXiv:2504.12762}.
\newblock
\begin{APACrefDOI} 10.48550/arXiv.2504.12762 \end{APACrefDOI}
\PrintBackRefs{\CurrentBib}

\bibitem [\protect \citeauthoryear {%
{Guolo}%
\ \protect \BOthers {.}}{%
{Guolo}%
\ \protect \BOthers {.}}{%
{\protect \APACyear {2024}}%
}]{%
guolo24}
\APACinsertmetastar {%
guolo24}%
\begin{APACrefauthors}%
{Guolo}, M.%
, {Pasham}, D\BPBI R.%
, {Zaja{\v{c}}ek}, M.%
\ et al.\end{APACrefauthors}%
\unskip\
\newblock
\APACrefYearMonthDay{2024}{{\APACmonth{03}}}{},
\newblock
\unskip
\newblock
\APACjournalVolNumPages{Nature Astronomy}{8}{}{347-358}.
\newblock
\begin{APACrefDOI} 10.1038/s41550-023-02178-4 \end{APACrefDOI}
\PrintBackRefs{\CurrentBib}

\bibitem [\protect \citeauthoryear {%
{Hern{\'a}ndez-Garc{\'\i}a}%
\ \protect \BOthers {.}}{%
{Hern{\'a}ndez-Garc{\'\i}a}%
\ \protect \BOthers {.}}{%
{\protect \APACyear {2025}}%
}]{%
hernandez25}
\APACinsertmetastar {%
hernandez25}%
\begin{APACrefauthors}%
{Hern{\'a}ndez-Garc{\'\i}a}, L.%
, {Chakraborty}, J.%
, {S{\'a}nchez-S{\'a}ez}, P.%
\ et al.\end{APACrefauthors}%
\unskip\
\newblock
\APACrefYearMonthDay{2025}{{\APACmonth{06}}}{},
\newblock
\unskip
\newblock
\APACjournalVolNumPages{Nature Astronomy}{9}{}{895-906}.
\newblock
\begin{APACrefDOI} 10.1038/s41550-025-02523-9 \end{APACrefDOI}
\PrintBackRefs{\CurrentBib}

\bibitem [\protect \citeauthoryear {%
{Hills}%
}{%
{Hills}%
}{%
{\protect \APACyear {1975}}%
}]{%
hills75}
\APACinsertmetastar {%
hills75}%
\begin{APACrefauthors}%
{Hills}, J\BPBI G.%
\end{APACrefauthors}%
\unskip\
\newblock
\APACrefYearMonthDay{1975}{{\APACmonth{03}}}{},
\newblock
\unskip
\newblock
\APACjournalVolNumPages{\nat}{254}{5498}{295-298}.
\newblock
\begin{APACrefDOI} 10.1038/254295a0 \end{APACrefDOI}
\PrintBackRefs{\CurrentBib}

\bibitem [\protect \citeauthoryear {%
{Hills}%
}{%
{Hills}%
}{%
{\protect \APACyear {1988}}%
}]{%
hills88}
\APACinsertmetastar {%
hills88}%
\begin{APACrefauthors}%
{Hills}, J\BPBI G.%
\end{APACrefauthors}%
\unskip\
\newblock
\APACrefYearMonthDay{1988}{{\APACmonth{02}}}{},
\newblock
\unskip
\newblock
\APACjournalVolNumPages{\nat}{331}{6158}{687-689}.
\newblock
\begin{APACrefDOI} 10.1038/331687a0 \end{APACrefDOI}
\PrintBackRefs{\CurrentBib}

\bibitem [\protect \citeauthoryear {%
{Hinkle}%
\ \protect \BOthers {.}}{%
{Hinkle}%
\ \protect \BOthers {.}}{%
{\protect \APACyear {2024}}%
}]{%
hinkle24}
\APACinsertmetastar {%
hinkle24}%
\begin{APACrefauthors}%
{Hinkle}, J\BPBI T.%
, {Auchettl}, K.%
, {Hoogendam}, W\BPBI B.%
\ et al.\end{APACrefauthors}%
\unskip\
\newblock
\APACrefYearMonthDay{2024}{{\APACmonth{12}}}{},
\newblock
\unskip
\newblock
\APACjournalVolNumPages{arXiv e-prints}{}{}{arXiv:2412.15326}.
\newblock
\begin{APACrefDOI} 10.48550/arXiv.2412.15326 \end{APACrefDOI}
\PrintBackRefs{\CurrentBib}

\bibitem [\protect \citeauthoryear {%
{Hjellming}%
\ \BBA {} {Webbink}%
}{%
{Hjellming}%
\ \BBA {} {Webbink}%
}{%
{\protect \APACyear {1987}}%
}]{%
hjellming87}
\APACinsertmetastar {%
hjellming87}%
\begin{APACrefauthors}%
{Hjellming}, M\BPBI S.%
\BCBT {}\ \BBA {} {Webbink}, R\BPBI F.%
\end{APACrefauthors}%
\unskip\
\newblock
\APACrefYearMonthDay{1987}{{\APACmonth{07}}}{},
\newblock
\unskip
\newblock
\APACjournalVolNumPages{\apj}{318}{}{794}.
\newblock
\begin{APACrefDOI} 10.1086/165412 \end{APACrefDOI}
\PrintBackRefs{\CurrentBib}

\bibitem [\protect \citeauthoryear {%
{Ingram}%
, {Motta}%
, {Aigrain}%
\BCBL {}\ \BBA {} {Karastergiou}%
}{%
{Ingram}%
\ \protect \BOthers {.}}{%
{\protect \APACyear {2021}}%
}]{%
ingram21}
\APACinsertmetastar {%
ingram21}%
\begin{APACrefauthors}%
{Ingram}, A.%
, {Motta}, S\BPBI E.%
, {Aigrain}, S.%
\BCBL {}\ \BBA {} {Karastergiou}, A.%
\end{APACrefauthors}%
\unskip\
\newblock
\APACrefYearMonthDay{2021}{{\APACmonth{05}}}{},
\newblock
\unskip
\newblock
\APACjournalVolNumPages{\mnras}{503}{2}{1703-1716}.
\newblock
\begin{APACrefDOI} 10.1093/mnras/stab609 \end{APACrefDOI}
\PrintBackRefs{\CurrentBib}

\bibitem [\protect \citeauthoryear {%
{Khokhlov}%
, {Novikov}%
\BCBL {}\ \BBA {} {Pethick}%
}{%
{Khokhlov}%
\ \protect \BOthers {.}}{%
{\protect \APACyear {1993}}%
}]{%
khokhlov93}
\APACinsertmetastar {%
khokhlov93}%
\begin{APACrefauthors}%
{Khokhlov}, A.%
, {Novikov}, I\BPBI D.%
\BCBL {}\ \BBA {} {Pethick}, C\BPBI J.%
\end{APACrefauthors}%
\unskip\
\newblock
\APACrefYearMonthDay{1993}{{\APACmonth{11}}}{},
\newblock
\unskip
\newblock
\APACjournalVolNumPages{\apj}{418}{}{181}.
\newblock
\begin{APACrefDOI} 10.1086/173380 \end{APACrefDOI}
\PrintBackRefs{\CurrentBib}

\bibitem [\protect \citeauthoryear {%
{King}%
}{%
{King}%
}{%
{\protect \APACyear {2020}}%
}]{%
king20}
\APACinsertmetastar {%
king20}%
\begin{APACrefauthors}%
{King}, A.%
\end{APACrefauthors}%
\unskip\
\newblock
\APACrefYearMonthDay{2020}{{\APACmonth{03}}}{},
\newblock
\unskip
\newblock
\APACjournalVolNumPages{\mnras}{493}{1}{L120-L123}.
\newblock
\begin{APACrefDOI} 10.1093/mnrasl/slaa020 \end{APACrefDOI}
\PrintBackRefs{\CurrentBib}

\bibitem [\protect \citeauthoryear {%
{Kochanek}%
}{%
{Kochanek}%
}{%
{\protect \APACyear {1992}}%
}]{%
kochanek92}
\APACinsertmetastar {%
kochanek92}%
\begin{APACrefauthors}%
{Kochanek}, C\BPBI S.%
\end{APACrefauthors}%
\unskip\
\newblock
\APACrefYearMonthDay{1992}{{\APACmonth{02}}}{},
\newblock
\unskip
\newblock
\APACjournalVolNumPages{\apj}{385}{}{604}.
\newblock
\begin{APACrefDOI} 10.1086/170966 \end{APACrefDOI}
\PrintBackRefs{\CurrentBib}

\bibitem [\protect \citeauthoryear {%
{Koenigsberger}%
\ \BBA {} {Estrella-Trujillo}%
}{%
{Koenigsberger}%
\ \BBA {} {Estrella-Trujillo}%
}{%
{\protect \APACyear {2024}}%
}]{%
koenigsberger24}
\APACinsertmetastar {%
koenigsberger24}%
\begin{APACrefauthors}%
{Koenigsberger}, G.%
\BCBT {}\ \BBA {} {Estrella-Trujillo}, D.%
\end{APACrefauthors}%
\unskip\
\newblock
\APACrefYearMonthDay{2024}{{\APACmonth{05}}}{},
\newblock
\unskip
\newblock
\APACjournalVolNumPages{\aap}{685}{}{A145}.
\newblock
\begin{APACrefDOI} 10.1051/0004-6361/202349075 \end{APACrefDOI}
\PrintBackRefs{\CurrentBib}

\bibitem [\protect \citeauthoryear {%
{Krolik}%
\ \BBA {} {Linial}%
}{%
{Krolik}%
\ \BBA {} {Linial}%
}{%
{\protect \APACyear {2022}}%
}]{%
krolik24}
\APACinsertmetastar {%
krolik24}%
\begin{APACrefauthors}%
{Krolik}, J\BPBI H.%
\BCBT {}\ \BBA {} {Linial}, I.%
\end{APACrefauthors}%
\unskip\
\newblock
\APACrefYearMonthDay{2022}{{\APACmonth{12}}}{},
\newblock
\unskip
\newblock
\APACjournalVolNumPages{\apj}{941}{1}{24}.
\newblock
\begin{APACrefDOI} 10.3847/1538-4357/ac9eb6 \end{APACrefDOI}
\PrintBackRefs{\CurrentBib}

\bibitem [\protect \citeauthoryear {%
{Kumar}%
\ \BBA {} {Goodman}%
}{%
{Kumar}%
\ \BBA {} {Goodman}%
}{%
{\protect \APACyear {1996}}%
}]{%
kumar96}
\APACinsertmetastar {%
kumar96}%
\begin{APACrefauthors}%
{Kumar}, P.%
\BCBT {}\ \BBA {} {Goodman}, J.%
\end{APACrefauthors}%
\unskip\
\newblock
\APACrefYearMonthDay{1996}{{\APACmonth{08}}}{},
\newblock
\unskip
\newblock
\APACjournalVolNumPages{\apj}{466}{}{946}.
\newblock
\begin{APACrefDOI} 10.1086/177565 \end{APACrefDOI}
\PrintBackRefs{\CurrentBib}

\bibitem [\protect \citeauthoryear {%
{Lin}%
\ \protect \BOthers {.}}{%
{Lin}%
\ \protect \BOthers {.}}{%
{\protect \APACyear {2024}}%
}]{%
lin24}
\APACinsertmetastar {%
lin24}%
\begin{APACrefauthors}%
{Lin}, Z.%
, {Jiang}, N.%
, {Wang}, T.%
\ et al.\end{APACrefauthors}%
\unskip\
\newblock
\APACrefYearMonthDay{2024}{{\APACmonth{08}}}{},
\newblock
\unskip
\newblock
\APACjournalVolNumPages{\apjl}{971}{1}{L26}.
\newblock
\begin{APACrefDOI} 10.3847/2041-8213/ad638e \end{APACrefDOI}
\PrintBackRefs{\CurrentBib}

\bibitem [\protect \citeauthoryear {%
{Linial}%
\ \BBA {} {Metzger}%
}{%
{Linial}%
\ \BBA {} {Metzger}%
}{%
{\protect \APACyear {2023}}%
}]{%
linial23}
\APACinsertmetastar {%
linial23}%
\begin{APACrefauthors}%
{Linial}, I.%
\BCBT {}\ \BBA {} {Metzger}, B\BPBI D.%
\end{APACrefauthors}%
\unskip\
\newblock
\APACrefYearMonthDay{2023}{{\APACmonth{11}}}{},
\newblock
\unskip
\newblock
\APACjournalVolNumPages{\apj}{957}{1}{34}.
\newblock
\begin{APACrefDOI} 10.3847/1538-4357/acf65b \end{APACrefDOI}
\PrintBackRefs{\CurrentBib}

\bibitem [\protect \citeauthoryear {%
{Linial}%
\ \BBA {} {Quataert}%
}{%
{Linial}%
\ \BBA {} {Quataert}%
}{%
{\protect \APACyear {2024}}%
}]{%
linial24}
\APACinsertmetastar {%
linial24}%
\begin{APACrefauthors}%
{Linial}, I.%
\BCBT {}\ \BBA {} {Quataert}, E.%
\end{APACrefauthors}%
\unskip\
\newblock
\APACrefYearMonthDay{2024}{{\APACmonth{01}}}{},
\newblock
\unskip
\newblock
\APACjournalVolNumPages{\mnras}{527}{2}{4317-4329}.
\newblock
\begin{APACrefDOI} 10.1093/mnras/stad3470 \end{APACrefDOI}
\PrintBackRefs{\CurrentBib}

\bibitem [\protect \citeauthoryear {%
{Liu}%
\ \protect \BOthers {.}}{%
{Liu}%
\ \protect \BOthers {.}}{%
{\protect \APACyear {2023}}%
}]{%
liu23}
\APACinsertmetastar {%
liu23}%
\begin{APACrefauthors}%
{Liu}, C.%
, {Mockler}, B.%
, {Ramirez-Ruiz}, E.%
\ et al.\end{APACrefauthors}%
\unskip\
\newblock
\APACrefYearMonthDay{2023}{{\APACmonth{02}}}{},
\newblock
\unskip
\newblock
\APACjournalVolNumPages{\apj}{944}{2}{184}.
\newblock
\begin{APACrefDOI} 10.3847/1538-4357/acafe1 \end{APACrefDOI}
\PrintBackRefs{\CurrentBib}

\bibitem [\protect \citeauthoryear {%
{Liu}%
, {Yarza}%
\BCBL {}\ \BBA {} {Ramirez-Ruiz}%
}{%
{Liu}%
\ \protect \BOthers {.}}{%
{\protect \APACyear {2025}}%
}]{%
liu25}
\APACinsertmetastar {%
liu25}%
\begin{APACrefauthors}%
{Liu}, C.%
, {Yarza}, R.%
\BCBL {}\ \BBA {} {Ramirez-Ruiz}, E.%
\end{APACrefauthors}%
\unskip\
\newblock
\APACrefYearMonthDay{2025}{{\APACmonth{01}}}{},
\newblock
\unskip
\newblock
\APACjournalVolNumPages{\apj}{979}{1}{40}.
\newblock
\begin{APACrefDOI} 10.3847/1538-4357/ad9b0b \end{APACrefDOI}
\PrintBackRefs{\CurrentBib}

\bibitem [\protect \citeauthoryear {%
{Liu}%
\ \protect \BOthers {.}}{%
{Liu}%
\ \protect \BOthers {.}}{%
{\protect \APACyear {2024}}%
}]{%
liu24}
\APACinsertmetastar {%
liu24}%
\begin{APACrefauthors}%
{Liu}, Z.%
, {Ryu}, T.%
, {Goodwin}, A\BPBI J.%
\ et al.\end{APACrefauthors}%
\unskip\
\newblock
\APACrefYearMonthDay{2024}{{\APACmonth{03}}}{},
\newblock
\unskip
\newblock
\APACjournalVolNumPages{\aap}{683}{}{L13}.
\newblock
\begin{APACrefDOI} 10.1051/0004-6361/202348682 \end{APACrefDOI}
\PrintBackRefs{\CurrentBib}

\bibitem [\protect \citeauthoryear {%
{Lu}%
\ \BBA {} {Quataert}%
}{%
{Lu}%
\ \BBA {} {Quataert}%
}{%
{\protect \APACyear {2023}}%
}]{%
lu23}
\APACinsertmetastar {%
lu23}%
\begin{APACrefauthors}%
{Lu}, W.%
\BCBT {}\ \BBA {} {Quataert}, E.%
\end{APACrefauthors}%
\unskip\
\newblock
\APACrefYearMonthDay{2023}{{\APACmonth{10}}}{},
\newblock
\unskip
\newblock
\APACjournalVolNumPages{\mnras}{524}{4}{6247-6266}.
\newblock
\begin{APACrefDOI} 10.1093/mnras/stad2203 \end{APACrefDOI}
\PrintBackRefs{\CurrentBib}

\bibitem [\protect \citeauthoryear {%
{Makrygianni}%
\ \protect \BOthers {.}}{%
{Makrygianni}%
\ \protect \BOthers {.}}{%
{\protect \APACyear {2025}}%
}]{%
makrygianni25}
\APACinsertmetastar {%
makrygianni25}%
\begin{APACrefauthors}%
{Makrygianni}, L.%
, {Arcavi}, I.%
, {Newsome}, M.%
\ et al.\end{APACrefauthors}%
\unskip\
\newblock
\APACrefYearMonthDay{2025}{{\APACmonth{07}}}{},
\newblock
\unskip
\newblock
\APACjournalVolNumPages{\apjl}{987}{1}{L20}.
\newblock
\begin{APACrefDOI} 10.3847/2041-8213/ade155 \end{APACrefDOI}
\PrintBackRefs{\CurrentBib}

\bibitem [\protect \citeauthoryear {%
{McMillan}%
, {McDermott}%
\BCBL {}\ \BBA {} {Taam}%
}{%
{McMillan}%
\ \protect \BOthers {.}}{%
{\protect \APACyear {1987}}%
}]{%
mcmillan87}
\APACinsertmetastar {%
mcmillan87}%
\begin{APACrefauthors}%
{McMillan}, S\BPBI L\BPBI W.%
, {McDermott}, P\BPBI N.%
\BCBL {}\ \BBA {} {Taam}, R\BPBI E.%
\end{APACrefauthors}%
\unskip\
\newblock
\APACrefYearMonthDay{1987}{{\APACmonth{07}}}{},
\newblock
\unskip
\newblock
\APACjournalVolNumPages{\apj}{318}{}{261}.
\newblock
\begin{APACrefDOI} 10.1086/165365 \end{APACrefDOI}
\PrintBackRefs{\CurrentBib}

\bibitem [\protect \citeauthoryear {%
{Miles}%
, {Coughlin}%
\BCBL {}\ \BBA {} {Nixon}%
}{%
{Miles}%
\ \protect \BOthers {.}}{%
{\protect \APACyear {2020}}%
}]{%
miles20}
\APACinsertmetastar {%
miles20}%
\begin{APACrefauthors}%
{Miles}, P\BPBI R.%
, {Coughlin}, E\BPBI R.%
\BCBL {}\ \BBA {} {Nixon}, C\BPBI J.%
\end{APACrefauthors}%
\unskip\
\newblock
\APACrefYearMonthDay{2020}{{\APACmonth{08}}}{},
\newblock
\unskip
\newblock
\APACjournalVolNumPages{\apj}{899}{1}{36}.
\newblock
\begin{APACrefDOI} 10.3847/1538-4357/ab9c9f \end{APACrefDOI}
\PrintBackRefs{\CurrentBib}

\bibitem [\protect \citeauthoryear {%
{Miniutti}%
, {Giustini}%
, {Arcodia}%
, {Saxton}%
, {Chakraborty}%
\BCBL {}\ \protect \BOthers {.}}{%
{Miniutti}%
, {Giustini}%
, {Arcodia}%
, {Saxton}%
, {Chakraborty}%
\BCBL {}\ \protect \BOthers {.}}{%
{\protect \APACyear {2023}}%
}]{%
miniutti23}
\APACinsertmetastar {%
miniutti23}%
\begin{APACrefauthors}%
{Miniutti}, G.%
, {Giustini}, M.%
, {Arcodia}, R.%
, {Saxton}, R\BPBI D.%
, {Chakraborty}, J.%
, {Read}, A\BPBI M.%
\BCBL {}\ \BBA {} {Kara}, E.%
\end{APACrefauthors}%
\unskip\
\newblock
\APACrefYearMonthDay{2023}{{\APACmonth{06}}}{},
\newblock
\unskip
\newblock
\APACjournalVolNumPages{\aap}{674}{}{L1}.
\newblock
\begin{APACrefDOI} 10.1051/0004-6361/202346653 \end{APACrefDOI}
\PrintBackRefs{\CurrentBib}

\bibitem [\protect \citeauthoryear {%
{Miniutti}%
, {Giustini}%
, {Arcodia}%
, {Saxton}%
, {Read}%
\BCBL {}\ \protect \BOthers {.}}{%
{Miniutti}%
, {Giustini}%
, {Arcodia}%
, {Saxton}%
, {Read}%
\BCBL {}\ \protect \BOthers {.}}{%
{\protect \APACyear {2023}}%
}]{%
miniutti23b}
\APACinsertmetastar {%
miniutti23b}%
\begin{APACrefauthors}%
{Miniutti}, G.%
, {Giustini}, M.%
, {Arcodia}, R.%
, {Saxton}, R\BPBI D.%
, {Read}, A\BPBI M.%
, {Bianchi}, S.%
\BCBL {}\ \BBA {} {Alexander}, K\BPBI D.%
\end{APACrefauthors}%
\unskip\
\newblock
\APACrefYearMonthDay{2023}{{\APACmonth{02}}}{},
\newblock
\unskip
\newblock
\APACjournalVolNumPages{\aap}{670}{}{A93}.
\newblock
\begin{APACrefDOI} 10.1051/0004-6361/202244512 \end{APACrefDOI}
\PrintBackRefs{\CurrentBib}

\bibitem [\protect \citeauthoryear {%
{Miniutti}%
\ \protect \BOthers {.}}{%
{Miniutti}%
\ \protect \BOthers {.}}{%
{\protect \APACyear {2019}}%
}]{%
miniutti19}
\APACinsertmetastar {%
miniutti19}%
\begin{APACrefauthors}%
{Miniutti}, G.%
, {Saxton}, R\BPBI D.%
, {Giustini}, M.%
\ et al.\end{APACrefauthors}%
\unskip\
\newblock
\APACrefYearMonthDay{2019}{{\APACmonth{09}}}{},
\newblock
\unskip
\newblock
\APACjournalVolNumPages{\nat}{573}{7774}{381-384}.
\newblock
\begin{APACrefDOI} 10.1038/s41586-019-1556-x \end{APACrefDOI}
\PrintBackRefs{\CurrentBib}

\bibitem [\protect \citeauthoryear {%
{Nicholl}%
\ \protect \BOthers {.}}{%
{Nicholl}%
\ \protect \BOthers {.}}{%
{\protect \APACyear {2024}}%
}]{%
nicholl24}
\APACinsertmetastar {%
nicholl24}%
\begin{APACrefauthors}%
{Nicholl}, M.%
, {Pasham}, D\BPBI R.%
, {Mummery}, A.%
\ et al.\end{APACrefauthors}%
\unskip\
\newblock
\APACrefYearMonthDay{2024}{{\APACmonth{10}}}{},
\newblock
\unskip
\newblock
\APACjournalVolNumPages{\nat}{634}{8035}{804-808}.
\newblock
\begin{APACrefDOI} 10.1038/s41586-024-08023-6 \end{APACrefDOI}
\PrintBackRefs{\CurrentBib}

\bibitem [\protect \citeauthoryear {%
{Nicholl}%
\ \protect \BOthers {.}}{%
{Nicholl}%
\ \protect \BOthers {.}}{%
{\protect \APACyear {2020}}%
}]{%
nicholl20}
\APACinsertmetastar {%
nicholl20}%
\begin{APACrefauthors}%
{Nicholl}, M.%
, {Wevers}, T.%
, {Oates}, S\BPBI R.%
\ et al.\end{APACrefauthors}%
\unskip\
\newblock
\APACrefYearMonthDay{2020}{{\APACmonth{11}}}{},
\newblock
\unskip
\newblock
\APACjournalVolNumPages{\mnras}{499}{1}{482-504}.
\newblock
\begin{APACrefDOI} 10.1093/mnras/staa2824 \end{APACrefDOI}
\PrintBackRefs{\CurrentBib}

\bibitem [\protect \citeauthoryear {%
{Nixon}%
\ \BBA {} {Coughlin}%
}{%
{Nixon}%
\ \BBA {} {Coughlin}%
}{%
{\protect \APACyear {2022}}%
}]{%
nixon22}
\APACinsertmetastar {%
nixon22}%
\begin{APACrefauthors}%
{Nixon}, C\BPBI J.%
\BCBT {}\ \BBA {} {Coughlin}, E\BPBI R.%
\end{APACrefauthors}%
\unskip\
\newblock
\APACrefYearMonthDay{2022}{{\APACmonth{03}}}{},
\newblock
\unskip
\newblock
\APACjournalVolNumPages{\apjl}{927}{2}{L25}.
\newblock
\begin{APACrefDOI} 10.3847/2041-8213/ac5118 \end{APACrefDOI}
\PrintBackRefs{\CurrentBib}

\bibitem [\protect \citeauthoryear {%
{Nixon}%
, {Coughlin}%
\BCBL {}\ \BBA {} {Miles}%
}{%
{Nixon}%
\ \protect \BOthers {.}}{%
{\protect \APACyear {2021}}%
}]{%
nixon21}
\APACinsertmetastar {%
nixon21}%
\begin{APACrefauthors}%
{Nixon}, C\BPBI J.%
, {Coughlin}, E\BPBI R.%
\BCBL {}\ \BBA {} {Miles}, P\BPBI R.%
\end{APACrefauthors}%
\unskip\
\newblock
\APACrefYearMonthDay{2021}{{\APACmonth{12}}}{},
\newblock
\unskip
\newblock
\APACjournalVolNumPages{\apj}{922}{2}{168}.
\newblock
\begin{APACrefDOI} 10.3847/1538-4357/ac1bb8 \end{APACrefDOI}
\PrintBackRefs{\CurrentBib}

\bibitem [\protect \citeauthoryear {%
{Norman}%
, {Nixon}%
\BCBL {}\ \BBA {} {Coughlin}%
}{%
{Norman}%
\ \protect \BOthers {.}}{%
{\protect \APACyear {2021}}%
}]{%
norman21}
\APACinsertmetastar {%
norman21}%
\begin{APACrefauthors}%
{Norman}, S\BPBI M\BPBI J.%
, {Nixon}, C\BPBI J.%
\BCBL {}\ \BBA {} {Coughlin}, E\BPBI R.%
\end{APACrefauthors}%
\unskip\
\newblock
\APACrefYearMonthDay{2021}{{\APACmonth{12}}}{},
\newblock
\unskip
\newblock
\APACjournalVolNumPages{\apj}{923}{2}{184}.
\newblock
\begin{APACrefDOI} 10.3847/1538-4357/ac2ee8 \end{APACrefDOI}
\PrintBackRefs{\CurrentBib}

\bibitem [\protect \citeauthoryear {%
{O'Connor}%
\ \protect \BOthers {.}}{%
{O'Connor}%
\ \protect \BOthers {.}}{%
{\protect \APACyear {2025}}%
}]{%
oconnor25}
\APACinsertmetastar {%
oconnor25}%
\begin{APACrefauthors}%
{O'Connor}, B.%
, {Gill}, R.%
, {DeLaunay}, J.%
\ et al.\end{APACrefauthors}%
\unskip\
\newblock
\APACrefYearMonthDay{2025}{{\APACmonth{09}}}{},
\newblock
\unskip
\newblock
\APACjournalVolNumPages{arXiv e-prints}{}{}{arXiv:2509.22787}.
\newblock
\begin{APACrefDOI} 10.48550/arXiv.2509.22787 \end{APACrefDOI}
\PrintBackRefs{\CurrentBib}

\bibitem [\protect \citeauthoryear {%
{Pan}%
, {Li}%
, {Cao}%
, {Liu}%
\BCBL {}\ \BBA {} {Yuan}%
}{%
{Pan}%
\ \protect \BOthers {.}}{%
{\protect \APACyear {2025}}%
}]{%
pan25}
\APACinsertmetastar {%
pan25}%
\begin{APACrefauthors}%
{Pan}, X.%
, {Li}, S\BHBI L.%
, {Cao}, X.%
, {Liu}, B.%
\BCBL {}\ \BBA {} {Yuan}, W.%
\end{APACrefauthors}%
\unskip\
\newblock
\APACrefYearMonthDay{2025}{{\APACmonth{08}}}{},
\newblock
\unskip
\newblock
\APACjournalVolNumPages{\apj}{989}{2}{196}.
\newblock
\begin{APACrefDOI} 10.3847/1538-4357/adf05d \end{APACrefDOI}
\PrintBackRefs{\CurrentBib}

\bibitem [\protect \citeauthoryear {%
{Pasham}%
, {Coughlin}%
, {Nixon}%
\BCBL {}\ \protect \BOthers {.}}{%
{Pasham}%
, {Coughlin}%
, {Nixon}%
\BCBL {}\ \protect \BOthers {.}}{%
{\protect \APACyear {2024}}%
}]{%
pasham24b}
\APACinsertmetastar {%
pasham24b}%
\begin{APACrefauthors}%
{Pasham}, D.%
, {Coughlin}, E.%
, {Nixon}, C.%
\ et al.\end{APACrefauthors}%
\unskip\
\newblock
\APACrefYearMonthDay{2024}{{\APACmonth{11}}}{},
\newblock
\unskip
\newblock
\APACjournalVolNumPages{arXiv e-prints}{}{}{arXiv:2411.05948}.
\newblock
\begin{APACrefDOI} 10.48550/arXiv.2411.05948 \end{APACrefDOI}
\PrintBackRefs{\CurrentBib}

\bibitem [\protect \citeauthoryear {%
{Pasham}%
, {Coughlin}%
, {Guolo}%
\BCBL {}\ \protect \BOthers {.}}{%
{Pasham}%
, {Coughlin}%
, {Guolo}%
\BCBL {}\ \protect \BOthers {.}}{%
{\protect \APACyear {2024}}%
}]{%
pasham24}
\APACinsertmetastar {%
pasham24}%
\begin{APACrefauthors}%
{Pasham}, D.%
, {Coughlin}, E\BPBI R.%
, {Guolo}, M.%
, {Wevers}, T.%
, {Nixon}, C\BPBI J.%
, {Hinkle}, J\BPBI T.%
\BCBL {}\ \BBA {} {Bandopadhyay}, A.%
\end{APACrefauthors}%
\unskip\
\newblock
\APACrefYearMonthDay{2024}{{\APACmonth{08}}}{},
\newblock
\unskip
\newblock
\APACjournalVolNumPages{\apjl}{971}{2}{L31}.
\newblock
\begin{APACrefDOI} 10.3847/2041-8213/ad57b3 \end{APACrefDOI}
\PrintBackRefs{\CurrentBib}

\bibitem [\protect \citeauthoryear {%
{Pasham}%
, {Kejriwal}%
\BCBL {}\ \protect \BOthers {.}}{%
{Pasham}%
, {Kejriwal}%
\BCBL {}\ \protect \BOthers {.}}{%
{\protect \APACyear {2024}}%
}]{%
pasham24c}
\APACinsertmetastar {%
pasham24c}%
\begin{APACrefauthors}%
{Pasham}, D.%
, {Kejriwal}, S.%
, {Coughlin}, E.%
\ et al.\end{APACrefauthors}%
\unskip\
\newblock
\APACrefYearMonthDay{2024}{{\APACmonth{10}}}{},
\newblock
\unskip
\newblock
\APACjournalVolNumPages{arXiv e-prints}{}{}{arXiv:2411.00289}.
\newblock
\begin{APACrefDOI} 10.48550/arXiv.2411.00289 \end{APACrefDOI}
\PrintBackRefs{\CurrentBib}

\bibitem [\protect \citeauthoryear {%
{Pasham}%
, {Coughlin}%
, {van Velzen}%
\BCBL {}\ \BBA {} {Hinkle}%
}{%
{Pasham}%
\ \protect \BOthers {.}}{%
{\protect \APACyear {2025}}%
}]{%
pasham25}
\APACinsertmetastar {%
pasham25}%
\begin{APACrefauthors}%
{Pasham}, D\BPBI R.%
, {Coughlin}, E.%
, {van Velzen}, S.%
\BCBL {}\ \BBA {} {Hinkle}, J.%
\end{APACrefauthors}%
\unskip\
\newblock
\APACrefYearMonthDay{2025}{{\APACmonth{02}}}{},
\newblock
\unskip
\newblock
\APACjournalVolNumPages{arXiv e-prints}{}{}{arXiv:2502.12078}.
\newblock
\begin{APACrefDOI} 10.48550/arXiv.2502.12078 \end{APACrefDOI}
\PrintBackRefs{\CurrentBib}

\bibitem [\protect \citeauthoryear {%
{Paxton}%
\ \protect \BOthers {.}}{%
{Paxton}%
\ \protect \BOthers {.}}{%
{\protect \APACyear {2011}}%
}]{%
paxton11}
\APACinsertmetastar {%
paxton11}%
\begin{APACrefauthors}%
{Paxton}, B.%
, {Bildsten}, L.%
, {Dotter}, A.%
, {Herwig}, F.%
, {Lesaffre}, P.%
\BCBL {}\ \BBA {} {Timmes}, F.%
\end{APACrefauthors}%
\unskip\
\newblock
\APACrefYearMonthDay{2011}{{\APACmonth{01}}}{},
\newblock
\unskip
\newblock
\APACjournalVolNumPages{\apjs}{192}{1}{3}.
\newblock
\begin{APACrefDOI} 10.1088/0067-0049/192/1/3 \end{APACrefDOI}
\PrintBackRefs{\CurrentBib}

\bibitem [\protect \citeauthoryear {%
{Paxton}%
\ \protect \BOthers {.}}{%
{Paxton}%
\ \protect \BOthers {.}}{%
{\protect \APACyear {2013}}%
}]{%
paxton13}
\APACinsertmetastar {%
paxton13}%
\begin{APACrefauthors}%
{Paxton}, B.%
, {Cantiello}, M.%
, {Arras}, P.%
\ et al.\end{APACrefauthors}%
\unskip\
\newblock
\APACrefYearMonthDay{2013}{{\APACmonth{09}}}{},
\newblock
\unskip
\newblock
\APACjournalVolNumPages{\apjs}{208}{1}{4}.
\newblock
\begin{APACrefDOI} 10.1088/0067-0049/208/1/4 \end{APACrefDOI}
\PrintBackRefs{\CurrentBib}

\bibitem [\protect \citeauthoryear {%
{Paxton}%
\ \protect \BOthers {.}}{%
{Paxton}%
\ \protect \BOthers {.}}{%
{\protect \APACyear {2015}}%
}]{%
paxton15}
\APACinsertmetastar {%
paxton15}%
\begin{APACrefauthors}%
{Paxton}, B.%
, {Marchant}, P.%
, {Schwab}, J.%
\ et al.\end{APACrefauthors}%
\unskip\
\newblock
\APACrefYearMonthDay{2015}{{\APACmonth{09}}}{},
\newblock
\unskip
\newblock
\APACjournalVolNumPages{\apjs}{220}{1}{15}.
\newblock
\begin{APACrefDOI} 10.1088/0067-0049/220/1/15 \end{APACrefDOI}
\PrintBackRefs{\CurrentBib}

\bibitem [\protect \citeauthoryear {%
{Paxton}%
\ \protect \BOthers {.}}{%
{Paxton}%
\ \protect \BOthers {.}}{%
{\protect \APACyear {2018}}%
}]{%
paxton18}
\APACinsertmetastar {%
paxton18}%
\begin{APACrefauthors}%
{Paxton}, B.%
, {Schwab}, J.%
, {Bauer}, E\BPBI B.%
\ et al.\end{APACrefauthors}%
\unskip\
\newblock
\APACrefYearMonthDay{2018}{{\APACmonth{02}}}{},
\newblock
\unskip
\newblock
\APACjournalVolNumPages{\apjs}{234}{2}{34}.
\newblock
\begin{APACrefDOI} 10.3847/1538-4365/aaa5a8 \end{APACrefDOI}
\PrintBackRefs{\CurrentBib}

\bibitem [\protect \citeauthoryear {%
{Payne}%
\ \protect \BOthers {.}}{%
{Payne}%
\ \protect \BOthers {.}}{%
{\protect \APACyear {2022}}%
}]{%
payne22}
\APACinsertmetastar {%
payne22}%
\begin{APACrefauthors}%
{Payne}, A\BPBI V.%
, {Shappee}, B\BPBI J.%
, {Hinkle}, J\BPBI T.%
\ et al.\end{APACrefauthors}%
\unskip\
\newblock
\APACrefYearMonthDay{2022}{{\APACmonth{02}}}{},
\newblock
\unskip
\newblock
\APACjournalVolNumPages{\apj}{926}{2}{142}.
\newblock
\begin{APACrefDOI} 10.3847/1538-4357/ac480c \end{APACrefDOI}
\PrintBackRefs{\CurrentBib}

\bibitem [\protect \citeauthoryear {%
{Payne}%
\ \protect \BOthers {.}}{%
{Payne}%
\ \protect \BOthers {.}}{%
{\protect \APACyear {2021}}%
}]{%
payne21}
\APACinsertmetastar {%
payne21}%
\begin{APACrefauthors}%
{Payne}, A\BPBI V.%
, {Shappee}, B\BPBI J.%
, {Hinkle}, J\BPBI T.%
\ et al.\end{APACrefauthors}%
\unskip\
\newblock
\APACrefYearMonthDay{2021}{{\APACmonth{04}}}{},
\newblock
\unskip
\newblock
\APACjournalVolNumPages{\apj}{910}{2}{125}.
\newblock
\begin{APACrefDOI} 10.3847/1538-4357/abe38d \end{APACrefDOI}
\PrintBackRefs{\CurrentBib}

\bibitem [\protect \citeauthoryear {%
{Peters}%
}{%
{Peters}%
}{%
{\protect \APACyear {1964}}%
}]{%
peters64}
\APACinsertmetastar {%
peters64}%
\begin{APACrefauthors}%
{Peters}, P\BPBI C.%
\end{APACrefauthors}%
\unskip\
\newblock
\APACrefYearMonthDay{1964}{{\APACmonth{11}}}{},
\newblock
\unskip
\newblock
\APACjournalVolNumPages{Physical Review}{136}{4B}{1224-1232}.
\newblock
\begin{APACrefDOI} 10.1103/PhysRev.136.B1224 \end{APACrefDOI}
\PrintBackRefs{\CurrentBib}

\bibitem [\protect \citeauthoryear {%
{Press}%
\ \BBA {} {Teukolsky}%
}{%
{Press}%
\ \BBA {} {Teukolsky}%
}{%
{\protect \APACyear {1977}}%
}]{%
press77}
\APACinsertmetastar {%
press77}%
\begin{APACrefauthors}%
{Press}, W\BPBI H.%
\BCBT {}\ \BBA {} {Teukolsky}, S\BPBI A.%
\end{APACrefauthors}%
\unskip\
\newblock
\APACrefYearMonthDay{1977}{{\APACmonth{04}}}{},
\newblock
\unskip
\newblock
\APACjournalVolNumPages{\apj}{213}{}{183-192}.
\newblock
\begin{APACrefDOI} 10.1086/155143 \end{APACrefDOI}
\PrintBackRefs{\CurrentBib}

\bibitem [\protect \citeauthoryear {%
{Price}%
\ \protect \BOthers {.}}{%
{Price}%
\ \protect \BOthers {.}}{%
{\protect \APACyear {2018}}%
}]{%
price18}
\APACinsertmetastar {%
price18}%
\begin{APACrefauthors}%
{Price}, D\BPBI J.%
, {Wurster}, J.%
, {Tricco}, T\BPBI S.%
\ et al.\end{APACrefauthors}%
\unskip\
\newblock
\APACrefYearMonthDay{2018}{{\APACmonth{09}}}{},
\newblock
\unskip
\newblock
\APACjournalVolNumPages{\pasa}{35}{}{e031}.
\newblock
\begin{APACrefDOI} 10.1017/pasa.2018.25 \end{APACrefDOI}
\PrintBackRefs{\CurrentBib}

\bibitem [\protect \citeauthoryear {%
{Quintin}%
\ \protect \BOthers {.}}{%
{Quintin}%
\ \protect \BOthers {.}}{%
{\protect \APACyear {2023}}%
}]{%
quintin23}
\APACinsertmetastar {%
quintin23}%
\begin{APACrefauthors}%
{Quintin}, E.%
, {Webb}, N\BPBI A.%
, {Guillot}, S.%
\ et al.\end{APACrefauthors}%
\unskip\
\newblock
\APACrefYearMonthDay{2023}{{\APACmonth{07}}}{},
\newblock
\unskip
\newblock
\APACjournalVolNumPages{\aap}{675}{}{A152}.
\newblock
\begin{APACrefDOI} 10.1051/0004-6361/202346440 \end{APACrefDOI}
\PrintBackRefs{\CurrentBib}

\bibitem [\protect \citeauthoryear {%
{Raj}%
\ \BBA {} {Nixon}%
}{%
{Raj}%
\ \BBA {} {Nixon}%
}{%
{\protect \APACyear {2021}}%
}]{%
raj21}
\APACinsertmetastar {%
raj21}%
\begin{APACrefauthors}%
{Raj}, A.%
\BCBT {}\ \BBA {} {Nixon}, C\BPBI J.%
\end{APACrefauthors}%
\unskip\
\newblock
\APACrefYearMonthDay{2021}{{\APACmonth{03}}}{},
\newblock
\unskip
\newblock
\APACjournalVolNumPages{\apj}{909}{1}{82}.
\newblock
\begin{APACrefDOI} 10.3847/1538-4357/abdc25 \end{APACrefDOI}
\PrintBackRefs{\CurrentBib}

\bibitem [\protect \citeauthoryear {%
{Ray}%
, {Kembhavi}%
\BCBL {}\ \BBA {} {Antia}%
}{%
{Ray}%
\ \protect \BOthers {.}}{%
{\protect \APACyear {1987}}%
}]{%
ray87}
\APACinsertmetastar {%
ray87}%
\begin{APACrefauthors}%
{Ray}, A.%
, {Kembhavi}, A\BPBI K.%
\BCBL {}\ \BBA {} {Antia}, H\BPBI M.%
\end{APACrefauthors}%
\unskip\
\newblock
\APACrefYearMonthDay{1987}{{\APACmonth{10}}}{},
\newblock
\unskip
\newblock
\APACjournalVolNumPages{\aap}{184}{1-2}{164-172}.
\PrintBackRefs{\CurrentBib}

\bibitem [\protect \citeauthoryear {%
{Rees}%
}{%
{Rees}%
}{%
{\protect \APACyear {1988}}%
}]{%
rees88}
\APACinsertmetastar {%
rees88}%
\begin{APACrefauthors}%
{Rees}, M\BPBI J.%
\end{APACrefauthors}%
\unskip\
\newblock
\APACrefYearMonthDay{1988}{{\APACmonth{06}}}{},
\newblock
\unskip
\newblock
\APACjournalVolNumPages{\nat}{333}{6173}{523-528}.
\newblock
\begin{APACrefDOI} 10.1038/333523a0 \end{APACrefDOI}
\PrintBackRefs{\CurrentBib}

\bibitem [\protect \citeauthoryear {%
{Sheng}%
\ \protect \BOthers {.}}{%
{Sheng}%
\ \protect \BOthers {.}}{%
{\protect \APACyear {2021}}%
}]{%
sheng21}
\APACinsertmetastar {%
sheng21}%
\begin{APACrefauthors}%
{Sheng}, Z.%
, {Wang}, T.%
, {Ferland}, G.%
, {Shu}, X.%
, {Yang}, C.%
, {Jiang}, N.%
\BCBL {}\ \BBA {} {Chen}, Y.%
\end{APACrefauthors}%
\unskip\
\newblock
\APACrefYearMonthDay{2021}{{\APACmonth{10}}}{},
\newblock
\unskip
\newblock
\APACjournalVolNumPages{\apjl}{920}{1}{L25}.
\newblock
\begin{APACrefDOI} 10.3847/2041-8213/ac2251 \end{APACrefDOI}
\PrintBackRefs{\CurrentBib}

\bibitem [\protect \citeauthoryear {%
{Shu}%
\ \protect \BOthers {.}}{%
{Shu}%
\ \protect \BOthers {.}}{%
{\protect \APACyear {2018}}%
}]{%
shu18}
\APACinsertmetastar {%
shu18}%
\begin{APACrefauthors}%
{Shu}, X\BPBI W.%
, {Wang}, S\BPBI S.%
, {Dou}, L\BPBI M.%
, {Jiang}, N.%
, {Wang}, J\BPBI X.%
\BCBL {}\ \BBA {} {Wang}, T\BPBI G.%
\end{APACrefauthors}%
\unskip\
\newblock
\APACrefYearMonthDay{2018}{{\APACmonth{04}}}{},
\newblock
\unskip
\newblock
\APACjournalVolNumPages{\apjl}{857}{2}{L16}.
\newblock
\begin{APACrefDOI} 10.3847/2041-8213/aaba17 \end{APACrefDOI}
\PrintBackRefs{\CurrentBib}

\bibitem [\protect \citeauthoryear {%
{Sniegowska}%
, {Czerny}%
, {Bon}%
\BCBL {}\ \BBA {} {Bon}%
}{%
{Sniegowska}%
\ \protect \BOthers {.}}{%
{\protect \APACyear {2020}}%
}]{%
sniegowska20}
\APACinsertmetastar {%
sniegowska20}%
\begin{APACrefauthors}%
{Sniegowska}, M.%
, {Czerny}, B.%
, {Bon}, E.%
\BCBL {}\ \BBA {} {Bon}, N.%
\end{APACrefauthors}%
\unskip\
\newblock
\APACrefYearMonthDay{2020}{{\APACmonth{09}}}{},
\newblock
\unskip
\newblock
\APACjournalVolNumPages{\aap}{641}{}{A167}.
\newblock
\begin{APACrefDOI} 10.1051/0004-6361/202038575 \end{APACrefDOI}
\PrintBackRefs{\CurrentBib}

\bibitem [\protect \citeauthoryear {%
{Somalwar}%
\ \protect \BOthers {.}}{%
{Somalwar}%
\ \protect \BOthers {.}}{%
{\protect \APACyear {2025}}%
}]{%
somalwar23}
\APACinsertmetastar {%
somalwar23}%
\begin{APACrefauthors}%
{Somalwar}, J\BPBI J.%
, {Ravi}, V.%
, {Yao}, Y.%
\ et al.\end{APACrefauthors}%
\unskip\
\newblock
\APACrefYearMonthDay{2025}{{\APACmonth{06}}}{},
\newblock
\unskip
\newblock
\APACjournalVolNumPages{\apj}{985}{2}{175}.
\newblock
\begin{APACrefDOI} 10.3847/1538-4357/adcc19 \end{APACrefDOI}
\PrintBackRefs{\CurrentBib}

\bibitem [\protect \citeauthoryear {%
{Sun}%
\ \protect \BOthers {.}}{%
{Sun}%
\ \protect \BOthers {.}}{%
{\protect \APACyear {2025}}%
}]{%
sun25}
\APACinsertmetastar {%
sun25}%
\begin{APACrefauthors}%
{Sun}, J.%
, {Guo}, H.%
, {Gu}, M.%
\ et al.\end{APACrefauthors}%
\unskip\
\newblock
\APACrefYearMonthDay{2025}{{\APACmonth{04}}}{},
\newblock
\unskip
\newblock
\APACjournalVolNumPages{\apj}{982}{2}{150}.
\newblock
\begin{APACrefDOI} 10.3847/1538-4357/adb724 \end{APACrefDOI}
\PrintBackRefs{\CurrentBib}

\bibitem [\protect \citeauthoryear {%
{Weinberg}%
, {Arras}%
, {Quataert}%
\BCBL {}\ \BBA {} {Burkart}%
}{%
{Weinberg}%
\ \protect \BOthers {.}}{%
{\protect \APACyear {2012}}%
}]{%
weinberg12}
\APACinsertmetastar {%
weinberg12}%
\begin{APACrefauthors}%
{Weinberg}, N\BPBI N.%
, {Arras}, P.%
, {Quataert}, E.%
\BCBL {}\ \BBA {} {Burkart}, J.%
\end{APACrefauthors}%
\unskip\
\newblock
\APACrefYearMonthDay{2012}{{\APACmonth{06}}}{},
\newblock
\unskip
\newblock
\APACjournalVolNumPages{\apj}{751}{2}{136}.
\newblock
\begin{APACrefDOI} 10.1088/0004-637X/751/2/136 \end{APACrefDOI}
\PrintBackRefs{\CurrentBib}

\bibitem [\protect \citeauthoryear {%
{Wevers}%
\ \protect \BOthers {.}}{%
{Wevers}%
\ \protect \BOthers {.}}{%
{\protect \APACyear {2023}}%
}]{%
wevers23}
\APACinsertmetastar {%
wevers23}%
\begin{APACrefauthors}%
{Wevers}, T.%
, {Coughlin}, E\BPBI R.%
, {Pasham}, D\BPBI R.%
\ et al.\end{APACrefauthors}%
\unskip\
\newblock
\APACrefYearMonthDay{2023}{{\APACmonth{01}}}{},
\newblock
\unskip
\newblock
\APACjournalVolNumPages{\apjl}{942}{2}{L33}.
\newblock
\begin{APACrefDOI} 10.3847/2041-8213/ac9f36 \end{APACrefDOI}
\PrintBackRefs{\CurrentBib}

\bibitem [\protect \citeauthoryear {%
{Yao}%
\ \BBA {} {Quataert}%
}{%
{Yao}%
\ \BBA {} {Quataert}%
}{%
{\protect \APACyear {2025}}%
}]{%
yao25}
\APACinsertmetastar {%
yao25}%
\begin{APACrefauthors}%
{Yao}, P\BPBI Z.%
\BCBT {}\ \BBA {} {Quataert}, E.%
\end{APACrefauthors}%
\unskip\
\newblock
\APACrefYearMonthDay{2025}{{\APACmonth{05}}}{},
\newblock
\unskip
\newblock
\APACjournalVolNumPages{arXiv e-prints}{}{}{arXiv:2505.10611}.
\newblock
\begin{APACrefDOI} 10.48550/arXiv.2505.10611 \end{APACrefDOI}
\PrintBackRefs{\CurrentBib}

\bibitem [\protect \citeauthoryear {%
{Zalamea}%
, {Menou}%
\BCBL {}\ \BBA {} {Beloborodov}%
}{%
{Zalamea}%
\ \protect \BOthers {.}}{%
{\protect \APACyear {2010}}%
}]{%
zalamea10}
\APACinsertmetastar {%
zalamea10}%
\begin{APACrefauthors}%
{Zalamea}, I.%
, {Menou}, K.%
\BCBL {}\ \BBA {} {Beloborodov}, A\BPBI M.%
\end{APACrefauthors}%
\unskip\
\newblock
\APACrefYearMonthDay{2010}{{\APACmonth{11}}}{},
\newblock
\unskip
\newblock
\APACjournalVolNumPages{\mnras}{409}{1}{L25-L29}.
\newblock
\begin{APACrefDOI} 10.1111/j.1745-3933.2010.00930.x \end{APACrefDOI}
\PrintBackRefs{\CurrentBib}

\end{thebibliography}
\end{document}